\RequirePackage{fix-cm}
\documentclass[usenatbib]{svjour3}   
\smartqed  
\usepackage{graphicx}
\usepackage{amssymb}
\usepackage{amsmath}
\usepackage{natbib}
\usepackage{appendix}
\usepackage{upgreek}
\usepackage[normalem]{ulem}
\usepackage{rotating}
\usepackage{footnote}
\usepackage[caption=false]{subfig}
\usepackage{wrapfig}
\usepackage{url}
\newcommand{\aofrb}{FRB~121102}
\def\cite{\citep}

\begin{document}

\title{Fast Radio Bursts}


\author{E. Petroff,
        J.W.T. Hessels \& 
        D.R. Lorimer
}


\institute{E.~Petroff \at
              University of Amsterdam / ASTRON \\
              \email{e.b.petroff@uva.nl}           
\and
J.~W.~T.~Hessels \at
University of Amsterdam / ASTRON
\and        
D.~R.~Lorimer \at
West Virginia University
}

\date{Received: date / Accepted: date}

\maketitle

\begin{abstract}
The discovery of radio pulsars over a half century ago was a seminal moment in astronomy. It demonstrated the existence of neutron stars, gave a powerful observational tool to study them, and has allowed us to probe strong gravity, dense matter, and the interstellar medium.  More recently, pulsar surveys have led to the serendipitous discovery of fast radio bursts (FRBs).  While FRBs appear similar to the individual pulses from pulsars, their large dispersive delays suggest that they originate from far outside the Milky Way and hence are many orders-of-magnitude more luminous.  While most FRBs appear to be one-off, perhaps cataclysmic events, two sources are now known to repeat and thus clearly have a longer-lived central engine.  Beyond understanding how they are created, there is also the prospect of using FRBs -- as with pulsars -- to probe the extremes of the Universe as well as the otherwise invisible intervening medium.  Such studies will be aided by the high implied all-sky event rate: there is a detectable FRB roughly once every minute occurring somewhere on the sky.  The fact that less than a hundred FRB sources have been discovered in the last decade is largely due to the small fields-of-view of current radio telescopes.  A new generation of wide-field instruments is now coming online, however, and these will be capable of detecting multiple FRBs per day.  We are thus on the brink of further breakthroughs in the short-duration radio transient phase space, which will be critical for differentiating between the many proposed theories for the origin of FRBs.  In this review, we give an observational and theoretical introduction at a level that is accessible to astronomers entering the field.

\end{abstract}

\setcounter{tocdepth}{3}
\tableofcontents

\section{Introduction}\label{sec:intro}

Astrophysical transients are events that appear and disappear on human-observable timescales, and are produced in a wide variety of physical processes. Longer-duration transients, on timescales of hours to decades, like fading supernovae, can emit incoherently from thermal electrons. Short-duration transients, however, with emission on timescales of seconds or less, are necessarily coherent in nature since the emission is too bright to be explained by individual electrons emitting separately. Whereas variable sources are characterized by occasional brightening and fading, often superimposed on a stable flux source, transients are often one-off events that fade when the emission mechanism turns off.
The processes that produce both fast and slow transients are some of the most energetic in the Universe.
The collapse of a massive star \citep{2014ARA&A..52..487S}, or the collision of two neutron stars \citep{2017ApJ...850L..39A}, injects massive amounts of energy and material into the surrounding environment, producing heavy elements and seeding further star formation in galaxies. These violent processes emit across the electromagnetic spectrum on various timescales -- from a few seconds of coherent gamma-ray emission from gamma-ray bursts \citep[GRBs;][]{2009ARA&A..47..567G} to the sometimes years-long incoherent thermal radio emission from expanding material after a supernova explosion or GRB \citep{2012ApJ...746..156C}. Binary neutron star mergers can now also be observed through gravitational radiation \citep{2017PhRvL.119p1101A}. The energetic remnants of stellar explosions such as neutron stars are also known to produce millisecond-duration radio pulses \citep{1968Natur.217..709H}. Studies of fast transients can provide new windows on the processes that fuel galaxy evolution \citep{2017PhRvL.119p1101A}, and the compact stellar remnants left behind \citep{1985MNRAS.214P...5H,2001MNRAS.321...67L}. Within this context, it is no surprise that the discovery of fast radio bursts (FRBs), bright and seemingly extragalactic radio pulses, in 2007 \citep{2007Sci...318..777L} presented a tantalizing opportunity to the astronomical community as a potential new window on energetic extragalactic processes. 

FRBs are one of the most exciting new mysteries of astrophysics. They are bright (50~mJy--100~Jy) pulses of emission at radio frequencies, with durations of order milliseconds or less. FRB emission has so far been detected between 400~MHz and 8~GHz. The origins of FRBs are still unknown and at present the source class is only defined observationally. In the following we provide some background on the FRB phenomenon, compare the observed population to other types of known transients, and describe our motivation for this review and its contents.

\subsection{A brief history}\label{sec:history}

The existence of coherent, short-duration radio pulses was predicted at least as early as the 1970s -- both from expanding supernova shells combing surrounding material in other galaxies \citep{1971ApJ...165..509C,1975ApJ...198..439C} and from small annihilating black holes \citep{1977Natur.266..333R}. These theories motivated early searches by, e.g.,  \citeauthor{1979Natur.277..117P} in 1979, who re-purposed data from the Arecibo telescope to search for pulses as short as 16~ms. Although limited in bandwidth and time resolution, these data represented one of the first sensitive high-time-resolution searches for extragalactic radio pulses. No astrophysical radio pulses were detected in this search, but they placed some of the first sensitive upper limits on short-duration radio pulses from other galaxies.
 
Several decades later, the first detections of FRBs \citep{2007Sci...318..777L} were made in surveys for radio pulsars, rapidly rotating neutron stars that emit beams of radio emission from the open magnetic field lines at their magnetic poles \citep[see][for more details]{2012hpa..book.....L}. The stable but extreme magnetic fields associated with radio pulsars make them natural and long-lived particle accelerators that produce coherent radio emission through an as-yet poorly understood process \citep{2017RvMPP...1....5M}. As the neutron star rotates, the beams at the magnetic poles sweep across the sky and are observed as periodic radio pulses, each pulse lasting approximately 0.1--1000~ms. The radio pulses from pulsars also experience a frequency-dependent time delay through the ionized interstellar medium (ISM), which is quantified by a dispersion measure (DM) that is proportional to the number of free electrons along the line of sight (see \S \ref{sec:observed} and \S \ref{sec:propagation} for more details).  This is useful for measuring the ionized content of the ISM as well as for estimating the source distance. In addition to `canonical' radio pulsar emission, some pulsars are also known to produce sporadic `giant pulses' (GPs), which can be much shorter duration and have much higher peak luminosity. Pulsar GPs can be as short as a few nanoseconds \citep{2003Natur.422..141H} and have been attributed to focused coherent emission by bunches of charged particles in the pulsar beam or magnetosphere \citep{2016JPlPh..82c6302E}. 

The first pulsars were found through their bright, single pulses at the Mullard Radio Observatory in 1967 \citep{1968Natur.217..709H}, and for the first few years after their discovery, single-pulse studies allowed for further understanding of the pulsar phenomenon \citep{1970Natur.227..692B,1970Natur.228.1297B,1970Natur.228...42B,1975A&A....43..395B}. However, given the highly periodic nature of pulsar signals, searches were soon optimized to take advantage of this property. As early as 1969, only two years after the discovery of the first pulsar, Fast Fourier Transforms (FFTs) and Fast Folding Algorithms (FFAs) were recognized as more efficient for discovering periodic signals appearing at multiple harmonics in the frequency domain --- resulting in the discovery of a larger number of Galactic pulsars, with diverse properties \citep{1969A&A.....2..280B}. These searches allowed for the discovery of fainter periodic signals, pulsars with millisecond rotational periods \citep{1982Natur.300..615B}, and pulsars in relativistic binary systems \citep{1975ApJ...195L..51H}. Periodicity searches have been highly successful, increasing the total pulsar population from a few tens in the first few years \citep{1969ApL.....3..205T} to over 2600 sources in 2018\footnote{All published pulsars are available through the pulsar catalogue: http://www.atnf.csiro.au/people/pulsar/psrcat/ \citep{2005AJ....129.1993M}}.

Modern surveys search for pulsars via their periodic emission as well as their sporadic, bright single pulses.  These searches are also well suited to FRB discovery due to their large time on sky and high time resolution, both of which are necessary for finding new and potentially rapidly rotating pulsars. The drive to find more millisecond pulsars (MSPs) pushed instrumentation towards the narrower frequency channels and higher time resolution required to find their signatures in the data. Improved frequency resolution in pulsar surveys also allowed more sensitive single-pulse searches up to higher DM values, including to DMs much larger than expected from the Galactic column of free electrons. Throughout the past 50 years, each new pulsar search has attempted to expand the phase space in which we search for new pulsars, expanding coverage along the axes of pulse duration, DM, duty cycle, spectrum, and acceleration in the case of pulsars in binary orbits. 

As many new pulsar searches focused on finding stable periodic sources, the parameter space of short-duration single event transients remained relatively unexplored. The study of the single pulses of known pulsars continued as an active area of research \citep[for a review, see][]{2003A&ARv..12...43R}. However, blind searches for new pulsars through their single pulses tapered off. Following a successful search for
single pulses in archival Arecibo data by \citet{1999ApJ...513..927N}, a return to
the single pulse search space was motivated by \citet{2003ApJ...596.1142C} and \citet{2003ApJ...596..982M}. In an effort to explore this parameter space within the Galaxy, \citet{2006Natur.439..817M} discovered 11 new sources identified through their bright, millisecond-duration radio pulses. These rotating radio transients (RRATs) were believed to be a subset of the radio pulsar population. Although RRATs had underlying periodicity, they were more readily discovered through single pulse searches, rather than through FFTs. Current observations probe only the tip of the pulse energy distribution \citep{2006ApJ...645L.149W} and some sources could be extreme examples of pulsars that exhibit various types of variable emission such as nulling, mode changing, and intermittency, as well as GPs. The first RRATs implied that a large population of bright single pulses might be hiding in existing radio survey data \citep{2011MNRAS.415.3065K}. 

\begin{figure}
\centering
\includegraphics[width=0.8\columnwidth]{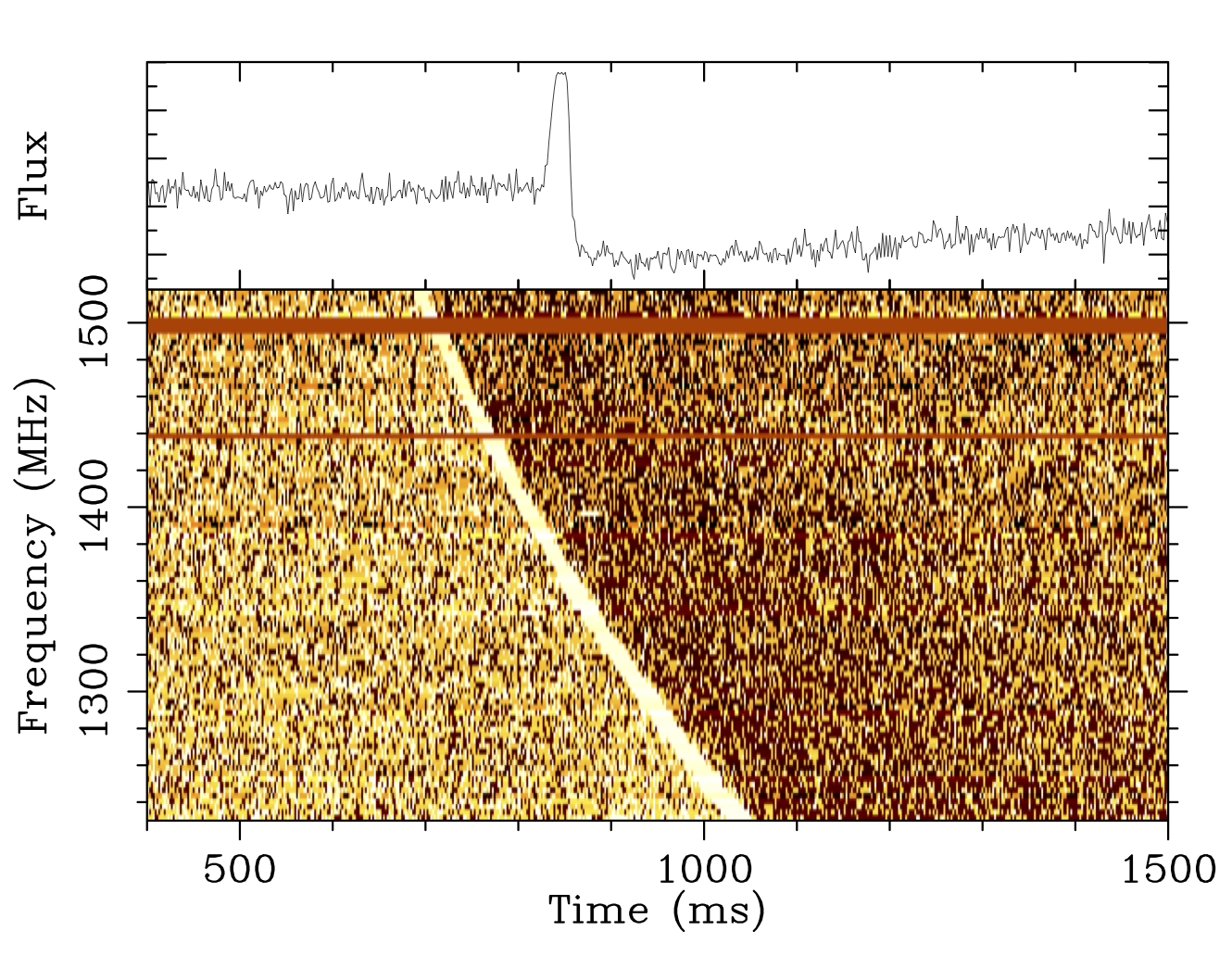}
\caption{The Lorimer burst \citep[][now also known as FRB~010724]{2007Sci...318..777L}, as seen in the beam of the Parkes multibeam receiver where it appeared brightest. These data have been one-bit digitized and contain 96 frequency channels sampled every millisecond. The burst has a DM of 375~cm$^{-3}$~pc. The pulse was so bright that it saturated the detector, causing a dip below the nominal baseline of the noise right after the pulse occurred. This signal was also detected in 3 other beams of the receiver.  The top panel shows the burst as summed across all recorded frequencies.  The bottom panel is the burst as a function of frequency and time (a `dynamic spectrum').  The red horizontal lines are frequency channels that have been excised because they are corrupted by RFI. \label{fig:LorimerBurst}}
\end{figure}

Single-pulse searches in archival data targeting the Small Magellanic Cloud (SMC), and taken with the Parkes telescope in 2001, revealed a single pulsar-like pulse, so bright it saturated the primary detection beam of the receiver and was originally estimated to have a peak flux density of $>30$~Jy \citep[Fig.~\ref{fig:LorimerBurst};][]{2007Sci...318..777L}. This pulse, which soon became known as the `Lorimer burst', was remarkable not only for its incredible brightness but also for its implied distance (see \S \ref{sec:LorimerBurst} for more details). The pulse's large dispersive delay was estimated to be roughly eight times greater than could be produced by the free electrons in the Milky Way (along this line of sight) or even in the circum-galactic medium occupying the space between the Milky Way and the SMC. Upon its discovery, the Lorimer burst suggested the existence of a population of bright, extragalactic radio pulses \citep{2007Sci...318..777L}. 

For several years after its discovery the Lorimer Burst remained the only known signal of its kind. A new pulse of potentially similar nature was discovered in 2011 by \citet{2011MNRAS.415.3065K}; however, this source was along a sight-line in the Galactic plane and thus a Galactic origin (like a RRAT) was also considered possible \citep[see \S \ref{sec:KeaneBurst}, and][]{2014MNRAS.440..353B}. Strong support in favor of the Lorimer burst as an astrophysical phenomenon came from \citet{2013Sci...341...53T}, who presented four high-DM pulses discovered in the High Time Resolution Universe survey at the Parkes telescope \citep[HTRU;][]{2010MNRAS.409..619K}.  The discoveries by \citet{2013Sci...341...53T} had similar characteristics to the Lorimer burst, and implied an all-sky population of extragalactic radio pulses, which they termed `Fast Radio Bursts', or FRBs. 

FRBs were immediately considered of great interest due to their large implied distances and the energies necessary to produce such bright pulses. As discussed further in \S \ref{sec:FRBproperties}, from the DMs of the four new FRB sources discovered by \citeauthor{2013Sci...341...53T} the bursts were estimated to have originated at distances as great as $z = 0.96$ (luminosity distance 6~Gpc). With peak flux densities of approximately 1~Jy, this implied an isotropic energy of $10^{32}$~J ($10^{39}$~erg) in a few milliseconds or a total power of 10$^{35}$~J~s$^{-1}$ ($10^{42}$~erg~s$^{-1}$).
The implied energies of these new FRBs were within a few orders of magnitude of those estimated for prompt emission from GRBs and supernova explosions, thereby leading to theories of cataclysmic and extreme progenitor mechanisms (see \S \ref{sec:progenitors}).

The excitement around the discovery by \citeauthor{2013Sci...341...53T} led to increased searches through new and archival data not just at the Parkes telescope \citep{2014ApJ...792...19B,2015ApJ...799L...5R,2016MNRAS.460L..30C}, but also at other telescopes around the world, resulting in FRB discoveries at the Arecibo Observatory \citep{2014ApJ...790..101S}, the Green Bank Telescope \citep[][]{2015Natur.528..523M}, the Upgraded Molonglo Synthesis Telescope \citep[UTMOST,][]{2016MNRAS.458..718C}, the Australian Square Kilometre Array Pathfinder \citep[ASKAP,][]{2017ApJ...841L..12B,2018Natur.562..386S}, and the Canadian Hydrogen Intensity Mapping Experiment \citep[CHIME,][]{2018ATel11901....1B,2019Natur.566..230C}. Since 2013, the discovery rate of FRBs has increased each year, with a doubling of the known population in the last 12-month period alone
\citep{2018Natur.562..386S,2019Natur.566..230C}.

Highlights from these discoveries have included the first two (so far) repeating FRB sources, FRB~121102 \citep{2016Natur.531..202S,2016ApJ...833..177S,2017Natur.541...58C} and FRB~180814.J0422+73
\citep{2019Natur.566..235C}, detections with interferometric techniques \citep{2016MNRAS.458..718C,2017ApJ...841L..12B,2017Natur.541...58C,2017ApJ...834L...8M}, and FRBs with measured polarization profiles \citep{2015MNRAS.447..246P,2015Natur.528..523M,2016Sci...354.1249R,2017MNRAS.469.4465P,2018Natur.553..182M,2018MNRAS.478.2046C}.

Searches through archival data in 2011 also revealed a peculiar class of artificial signal at Parkes that mimicked the dispersive sweep of a genuine astrophysical signal, but through multi-beam coincidence was thought to be local in origin \citep{2011ApJ...727...18B}. These signals, dubbed `Perytons', remained a curiosity and source of controversy in the field of FRBs for several years.  Because of the Perytons, some astronomers speculated that perhaps all FRBs were artificial in origin.  Further investigation of the Peryton phenomenon with a larger population of events and upgraded RFI monitoring at the Parkes telescope subsequently pinpointed their source to microwave ovens being operated at the site \citep{2015MNRAS.451.3933P}. Their identification as spurious RFI put the Peryton mystery to bed and allowed for further progress on the study of genuine astrophysical FRBs.

The discovery of FRBs as an observational class has also prompted re-examination of previously published transients surveys such as the reported discovery of highly dispersed radio pulses from M87 in the Virgo cluster in 1980 \citep{1980ApJ...236L.109L} and the 1989 sky survey with the Molonglo Observatory Synthesis Telescope by \citet{1989PASAu...8..172A}, which discovered an excess of non-terrestrial short-duration bursts ($1~\upmu$s to 1~ms) in 4000 hours of observations. These unexplained bursts showed no clustering in time or position and were not associated with known Galactic sources. Building on the searches by \citet{1979Natur.277..117P}, these may have been the first reported detections of FRBs; however, the limited bandwidth and time resolution of these instruments hampered further classification of the events. 

\subsection{The FRB population}\label{sec:FRBpopSummary}

\begin{figure}
\centering
\includegraphics[width=\columnwidth]{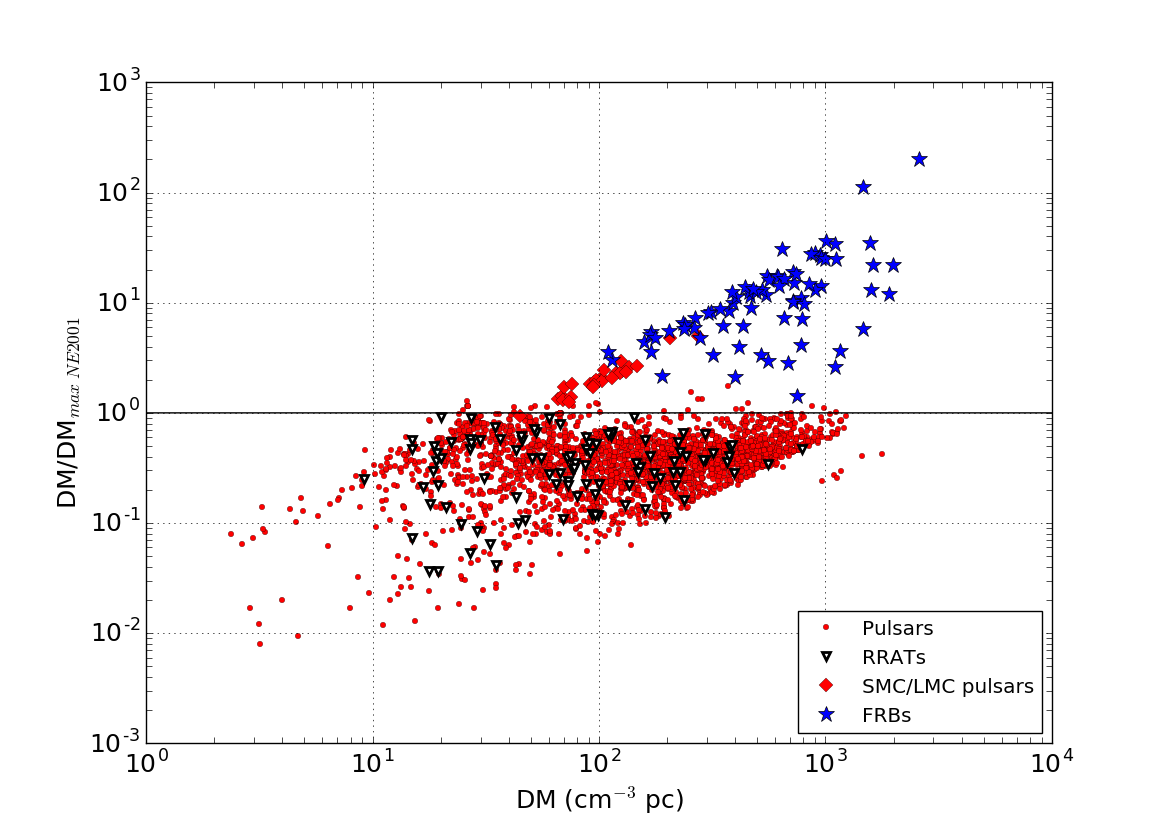}
\caption{The dispersion measures (DMs) of Galactic radio pulsars, Galactic rotating radio transients (RRATs), radio pulsars in the Small and Large Magellanic Clouds (SMC \& LMC), and published FRBs, relative to the modeled maximum Galactic DM along the line of sight from the NE2001 model \citep{2002astro.ph..7156C}. Sources with $\mathrm{DM}/\mathrm{DM}_\mathrm{max} > 1$ are thought to originate at extragalactic distances and accrue additional DM from the intergalactic medium and their host galaxy.  This figure is based on an earlier version presented in \citet{2014ApJ...790..101S}. 
\label{fig:DM_DMmax}}
\end{figure}

Currently, the research community has no strict and standard formalism for defining an FRB, although attempts to formalize FRB classification are ongoing \citep{2018MNRAS.481.2612F}.  In practice, we identify a signal as an FRB if it matches a set of loosely defined criteria.  These criteria include the pulse duration, brightness, and broadbandedness, and in particular whether the DM is larger than expected for a Galactic source.  For signals where the DM is close to the expected maximum Galactic contribution along the line of sight there is ambiguity as to whether the source is a Galactic pulsar/RRAT or an extragalactic FRB (Fig.~\ref{fig:DM_DMmax}).

As a population, FRBs have not yet been linked to any specific progenitors, although dozens of theories exist (see \citet{2018arXiv181005836P} and \S \ref{sec:progenitors}).  As of the writing of this review, the known population of FRBs consists of more than 60 independent sources detected at 10 telescopes and arrays around the world\footnote{All published FRBs are available via the FRB Catalogue (FRBCAT) \url{www.frbcat.org}.} \citep{2016PASA...33...45P}. The observed population spans a large range in DM, pulse duration, and peak flux density, as well as detected radio frequency. Two sources have been found to repeat \citep{2016Natur.531..202S,2019Natur.566..235C} and over 10 have now been discovered in real-time and followed up across the electromagnetic spectrum \citep{2015MNRAS.447..246P,2016Natur.530..453K,2017MNRAS.469.4465P,2018MNRAS.475.1427B}. The properties of the observed FRB population are discussed in \S \ref{sec:popFRBs}.

The estimated rate is roughly $\gtrsim 10^{3}$ FRBs detectable over the whole sky every day with large radio facilities \citep[e.g.][]{2016MNRAS.460L..30C}. Even for a cosmological distribution, if FRBs are generated in one-off cataclysmic events their sources must be relatively common and abundant. The redshift distribution is poorly known; however, the rate is higher than some sub-classes of supernovae, although lower than the overall core-collapse supernova (CCSN) rate by two orders of magnitude. A more detailed discussion of the FRB rate is presented in \S \ref{sec:IntrinsicPopDist}.

At the time of this review the progenitor(s) of FRBs remain unknown. Many theories link FRBs to known transient populations or to new phenomena not observable at other wavelengths.  Emission and progenitor theories are discussed in \S \ref{sec:emission} and \S \ref{sec:progenitors} \citep[see also][for a living catalog of theories]{2018arXiv181005836P}. 

\subsection{Motivation for this review}\label{sec:reviewMotivation}

Because of the rapid expansion of the research related to FRBs, and the many new discoveries reported each year, we feel that now is the ideal time for a review that covers these topics. The growing population of FRBs is also expected to bring a larger population of researchers to the field. We intend this review as a resource for researchers entering the field, as well as its growing list of practitioners. 

The timing of this review is such that we hope to encapsulate the field as it stands at the beginning of 2019, with close to a hundred sources discovered but many questions left unanswered. It is our hope that many questions related to the  origins and physics of FRBs will be understood as a larger population is discovered in the next few years with large instruments like CHIME, FAST, ASKAP, APERTIF, UTMOST and MeerKAT. These and many other telescopes are expected to cumulatively find hundreds of FRBs per year.

The outline of the remainder of this review is as follows: in \S \ref{sec:FRBproperties} we introduce the observed and derived properties of FRBs. In \S \ref{sec:propagation} we detail the propagation effects that act on an FRB as it travels through the intervening magnetized and ionized medium. In \S \ref{sec:obsTechniques} we summarize the current observational techniques used for finding FRBs, including search pipelines and single dish and interferometric methods. \S \ref{sec:indivFRBs} discusses some of the landmark FRB discoveries from the past decade. \S \ref{sec:popFRBs} discusses the FRB population in terms of the distributions of observed parameters such as width, DM, and sky position. In \S \ref{sec:IntrinsicPopDist} we extrapolate these observed distributions and speculate as to the intrinsic population distributions. \S \ref{sec:emission} details some of the proposed mechanisms for generating FRB emission, and \S \ref{sec:progenitors} more generally discusses the progenitor theories proposed for FRBs. We summarize the review in \S \ref{sec:summary} and conclude with predictions for the next five years in \S\ref{sec:predictions}.


\section{Properties of  FRBs}\label{sec:FRBproperties}

\begin{figure}[hbt]
\includegraphics[width=\textwidth]{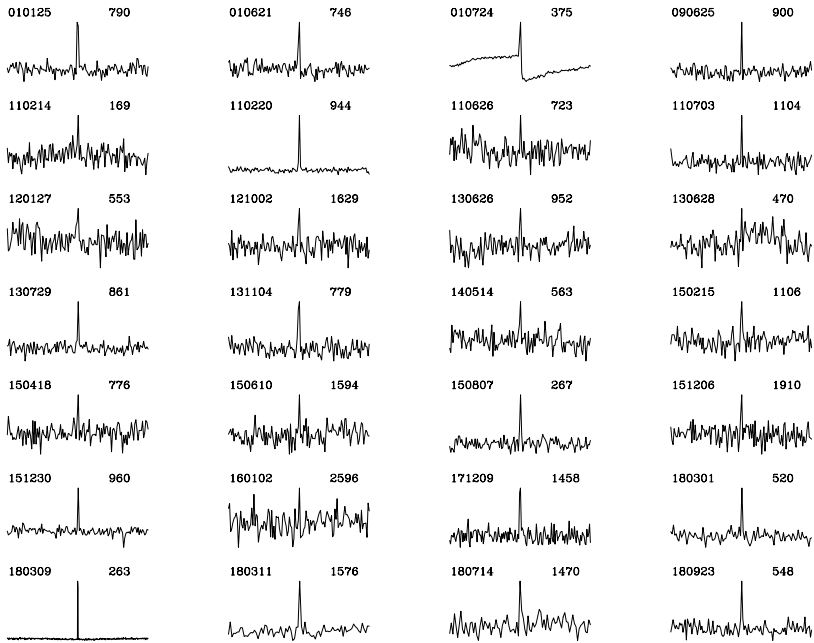}
\caption{\label{fig:profiles}
Compilation showing the first twenty-eight FRBs discovered using the Parkes telescope.  The detections are arranged in order of date. Each light curve shows a 2-s window around the pulse.  Following gamma-ray burst notation, the FRBs are named in YYMMDD format to indicate the year (YY) month (MM) and day (DD) on which the burst was detected. Also listed to the right of each pulse are the observed dispersion measures (DMs) in units of cm$^{-3}$~pc.
}
\end{figure}

Following an introduction to the observed properties of FRBs, we discuss some basic physical inferences that can be made from the most readily observable parameters. A selection of the current sample of FRBs is shown in Fig.\ref{fig:profiles}, which displays all those found with the Parkes 
telescope to date.

\subsection{Observed properties}\label{sec:observed}

The FRB search process is described in detail in \S \ref{sec:obsTechniques}. In brief, it consists of looking for dispersed pulses like the one shown in Fig.~\ref{fig:LorimerBurst} in radio astronomical data that is sampled in frequency and
time. Searches are most commonly done by forming
a large number of time series corresponding to
different amounts of dispersion over a wide 
range. The amount of dispersion is quantified
by the time delay of the pulse between the highest and lowest radio frequencies of the observation, $\nu_{\rm hi}$ and $\nu_{\rm lo}$ are the high, respectively, as
\begin{equation}
\label{equ:defdt}
 \Delta t = \frac{e^2}{2 \pi m_e c} 
  (\nu_{\rm lo}^{-2} - \nu_{\rm hi}^{-2}) \, \,
  {\rm DM} \approx 4.15 \, (\nu_{\rm lo}^{-2} - \nu_{\rm hi}^{-2}) \,\, {\rm DM}~{\rm ms}
\end{equation}
where $m_e$ is the mass of the electron, and $c$ is the speed of light. The second approximate equality holds when $\nu_{\rm lo}$ and $\nu_{\rm hi}$ are in units of GHz. The dispersion measure is given as
\begin{equation}
\label{eq:dm_def}
\mbox{DM} = \int_{0}^{d} n_e(l) \,{\rm d}l.
\end{equation}
In this expression, $n_e$ is the electron number density, $l$ is a path length
and $d$ is the distance to the FRB, which we will estimate below. Note that, as in pulsar astronomy, DM is typically quoted in units of cm$^{-3}$~pc.  This makes the numerical value of DM more easy to quote compared to using column density units of, e.g., cm$^{-2}$.  In practice, depending on the observational setup
and signal-to-noise ratio (S/N), the DM can be 
measured with a precision of about 0.1~cm$^{-3}$~pc.

The process for finding the optimum DM of a pulse is described in \S\ref{sec:pipelineSteps}.
Once the DM value has been optimised, a  
de-dispersed time series can be formed in which the
pulse S/N is maximized. If this time series
can be calibrated such that intensity can be converted to flux density as a function of time, $S(t)$, the pulse
can be characterized in terms of its width and peak
flux density, $S_{\rm peak}$. In practice, the
calibration process is approximated from a measurement of the root-mean-square (rms) fluctuations in the dedispersed time series,
$\sigma_S$. From radiometer noise considerations \citep[see, e.g.,][]{2012hpa..book.....L}, 
\begin{equation}
\sigma_S = \frac{T_{\rm sys}}{G \sqrt{2\, \Delta\nu \, t_{\rm samp}}},
\end{equation}
where $T_{\rm sys}$ is the system temperature,
$G$ is the antenna gain, $\Delta \nu$ is the receiver
bandwidth and $t_{\rm samp}$ is the data sampling interval.

For each FRB, the observed pulse width, $W$, is typically thought of as a combination of an intrinsic pulse of width $W_{\rm int}$ and instrumental broadening contributions.
In general, for a top-hat pulse,
\begin{equation}
W = \sqrt{W_{\rm int}^2 + t_{\rm samp}^2 +  \Delta t_{\rm DM}^2 + \Delta t_{\rm DMerr}^2 + \tau_{\rm s}^2},
\end{equation}
where $t_{\rm samp}$ is the sampling time as above, $\Delta t_{\rm DM}$ is the dispersive
delay across an individual frequency channel
and
$\Delta t_{\rm DM_{\rm err}}$ represents
the dispersive delay due to de-dispersion at
a slightly incorrect DM.
FRB pulses
can also be temporally broadened 
by multi-path propagation
through a turbulent medium. The
so called `scattering time-scale'
$\tau_{\rm s}$ due to this effect is discussed in detail in 
\S \ref{sec:scattering}. 

Pulse width is often measured at 50\% and 10\% of the peak \citep{2012hpa..book.....L}; however, for a pulse of arbitrary shape, it is also common 
to quote the equivalent width $W_{\rm eq}$ of
a top-hat pulse with the same $S_{\rm peak}$. Such a pulse has an  
energy or fluence
\begin{equation}
{\cal F}=S_{\rm peak} W_{\rm eq} = \int_{\rm pulse} S(t) \, {\rm d}t.
\end{equation}
A complicating factor with quoting flux density or fluence values is the fact that, for many FRBs, the true sky position is not known well enough to uniquely pinpoint the source to a position in the beam. Here, `beam' is defined as the field of view of the radio telescope, which is typically diffraction limited, as discussed more in \S\ref{sec:singleDish}. The sensitivity across this beam is not uniform, with the response as a function of angular distance from the center being approximately Gaussian, in most cases. As a result, with the exception of the ASKAP FRBs \citep{2017ApJ...841L..12B,2018Natur.562..386S}
most one-off FRB fluxes and fluences determined so far are lower limits. In addition, the limited angular resolution of most FRB searches so far leads to typical positional uncertainties that are on the order of a few arcminutes.

As is commonly done for other radio sources, measurements of the flux density spectrum of
FRBs as described by $S_{\nu} \propto \nu^{\alpha}$, where
$\alpha$ is the spectral index, are typically complicated by the
small available observing bandwidth. As a result, $\alpha$ is
usually rather poorly constrained. An 
additional complication also arises from
the poor localization of FRBs within the telescope beam, where the uncertain positional offset and variable beam response with radio frequency can lead to significant variations in measured
$\alpha$ values. We also note that a simple power-law spectral model may not be an optimal model of the intrinsic FRB emission process \citep[e.g.,][]{2018arXiv181110748H}. As discussed in \S \ref{sec:propagation}, the spectrum can also be modified by propagation effects.

One exception to these positional uncertainty
limitations
is the repeating source FRB~121102,
which is discussed further below (\S \ref{sec:FRB121102}). We
note here that flux density $S(t)$
defined above is the integral of the flux
per unit frequency interval over some
observing band from
$\nu_{\rm lo}$ to $\nu_{\rm hi}$. For the purposes of
the discussion below, and in the absence
of any spectral information, we assume
$\alpha=0$ so that
\begin{equation}
\label{eq:flatspec}
S(t) = \int_{\nu_{\rm lo}}^{\nu_{\rm hi}} S_{\nu} {\rm d}\nu = (\nu_{\rm hi}-\nu_{\rm lo}) S_{\nu}.
\end{equation}

For a few FRBs, measurements of polarized flux are also available \citep[see, e.g.,][]{2015MNRAS.447..246P,2015Natur.528..523M,2016Sci...354.1249R,2018Natur.553..182M}.
In these cases it is often possible to measure the change in the 
position angle of linear polarization, which
scales with wavelength squared. As discussed in \S \ref{sec:faraday}, the constant
of proportionality for this scaling is the
rotation measure (RM), which probes the
magnetic
field component 
along the line of sight,
weighted by electron density.

\subsection{Basic derived properties}\label{sec:derived}

For most FRBs, the only observables are position, flux density,
pulse width, and DM. We now provide the simplest set
of derived expressions that can be used to estimate relevant physical parameters for FRBs.

\subsubsection{Distance constraints}

Starting with the observed DM, we follow what is now tending towards standard practice
\citep[see, e.g.,][]{2014ApJ...783L..35D}
and define the dispersion measure
excess
\begin{equation}\label{eq:DMbreakdown}
{\rm DM}_{\rm E} = {\rm DM} - {\rm DM}_{\rm MW} = {\rm DM}_{\rm IGM} + \left(\frac{{\rm DM}_{\rm Host}}{1+z}\right),
\end{equation}
where DM$_{\rm MW}$ is the Galactic (i.e.~Milky Way) contribution from this
line of sight, typically obtained
from electron density models such as NE2001 \citep{2002astro.ph..7156C} or YMW16 \citep{2017ApJ...835...29Y}, DM$_{\rm IGM}$ is
the contribution from the 
intergalactic medium (IGM) and
DM$_{\rm Host}$ is the contribution from the host galaxy. The $(1+z)$ factor accounts for
cosmological time dilation for a source at redshift $z$.
The last term on the right-hand side of Eq.~\ref{eq:DMbreakdown} 
could be further broken down into host galaxy free electrons and local source terms, as needed. In any case, DM$_{\rm E}$ provides an upper limit for DM$_{\rm IGM}$, and most conservatively ${\rm DM}_{\rm IGM} < {\rm DM}_{\rm E}$. We note that DM$_{\rm MW}$ is likely uncertain at least at the tens of percent level, but could in rare cases be quite far off if there are unmodelled H\textsc{ii} regions along the line of sight \citep{2014MNRAS.440..353B}.

To find a relationship between DM and $z$, following, e.g.,
\citet{2014ApJ...783L..35D}, one can
assume all baryons are homogeneously distributed and ionized with an ionization fraction $x(z)$. In this case, 
the mean contribution from the IGM,
\begin{equation}
\langle\mathrm{DM_{IGM}}\rangle = \int n_{\rm e, IGM} \, dl
 = K_{\rm IGM} \int\limits_{0}^{z}  \frac{(1+z) x(z) \, dz}{\sqrt{\Omega_m (1+z)^3 + \Omega_{\Lambda}}},
\end{equation}
where the constant $K_{\rm IGM} = 933$~cm$^{-3}$~pc assumes standard
Planck cosmological parameters\footnote{For details, see Eq.~6 of Yang \& Zhang (2016).} and
a baryonic mass fraction of 83\%
\cite{2016ApJ...830L..31Y} and
$\Omega_m$ and $\Omega_{\Lambda}$ are,
respectively,
the energy densities of matter
and dark energy.
At low redshifts, the ionization fraction
$x(z) \simeq 7/8$, and we find \citep[see Fig.~1a of][]{2016ApJ...830L..31Y} $\mathrm{DM_{IGM}} \simeq z~1000$~cm$^{-3}$~pc. For
a given FRB with a particular observed DM, a very crude but commonly used rule of thumb is to estimate redshift as $z < {\rm DM}/1000$~cm$^{-3}$~pc.

Finally, to convert this redshift estimate to 
a luminosity distance, $d_L$, we can make use
of the approximation\footnote{This result is not widely used, but can be easily verified by numerical integration.} $d_L \simeq 2z(z+2.4)$~Gpc, which is valid for $z<1$.
In this case, for the most conservative
assumption, we find that
\begin{equation}
\label{equ:dl}
  d_L < \left(\frac{\rm DM}{500~{\rm cm}^{-3}~{\rm pc}}\right) \, \left[
  \left(\frac{\rm DM}{1000~{\rm cm}^{-3}~{\rm pc}}\right) + 2.4 \right] \, {\rm Gpc}.
\end{equation}
For the repeating FRB~121102, where $d_L$ can be 
inferred directly from the measured redshift of the host galaxy,
and constraints on dispersion in the host galaxy can be made, these expressions can be used instead to place constraints on DM$_{\rm IGM}$, as discussed in \S\ref{sec:FRB121102}.

\subsubsection{Source luminosity}

Having obtained a distance limit, for an FRB observed over
some bandwidth $\Delta \nu$, we can
place constraints on the isotropic
equivalent source luminosity
\begin{equation}
\label{equ:lum}
	L = \frac{4 \pi d_L^2 S_{\nu} \Delta \nu}{(1+z)}.
\end{equation}
In arriving at this expression, we
have started from the differential flux
per unit logarithmic frequency interval,
$S_{\nu} \Delta \nu$ \citep[see, e.g. Eq.~24 of][]{1999astro.ph..5116H} in the simplest 
case of a flat spectrum source (i.e.~constant
$S_{\nu}$, see Eq.~\ref{eq:flatspec}). The $(1+z)$ factor accounts
for the redshifting of the frequencies between
the source and observer frames. We 
also note that replacing $S_\nu$ with 
fluence ${\cal F}$ in the above expression yields the equivalent isotropic
energy release for a flat spectrum source.

As an example, we apply Eq.~\ref{equ:dl} to a typical FRB (FRB~140514) with a 
DM of 563~cm$^{-3}$~pc and a
peak flux density of 0.5~Jy. The limiting luminosity distance $d_L<3.3$~Gpc, i.e.~$z<0.56$. The limiting luminosity $L<44$~Jy~Gpc$^2$
per unit bandwidth. Assuming a 300-MHz
bandwidth, this translates to a
luminosity release of approximately
$10^{17}$~W ($10^{24}$~erg s$^{-1}$).

\subsubsection{DM--flux relationship}

As shown by \citet{2017ApJ...839L..25Y},
for $z<1$, the luminosity distance can
be directly related to the IGM DM as follows:
\begin{equation}
\label{eq:lumdist}
d_L \propto \langle{\rm DM}_{\rm IGM}\rangle/ (K_{\rm IGM} x(z)).
\end{equation}
\citet{2017ApJ...839L..25Y} 
find the following useful approximate relationship:
\begin{equation}
\langle {\rm DM}_{\rm E} \rangle \simeq
K \sqrt{L/S} + 
\langle {\rm DM}_{\rm Host} \rangle,
\end{equation}
where the constant $K$ can be computed in
terms of the assumed values of the constants 
in Eq.~\ref{eq:lumdist} at a particular
observing frequency \citep[for details, see][]{2017ApJ...839L..25Y}. Such a trend is 
apparent in the observed sample, albeit with a considerable amount of scatter. Applying this model to the FRBs found with the Parkes telescope, the authors constrain
host galaxy DMs to have a broad
distribution with a mean value
$\langle {\rm DM}_{\rm Host} \rangle = 270^{+170}_{-110}$~cm$^{-3}$~pc and
$L \sim 10^{36}$~W ($\sim 10^{43}$ erg s$^{-1}$).

\subsubsection{Brightness temperature}

As in the case of other radio sources,
where the emission mechanism is likely
to be non-thermal in origin, it is often
useful to quote the brightness temperature
inferred from the source, $T_{\rm B}$, which
is defined as the thermodynamic temperature of
a black body of equivalent luminosity. Making
similar arguments as is commonly done for
pulsars \citep[see, e.g., Section 3.4 of][]{2012hpa..book.....L}, we find 
\begin{eqnarray}
T_{\rm B} 
&\simeq& 10^{36}\;\mbox{\rm K}
\left( \frac{S_{\rm peak}}{\mbox{Jy}} \right)
\left( \frac{\nu}{\mbox{GHz}} \right)^{-2}\;
\left( \frac{W }{\rm ms} \right)^{-2}\;
\left( \frac{d_L}{\mbox{Gpc}} \right)^2\!.
\end{eqnarray}
 Again evaluating this for our
example FRB~140514 from the
previous section, where the pulse
width $W=2.8$~ms, we find
$T_B<3.5 \times 10^{35}$~K.

\section{Propagation effects}
\label{sec:propagation}

To date, FRBs have only been detected in the radio band\footnote{\citet{2016ApJ...832L...1D} claim the detection of a contemporaneous gamma-ray counterpart to FRB~131104.  However, given the low signal-to-noise and the fact that they needed to search a large positional uncertainty region, the association appears only tentative.}; no contemporaneous optical, X-ray or gamma-ray flash has been detected \citep[e.g.,][]{2017ApJ...846...80S,2017MNRAS.472.2800H}.  This currently leaves us in the situation where we need to maximize what we can learn from the properties of the radio pulses themselves.

In \S \ref{sec:FRBproperties}, we presented the basic observed properties of FRBs -- i.e. the parameters we use to characterize individual bursts.  Propagation effects in the intervening material between source and observer lead to many of the important observed properties of FRBs, as well as their derived properties, and we discuss them in more detail here.

The signal from an extragalactic FRB will pass through material in the direct vicinity of the source (e.g., a supernova remnant or pulsar/magnetar wind nebula in some models), the interstellar medium of its host galaxy (ISM$_{\rm Host}$), the intergalactic medium (IGM), and finally through the interstellar medium of our own galaxy (ISM$_{\rm MW}$) before reaching our radio receivers\footnote{Here we ignore any potential effects from the interplanetary medium of our Solar System or the Earth's ionosphere, both of which produce only very subtle effects compared to those imparted in the ISM$_{\rm Host}$, IGM and ISM$_{\rm MW}$.}.  This intervening material can be ionized, magnetized, and clumpy on a range of scales.

Radio waves can be diffracted, refracted, absorbed and have their polarization state changed by the material along the line-of-sight between observer and astronomical source.  Such propagation effects play an important role in our understanding of FRBs.  

While searching a range of trial DMs increases the computational load of FRB surveys (\S \ref{sec:pipelineSteps}), without this dispersive delay it would be even more challenging to separate astrophysical signals from human-generated RFI (which itself already presents significant limits to survey sensitivity).  As already discussed, DM is also a vital -- though nevertheless rough -- proxy for estimating Galactic distances and the redshift to extragalactic sources.  Indeed, this was the original -- and for all but one published FRB, the only -- evidence that FRBs originate at extragalactic distances; first and foremost, it is what separates them observationally from sporadically emitting pulsars (e.g. Fig.~\ref{fig:DM_DMmax}).

Beyond dispersive delay, and as with radio pulsars, FRB pulses can also show other propagation effects: e.g., scintillation, scattering and Faraday rotation.  All of these effects carry important clues about the local environments and galactic hosts of FRBs.  At the same time, we need to disentangle these effects to recover information about the intrinsic signal produced by the FRB source itself.    

We record FRB data using the widest possible range of radio frequencies (a bandwidth, $\Delta \nu$), in order to improve sensitivity.  Nominally, the sensitivity scales as $\sqrt{\Delta\nu}$, but a wider frequency range has the added advantage of detecting signals that peak in brightness at particular frequencies, as opposed to following a broadband power-law \citep[e.g.,][]{2016Natur.531..202S,2018ApJ...863....2G,2018arXiv181110748H}.  Additionally, these propagation effects have strong frequency dependencies (becoming much stronger at lower radio frequencies), and mapping their evolution across the widest-possible range can help in disentangling extrinsic propagation effects from the intrinsic signal properties.

Here we outline these various propagation effects, paying particular attention to how they are relevant to FRB observations and the scientific interpretation of the signals.  A much more detailed and fundamental description of propagation effects in radio astronomy, in general, can be found in reviews such as \citet{1977ARA&A..15..479R,1990ARA&A..28..561R}.  An overview in the context of pulsar observations can be found in \citet{2002astro.ph..7156C} and Chapter 4 of the Pulsar Handbook \citep{2012hpa..book.....L}, where -- presumably unlike FRBs -- the velocity of the source produces significant proper motion and leads to changing propagation effects with time.

\subsection{Dispersion}

In a dispersive medium, the velocity of light is frequency dependent.  The ionized interstellar and intergalactic media are dispersive, and for a typical FRB DM$ = 500$~cm$^{-3}$~pc (Eq.~\ref{eq:dm_def}) and observing frequency of 1.4~GHz this delays the signal by approximately one second compared with infinite frequency: 

\begin{equation}
\label{eq:dm500_delay}
1.06
\left(\frac{\rm DM}{500~\rm{cm}^{-3}~pc}\right)\left(\frac{\nu}{1.4~\rm{GHz}}\right)^{-2}~{\rm s}. 
\end{equation}
When considering the observed DM of an FRB, the contributions from different components along the line of sight from Eq.~\ref{eq:DMbreakdown} can be further separated as:

\begin{equation}
\label{eq:DMbreakdown_detail}
\resizebox{.9\hsize}{!}{${\rm DM}_{\rm FRB} = {\rm DM}_{\rm Iono} + {\rm DM}_{\rm IPM} + {\rm DM}_{\rm ISM} + {\rm DM}_{\rm IGM} + \left( \frac{{\rm DM}_{\rm Host} + {\rm DM}_{\rm Local}}{1+z} \right)$},
\end{equation}

\noindent where the contributions to the DM from these various ionized regions are summarized in Table~\ref{tab:DMcontributions}. Note that the expected ${\rm DM}_{\rm Host}$ and ${\rm DM}_{\rm Local}$ depends strongly on host galaxy type and local environment, and thus can serve to distinguish between progenitor models.

\begin{table}[]
\caption{Various contributions to the total dispersion measure of an FRB from Eq.~\ref{eq:DMbreakdown_detail}. \label{tab:DMcontributions}}
\begin{tabular}{llc}
\hline \hline 
Variable & Type & DM contribution (cm$^{-3}$ pc) \\
 \hline
${\rm DM}_{\rm Iono}$ & Earth ionosphere  & $\sim 10^{-5}$ \\
${\rm DM}_{\rm IPM}$ & Interplanetary medium of Solar System  & $\sim 10^{-3}$ \\
${\rm DM}_{\rm ISM}$ & Galactic interstellar medium & $\sim 10^{0}-10^{3}$ \\
${\rm DM}_{\rm IGM}$ & Intergalactic medium & $\sim 10^{2}-10^{3}$ \\
${\rm DM}_{\rm Host}$ & Host galaxy interstellar medium & $\sim 10^{0}-10^{3}$ \\
${\rm DM}_{\rm Local}$ & Local FRB environment & $\sim 10^{0}-10^{3}$ \\
\hline
\end{tabular}
\end{table}

Unfortunately, since the observed DM$_{\rm FRB}$ is the sum of these contributions, it is only possible to estimate the separate contributions by using models of the Galactic and extragalactic contribution, along with complementary information about the properties of the host galaxy and the FRB's local environment \citep[e.g.,][]{2017ApJ...834L...7T,2017ApJ...843L...8B}.  Ultimately, the accuracy of these models and assumptions will likely limit our ability to use FRBs as probes of the intergalactic medium, unless such complicating factors can be overcome by having statistics from a very large population of observed sources \citep[][and references therein]{2015aska.confE..55M}. 

Unlike with Galactic pulsars, cosmological redshift corrections are also relevant (see \S \ref{sec:FRBproperties}). At a more subtle level, determining an accurate FRB DM can be more challenging if the pulse shape changes with radio frequency.  Metrics that aim to maximize pulse structure as opposed to band-averaged peak signal to noise will lead to different conclusions about the DM and the finest-time-scale pulse structure \citep{2018ApJ...863....2G,2018arXiv181110748H}.  While pulsars show DM variations, this is dominated by the source's proper motion, which is expected to be negligible in the case of the much more distant FRBs.  Nonetheless, in the case of repeating FRBs, DM variations could be expected in a dense, dynamic environment like that of a surrounding, expanding supernova remnant \citep{2017ApJ...847...22Y,2018ApJ...861..150P}.

\subsection{Scintillation}
\label{sec:scint}

\begin{figure}[hbt]
\centering
\includegraphics[width=0.9\textwidth,angle=-90]{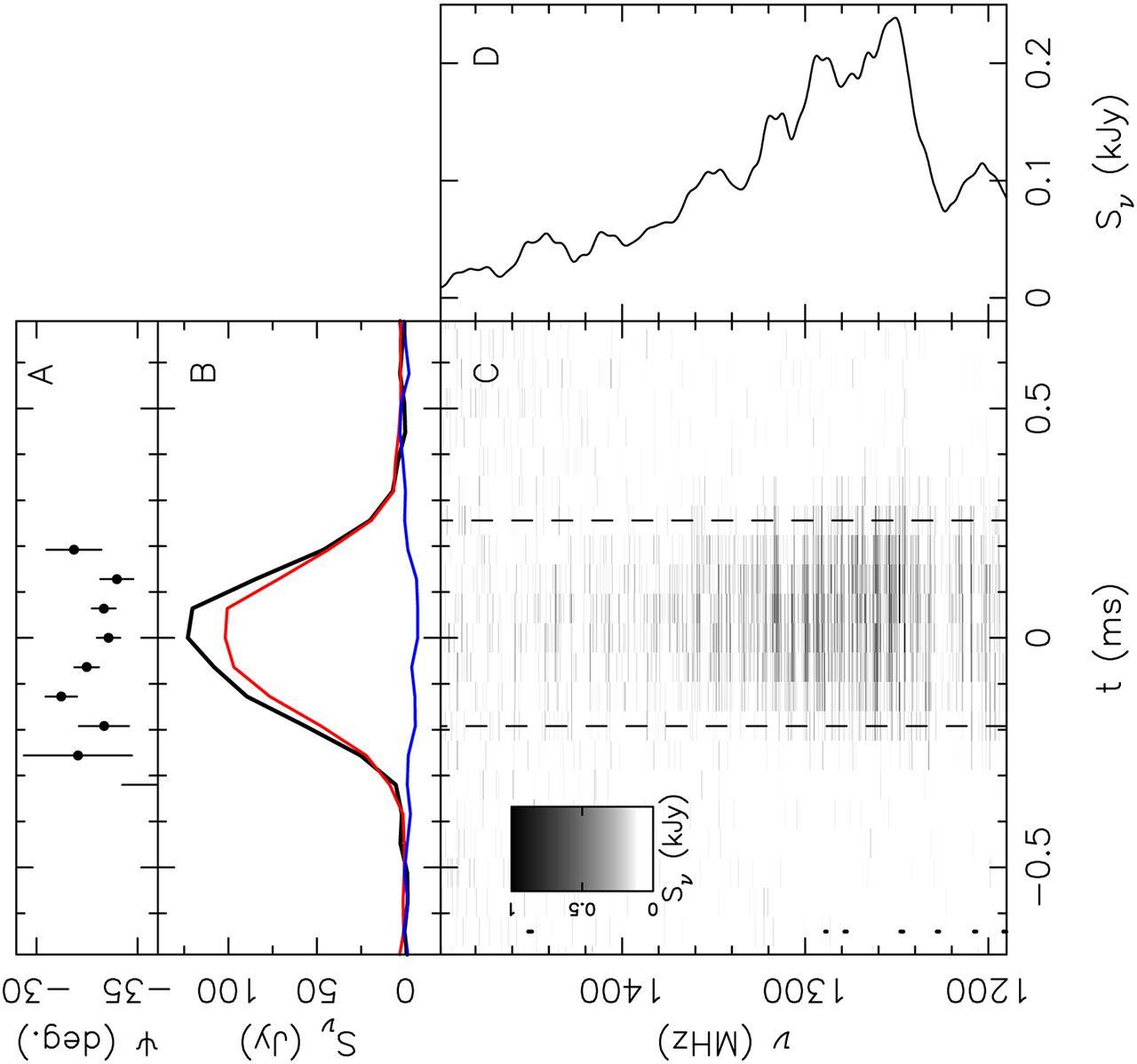}
\caption{\label{fig:scintillation}
Apparent scintillation seen in FRB~150807.  Panel C shows a dedispersed dynamic spectrum of the burst at 390-kHz spectral resolution.  The inferred scintillation bandwidth is $100 \pm 50$\,kHz.  Panel B shows the frequency-averaged burst profile with total intensity (black), linearly polarized signal (red), and circularly polarized signal (blue).  Panel A shows the polarization angle across the burst, and Panel D shows a smoothed version of the burst spectrum.  Fig.~1 from \citet{2016Sci...354.1249R}.
}
\end{figure}

Given their implied small emitting regions and large distances \citep{2018Natur.553..182M,2017ApJ...834L...7T}, FRBs should be perfect point sources, and thus scintillate (unless there is significant angular broadening of the source).

Scintillation is caused by refractive and diffractive effects as the signal passes through the clumpy and turbulent intervening material, which has electron density variations on a variety of length scales.  Delays imparted on the signal can cause destructive or constructive interference when these waves come back together.  In the plane of the observer, this creates a complex frequency structure that varies with time.  The relative motion between observer, source, and scattering medium dominates the time variability of the scintillation pattern observed at Earth.  Examples of such dynamic spectra showing scintillation in pulsars can be found in many places, e.g. in Fig.~3 of \citet{2014ApJ...794...21D}.  The characteristic frequency scale is called the scintillation bandwidth, while the characteristic timescale for a scintle to persist is called the scintillation time.  The scintillation bandwidth scales strongly with radio frequency:

\begin{equation}
\label{eq:scint_bw}
\Delta\nu_{\rm scint} \propto \nu^4
\end{equation}

Although scintillation is expected, care is needed when interpreting spectral features in an FRB to differentiate which signal effects are plausibly due to propagation, and which might be intrinsic to the emission mechanism.  The presence of RFI can also complicate the interpretation of fine-scale frequency structure.

Apparent scintillation\footnote{There are also other possible interpretations for fine spectral structure; e.g., see \citet{2016Sci...354.1249R}.} has been detected in bright FRBs like FRB~150807 \citep[Fig.~\ref{fig:scintillation}; and][]{2016Sci...354.1249R}, where its origin is plausibly from weak scattering in the IGM or host galaxy.  In the case of FRB~121102, fine-scale frequency structure has been ascribed to scintillation from the Milky Way \citep{2018ApJ...863....2G} because the observed scintillation bandwidth matches well with the prediction from the Galactic electron density model NE2001 \citep{2002astro.ph..7156C}.  If so, this means that the source was not significantly angularly broadened \citep{2017ApJ...834L...8M} and still appeared point-like when it arrived at the Milky Way.

It is also interesting to consider whether scintillation has a significant influence on the detectability of FRBs and the overall inferred event rate.  \citet{2015MNRAS.451.3278M} invoked Galactic scintillation as a possible explanation for an apparent Galactic-latitude dependence in the FRB rate \citep{2014ApJ...789L..26P}, but this has been debated.  Given that typical FRB search experiments record several hundred megahertz of bandwidth, and the expected Galactic scintillation bandwidth is $\lesssim 10$\,MHz (at 1.4\,GHz) for most lines of sight, it is likely that Galactic scintillation is always averaged out and will not be a deciding factor in whether an FRB is detectable.

For FRBs with very high signal-to-noise ratios, it may be possible to study the time-frequency structure using the secondary spectrum method in which scintillation arcs are visible \citep{2001ApJ...549L..97S}.  Though this is unlikely to provide much insight into the FRB itself, it may be an interesting method for probing the properties of the intervening material.

While the picture we sketch above is typically termed `diffractive scintillation', refraction associated with larger scales in the scattering screen could also cause broad focusing and defocusing of the FRB signal and result in smaller-amplitude intensity variations.  This may be relevant for understanding the periods of apparent activity and quiescence in repeating FRBs, where refractive scintillation could play a role in pushing the source brightness above the instrumental detection level on timescales of weeks to months \citep{2016ApJ...833..177S}.

\subsection{Scattering}\label{sec:scattering}

\begin{figure}[hbt]
\centering
\includegraphics[width=0.7\textwidth]{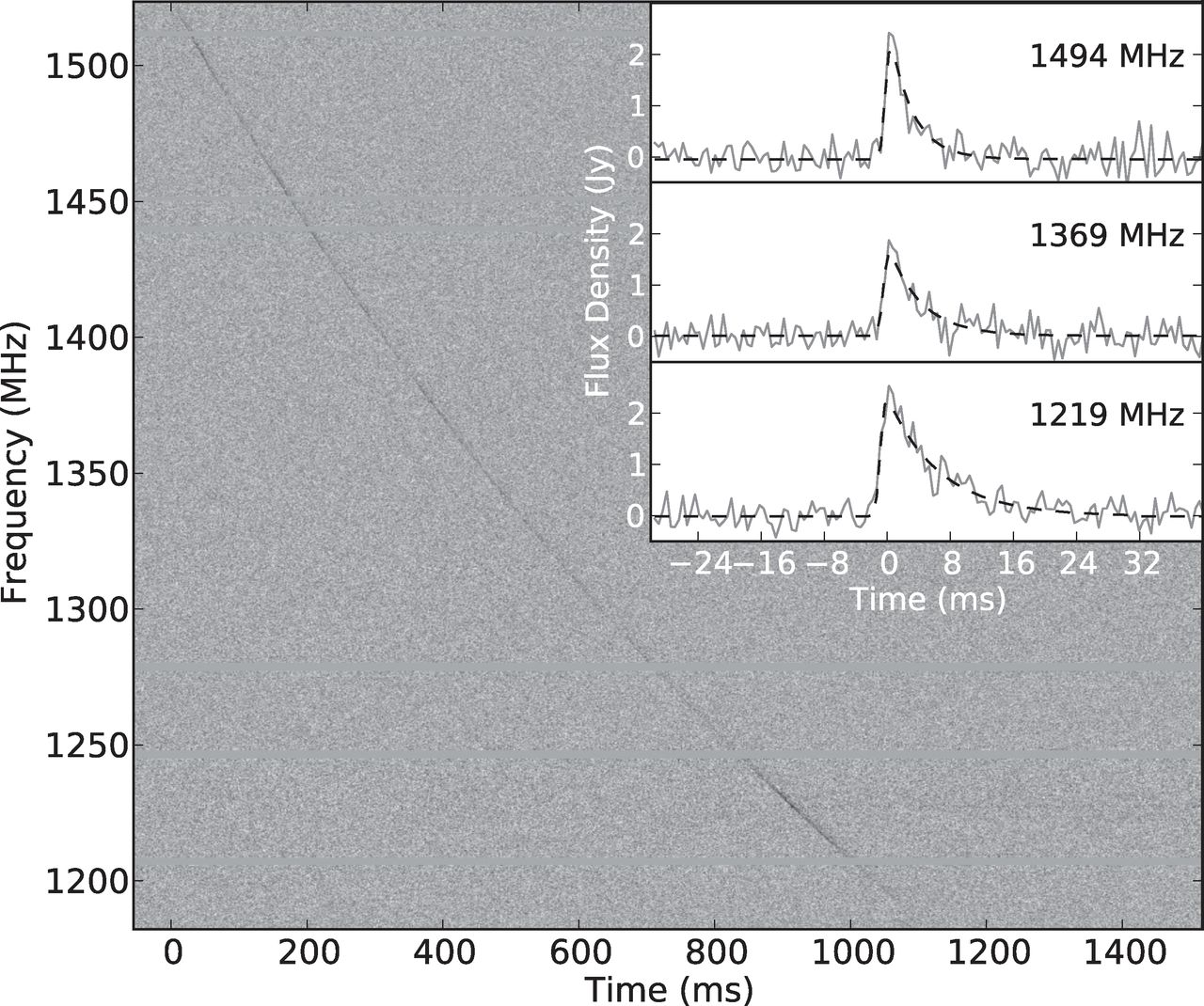}
\caption{\label{fig:scattering}
Scattering seen in FRB~110220.  The main panel shows the dynamic spectrum of the burst and its dispersive sweep.  The inset shows how the burst becomes asymmetrically broadened towards lower radio frequencies.  Fig.~2 from \citet{2013Sci...341...53T}.
}
\end{figure}

FRBs can be temporally broadened by scattering, which induces multi-path propagation and thus a later arrival time for parts of the signal that travel along longer path lengths. In the simple case of a thin, and infinitely extended scattering screen this effectively convolves the FRB pulse with a one-sided exponential decay.  In this simple picture, the decay time of this exponential tail scales strongly with frequency, as: 

\begin{equation}
\tau \propto \nu^{-4}.
\end{equation}

Scattering can also cause a detectable angular broadening of the source, which is observable using Very Long Baseline Interferometry (VLBI) \citep{2017ApJ...834L...8M}. One of the clearest examples of temporal scattering in an FRB is FRB~110220 (Fig.~\ref{fig:scattering}), where an exponential tail increasing as $\nu^{-4.0 \pm 0.4}$ was measured \citep{2013Sci...341...53T}. While DM quantifies the column density of free electrons along the line-of-sight, the scattering measure (SM) describes their distribution:

\begin{equation}
{\rm SM} = \int_{0}^{d} C^2_{{\rm n}_{\rm e}}(l)~dl, 
\end{equation}

\noindent where $C^2_{{\rm n}_{\rm e}}(l)$ indicates the strength of the fluctuations along the line-of-sight.

The SM can be determined empirically using scintillation measurements, pulse broadening from scattering, and angular broadening.  However, these different methods can lead to disparate SMs because of different line-of-sight weighting for $C^2_{{\rm n}_{\rm e}}(l)$.

\subsection{Faraday rotation}
\label{sec:faraday}

\begin{figure}[hbt]
\includegraphics[width=\textwidth]{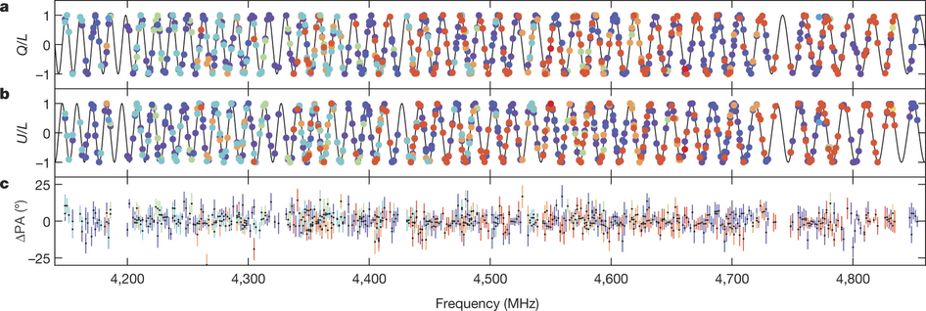}
\caption{\label{fig:faraday}
Faraday rotation seen in FRB~121102.  Panels a and b show the values of the Stokes Q and U parameters across the measured frequency range, normalized to the total linear intensity.  Panel c shows the residuals compared to a best-fit Faraday rotation model.  The various colors represent measurements from separate bursts detected in the same observing session.  Fig.~2 from \citet{2018Natur.553..182M}.
}
\end{figure}

If one considers a transverse electromagnetic wave decomposed into right- and left-hand circularly polarized components, then electrons interacting with a magnetic field component along the direction of the traveling wave will cause the right-hand component to propagate faster. A polarized signal will have a linear polarization position angle $\Theta$ that changes with wavelength as:

\begin{equation}\label{eq:PA}
\Theta = {\rm RM}~\lambda^2, 
\end{equation}

\noindent where RM is the Faraday rotation measure.  The relation between RM and physical parameters along the line of sight is given by:

\begin{equation}
\label{eq:rm_def}
{\rm RM} = -0.81 \int_{\rm 0}^{d} B(l)_{\parallel} n_e(l) dl,
\end{equation}

\noindent where $B(l)_{\parallel}$ is the magnetic field parallel to the line of sight. This is particularly nicely illustrated in Fig.~\ref{fig:faraday}, which shows the change in linear polarization angle for pulses from FRB~121102, an FRB with an extremely large ($\sim 10^5$ rad m$^{-2}$) rotation measure. The sign of the RM gives the direction, where a positive RM indicates a magnetic field directed towards the observer. In a situation where the Faraday rotation is believed to originate predominantly in the local environment of the source and its distant host galaxy \citep[e.g.,  ][]{2015Natur.528..523M,2018Natur.553..182M}, then a redshift correction should also be made:

\begin{equation}
{\rm RM}_{\rm src} = {\rm RM}_{\rm obs}(1+z)^2
\end{equation}

As Eq.~\ref{eq:rm_def} shows, the measured RM is the sum of all contributions along the line of sight, and different Faraday regions along the way can have different directionality and add to or cancel each other out.  Disentangling these various contributions is non-trivial, though it is likely that any observed RM variability (in the absence of equivalent DM variability) is from material local to the source \citep{2018Natur.553..182M}.  The RM contribution from the IGM may be very small ($< 10$\,rad\,m$^{-2}$) in many cases \citep{2016Sci...354.1249R}, though if the burst passes through the hot medium of a galaxy cluster this can introduce a more sizable ($\sim 50$\,rad\,m$^{-2}$) contribution \citep{2016ApJ...824..105A}.  Like DM, there is a Galactic foreground that should be considered, and models exist to estimate this contribution for any particular line-of-sight \citep{2015A&A...575A.118O}.  

Given that FRBs are likely produced in small emission regions viewed behind a number of distinct Faraday regions, it is reasonable to expect that -- like pulsars \citep[e.g.,][]{2019MNRAS.484.3646S} -- they will have Faraday thin spectra (the burst is a single pierce point through these regions). Rotation measure synthesis \citep{2005A&A...441.1217B} combined with the `rmclean' deconvolution method \citep[e.g., ][]{2009A&A...503..409H} can indicate whether there is more complicated Faraday structure due to emission at a range of Faraday depths \citep[for an application see, e.g.,][]{2018Natur.553..182M}.
Furthermore, it has been proposed that Faraday conversion -- in which linear polarization can convert to circular polarization (and vice versa) as a function of radio frequency -- may be detectable in FRBs \citep{2019MNRAS.tmpL..41V,2019arXiv190201485G}.  If so, this could provide a powerful diagnostic of the magnetic field structure and medium surrounding the source.

If both DM and RM are measured, then one can infer the average line-of-sight magnetic field strength, weighted by electron density:

\begin{equation}
<B_{\parallel}> = \frac{{\rm RM}}{0.81 {\rm DM}}
\end{equation}

However, care is required here because the DM and RM need to be associated with the same region of magneto-ionic material, which may not be the case for many FRBs.

\subsection{Plasma lensing}

Any refractive medium can act as a lens, including plasma. Radio waves passing through a plasma are bent; in the plane of the observer these rays can overlap, causing bright caustic spots \citep{1998ApJ...496..253C}.  The effect is highly chromatic, meaning that the brightening occurs in specific frequency ranges, and can be time variable given that the source, lens, and observer are all moving with respect to each other and small relative motion can produce large brightness variations.

As dispersion demonstrates, FRBs travel through plasma in many distinct regions on their way to Earth, but there are also reasons to expect that there may be local, high-density plasma associated with FRBs.  For example, if FRBs originate from particularly young neutron stars, then they may be embedded in nebulae or supernova remnants.  As the Crab pulsar has demonstrated, plasma prisms or dense linear filaments can alter the shape of the observed pulse profiles, creating highly chromatic echoes \citep{2000ApJ...543..740B,2011MNRAS.410..499G}.  More recently, plasma lensing has been convincingly demonstrated in the original Black Widow pulsar B1957+20, where the individual pulses can be amplified by factors up to 70 \citep{2018Natur.557..522M}.  This effect is again highly chromatic, and the observed spectra of the pulses can vary on timescales comparable to the 1.6-ms pulse period.  PSR~B1957+20 is eclipsed by intra-binary plasma that has been blown off the companion star by the pulsar's wind.  The lensing events seen in PSR~B1957+20 occur specifically around eclipse ingress and egress, suggesting that it is clumps in this intra-binary material that are acting as lenses.

\citet{2017ApJ...842...35C} consider the relevance of plasma lensing for understanding both the spectra and apparent luminosities of FRBs.  Plasma lensing could explain the highly variable radio spectra seen in the repeating FRB~121102, and in a more general sense it could potentially decrease the required energy per burst.  The time-frequency pulse structure seen in FRB~121102 \citep{2018arXiv181110748H} is also potentially explained by plasma lensing, which can create multiple images that will interfere with each other if the differential delay is within a wavelength.

These ideas will be best tested by ultra-wide-band observations that can map the spectra of FRB from $\sim 0.1 - 10$\,GHz.  Plasma lensing may be occurring at some level, but the question remains how relevant this effect is for interpreting the properties of individual FRBs and the distribution of the population as a whole.

\subsection{H\textsc{i} absorption}

Dispersive delay is instrumental to the argument that FRBs are extragalactic in origin. Without a precise localization of the burst, it is the only proxy for distance that we have.  Like for Galactic pulsars, measuring H\textsc{i} absorption can provide complementary information to DM.  It could conceivably also provide an independent confirmation of an FRB's extragalactic nature.  H\textsc{i} absorption comes from fine structure in the hydrogen atom's quantum states, where the electron and proton spins can be aligned or anti-aligned.  The corresponding absorption feature occurs at 1420.4\,MHz, and frequency shifts of this line encode valuable kinematic information about the intervening gas.

\citet{2015MNRAS.451L..75F} consider H\textsc{i} absorption in FRB bursts imparted by the Galactic spiral arms or extragalactic clouds.  Detection of H\textsc{i} absorption can set a firm lower limit on distance.  However, H\textsc{i} absorption is only detectable for very high signal-to-noise bursts passing through a high column density of neutral hydrogen. Existing telescopes might just barely be able to detect H\textsc{i} absorption for bright FRBs at low Galactic latitudes. If we ever hope to detect absorption from extragalactic H\textsc{i} clouds, then much higher sensitivities (like those provided by SKA) are going to be necessary.  Because the H\textsc{i} line is intrinsically very narrow and only somewhat broadened by kinematic effects, very high spectral resolution (ideally baseband) data will be needed to detect this signature in FRB data.  It is likely worth the effort: \citet{2016MNRAS.460L..25M} find that there is a $\sim 10$\% chance that neutral material in an FRB host galaxy will produce a detectable H\textsc{i} absorption signature that can be used to infer the redshift directly from the FRB pulse.

\subsection{Free-free absorption}

If FRBs are found in dense environments (like a supernova remnant or active star-forming region), then their detectability at low radio frequencies ($< 1$\,GHz) may be limited by free-free absorption.  For fixed temperature and electron density, the opacity of an H\textsc{ii} region scales as $\nu^{-2.1}$.

The large event rate of FRBs, coupled with the large fields-of-view of low-frequency radio telescopes -- especially aperture arrays like LOFAR and MWA -- led to some early predictions that these should be phenomenal FRB-finding machines \citep{2013MNRAS.436..371H}.  However, as yet no FRB has been detected below $\sim 400$\,MHz \citep{2019Natur.566..230C}, despite concerted efforts with GBT \citep{2017ApJ...844..140C}, Arecibo \citep{2016ApJ...821...10D}, LOFAR \citep{2014A&A...570A..60C,2015MNRAS.452.1254K}, and MWA \citep{2018ApJ...867L..12S}. While the intrinsic spectra of FRBs or temporal broadening from scattering may explain the dearth of detected FRBs at low frequencies, free-free absorption is potentially another contributing factor. Early detections from CHIME down to 400~MHz indicate that FRBs may indeed be detectable at lower frequencies, but a larger sample at these frequencies is needed to clarify the relevance of temporal scattering and free-free absorption, and whether the observed rate is lower at longer wavelengths.

\section{Observational Techniques}\label{sec:obsTechniques}

In previous sections we defined FRBs and their obserational properties. In the following, we delve into the details of how we search for and discover FRBs using single dish and interferometric radio telescopes.

\subsection{Searching for FRBs}\label{sec:pipelineSteps}

Radio telescopes typically consist of an aperture that brings electromagnetic signals from the sky to a focus so that they can be measured as a function of time using feeds \citep[for an introduction to radio astronomy, see e.g.][]{2016era..book.....C}. The antenna and feed response is typically measured over a range of radio frequencies, i.e. a bandwidth, which is amplified and discretely sampled by a number of frequency channels. High-time-resolution observations, like those used to search for FRBs and pulsars, record the stream of voltages in each channel over time, sampling the voltage stream at some finite time resolution. These data can be saved to disk in the native voltage data format, or further compressed (i.e. downsampled), by summing adjacent time or frequency channels, which decreases the resolution. If there are multiple polarizations recorded, in the case of two orthogonal antennas in the receiver, they may also be summed at this stage. The resulting data cube of intensities at each time and frequency channel can be saved to disk as a `filterbank' file. 

Searching for dispersed pulses in these data cubes requires several steps. In some cases there is a pre-processing step to sum the polarizations, if they are recorded separately. The total intensity data are then analyzed to produce a list of candidate FRB signals. Each step is described briefly below.

\begin{figure}
\centering
\includegraphics[width=0.5\textwidth]{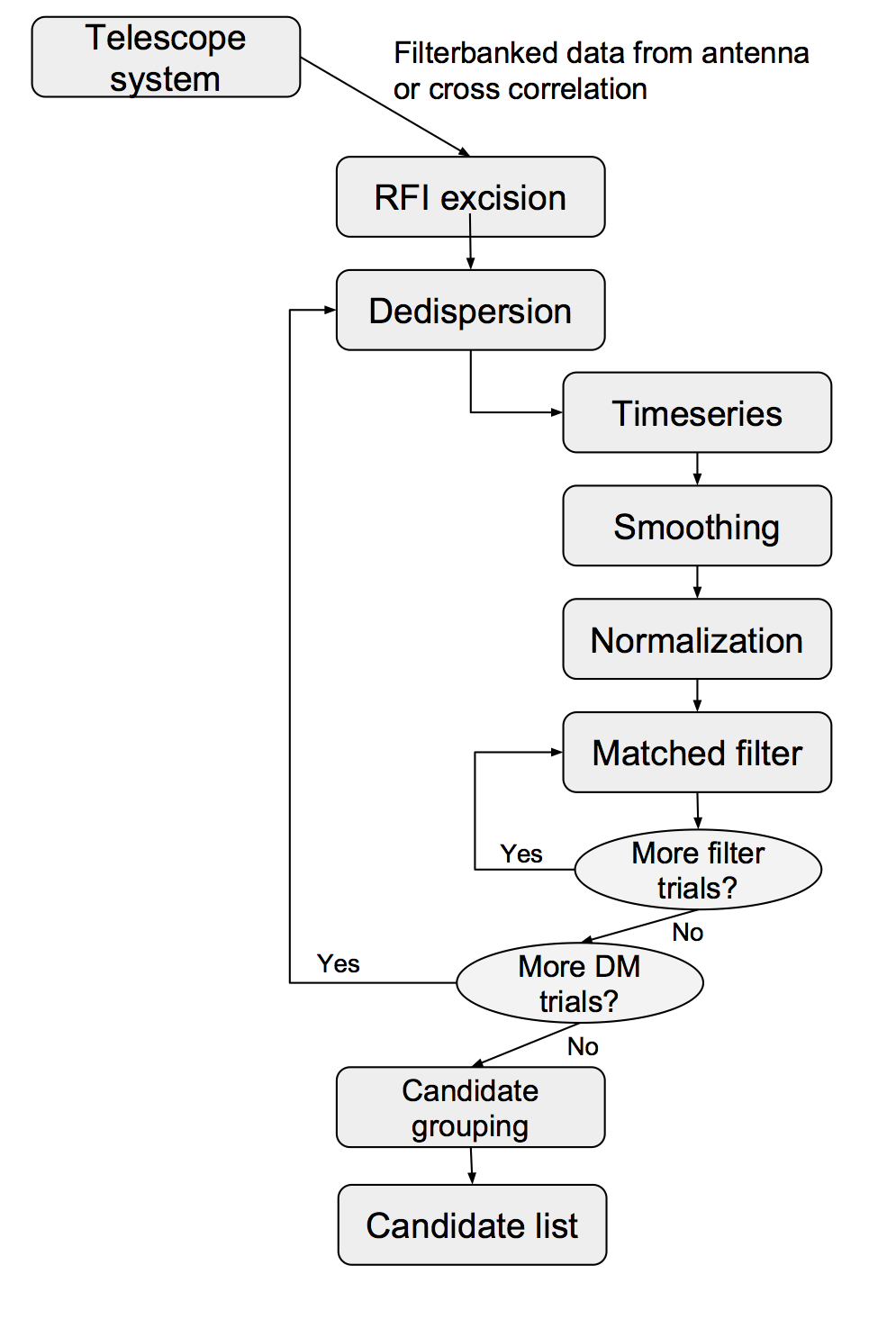}
\caption{A block diagram summarizing the analysis procedure discussed in \S \ref{sec:pipelineSteps}. \label{fig:DataBlockDiagram}}
\end{figure}

\subsubsection{Preliminary radio frequency interference excision} 

Artificial radio frequency interference (RFI) is ubiquitous in radio astronomical data. RFI can be persistent or impulsive as well as broad- or narrow-band.  It can overwhelm the intensity of astrophysical signals and, in pernicious cases, masquerade as an astrophysical signal by matching some of the expected properties \citep[e.g., a frequency-dependent sweep in time that looks like astrophysical dispersion, see][]{2018MNRAS.481.2612F}.  In most FRB searches, an initial attempt is made to remove or mitigate RFI before the data are searched for pulses. The most common approaches involve masking time samples and frequency channels.  If there are known in-band artificial emitters, the corresponding frequency channels can be automatically masked. Additionally, the data cube can be searched for impulsive RFI by looking for peaks in the $\mathrm{DM} = 0$ cm$^{-3}$ pc time series (where dispersed astrophysical bursts should be smeared out) and masking the contaminated time samples \citep{2012MNRAS.420..271K}.  One can also subtract the $\mathrm{DM} = 0$ cm$^{-3}$ pc time series from the time series at higher DM trials \citep[][but note that this will alter the pulse shapes]{2009MNRAS.395..410E}.
Spectral kurtosis \citep{2010MNRAS.406L..60N} can also be used to clean the data. The goal is to mask as much RFI as possible, without removing any astronomical signals.

\subsubsection{De-dispersion}

Since the DM of a new FRB is not known \textit{a priori}, a large number of DM trials must be searched. Narrow pulses could be missed if the DM is not sufficiently close to one of the trial DM values, so a fine spacing of trials is necessary. Instrumental broadening (or smearing) of the pulse within a single frequency channel can be calculated as 
\begin{equation}\label{eq:smearing}
\Delta t_\mathrm{DM} = 8.3 \times 10^{6} \, \mathrm{DM} \, \Delta \nu_\mathrm{ch} \, \nu^{-3} \; \mathrm{ms},
\end{equation}
 
\noindent where observing frequency $\nu$ and channel bandwidth $\Delta \nu_\mathrm{ch}$ are both in MHz. The next DM trial in the sequence should be chosen such that sensitivity to a dispersed pulse never drops below a specified level. Thus, more closely spaced DM trials provide higher sensitivity to narrow pulses, but this comes with an added computational cost and may slow down the search to less than real-time.

The de-dispersion process, correcting for the DM to maximize the S/N of the pulse, is the most computationally expensive step in a single-pulse search and reducing the computational complexity of this task is a continuing goal, often involving parallelization of code on graphics processing units (GPUs) or highly optimised algorithms running on CPUs \citep{2012MNRAS.422..379B,2016A&C....14....1S,2017ApJ...835...11Z,2018ApJ...863...48C}. There are several implementations of dedispersion algorithms that are commonly used. Here we group them into two main categories: incoherent and coherent dedispersion. \\

\noindent \textbf{Incoherent dedispersion} Incoherent dedispersion applies time-delay corrections to individual frequency channels. The dispersion delay across a bandwidth for a given DM can be calculated using Eq.~\ref{equ:defdt} and the delay is subtracted from each frequency channel to arrive at a channelized dataset with propagation delays removed. The accuracy of incoherent dedispersion depends on the bandwidth of individual frequency channels. Wide channels make it impossible to adequately remove dispersion effects. Incoherent dedispersion trials are often performed when the DM of the pulsed signal is not known \textit{a priori}, such as in blind FRB searches that search thousands of DM trials. 

In FRB searches, the incoherent dedispersion operation over several trial steps occupies the majority of the processing time. Brute force dedispersion applies delays to all frequency channels for each DM trial. This method is computationally expensive ($\mathcal{O}\left[N_{t} N_\nu N_\mathrm{DM}\right]$), however recent implementations on GPUs have accelerated these searches to real-time performance\footnote{\url{https://code.google.com/archive/p/dedisp/} and \url{https://sourceforge.net/projects/heimdall-astro/}}. Tree dedispersion \citep{1974A&AS...15..367T} instead integrates over straight-line paths through $\nu$ and $t$, for lower computational complexity ($\mathcal{O}\left[N_t N_\nu \mathrm{log} N_\nu\right]$). Sub-band dedispersion implements tree dedispersion over sub-bands of the total bandwidth\footnote{For example, \texttt{prepsubband} in \textsc{presto}, \url{https://github.com/scottransom/presto}} \citep{2011ascl.soft07017R}. More recently, fast discrete dispersion measure transforms (FDMT) have been implemented \citep{2017ApJ...835...11Z}, which use the two-dimensional array of intensities in frequency and time to calculate integrals over quadratic curves, reducing the computational complexity of the dedispersion algorithm by two orders of magnitude\footnote{See for example \url{https://github.com/iansbrown/FRB-FDMT-Search/blob/master/FDMT\_functions.py} or a GPU implementation at \url{https://github.com/ledatelescope/bifrost/blob/master/src/fdmt.cu}} ($\mathcal{O}\left[\mathrm{max}\{N_t N_\mathrm{DM} \mathrm{log}_2 N_\nu,\, 2 N_\nu N_t\}\right]$). The preferred choice of dedispersion algorithm used may depend on the computer architecture (GPU, CPU, combination) and pipeline design. \\

\noindent \textbf{Coherent dedispersion} In contrast to incoherent dedispersion, coherent dedispersion more precisely recovers the intrinsic pulse shape (assuming that there is no significant scattering). This is achieved by operating on raw voltage data. The ISM effects on the signal can be modeled as a filter, and the reverse filtering operation can be applied in the Fourier domain \citep{1975MComP..14...55H}. In this way, the high-resolution pulsed signal can be recovered \citep{1987ApJ...315..149H}. The impulse response of the ISM filter depends on the bandwidth of the observations as well as the DM of the signal, thus for high-DM pulses, such as those from FRBs, coherent dedispersion can be computationally complex and slow. Typically coherent dedispersion is only performed for a single value, when the DM of the source is already known. In the case of FRBs, this can be useful for a repeating source (see \S \ref{sec:FRB121102} and \citet{2018Natur.553..182M}) but does not yet hold much practicality in blind searches. \\

\noindent \textbf{Semi-coherent dedispersion} A compromise approach between incoherent and coherent dedispersion, called semi-coherent dedispersion, has been used in pulsar searches by \citet{2017A&C....18...40B} \citep[see also the techniques and discussion in][]{2017ApJ...835...11Z}. In this implementation, the data are coherently dedispersed to a select few trial DMs and the output of this process is then searched incoherently around the coherent dedispersion value\footnote{\url{https://github.com/cbassa/cdmt}}. This approach, while still computationally expensive due to coherent dedispersion, allows for a much more sensitive search than incoherent methods alone, particularly in cases where the intra-channel dispersive smearing is large, such as at low radio frequencies.

\subsubsection{Extracting a time series}

For each DM trial of the incoherent brute force and tree de-dispersion methods, the data are summed over all frequencies in a way that follows the dispersive sweep. For coherently dedispersed data, the data are summed in each time sample. The resulting integrated intensity is a one-dimensional array of total signal versus time, called the \textit{time series}. The time series can then be searched for astrophysical pulses. In other cases, such as with FDMT, the data are searched directly in the dynamic spectra (frequency-time plane).

\subsubsection{Baseline estimation or smoothing}

The mean signal level in an observation can vary more slowly than the signals being searched for (over seconds to minutes) due to instrumental effects and RFI. This can result in a non-uniform baseline in the time series, making it difficult to extract astrophysical pulses from the noise. Typically, a stable baseline is removed from the time series before it is searched for pulses. The baseline can be measured by calculating the running median (or mean) of the time series, clipping outliers above a specified threshold, and then re-calculating the median \citep{2012PhDT.......418B}  A suitable time window should be chosen for this smoothing.  

\subsubsection{Normalization}

In order to derive a pulse's signal-to-noise ratio, the noise properties must first be estimated. Some FRB search codes calculate the rms by first calculating the median absolute deviation (MAD) and then estimating the noise as rms = $k$ $\times$ MAD, where the scale factor $k$ is $\simeq$1.4826 for normally distributed data. This assumption holds for Gaussian noise, which is typically true of radio data in the absence of strong RFI. The signal-to-noise ratio can then be calculated in a single time sample $x$ as S/N~=~$\mathrm{timeseries}(x) / \mathrm{rms}$.

\subsubsection{Matched filtering}

To find pulses in the data wider than a single time sample, the time series are convolved with boxcar functions of width $W$ for multiple trial pulse durations. In the case of a pulse duration greater than a single time sample, the signal-to-noise ratio must be normalized by the boxcar width such that S/N~=~$\mathrm{timeseries}(x) / (\mathrm{rms} \times \sqrt{W})$. Peaks in the de-dispersed, normalized, and boxcar-convolved time series are typically reported as candidates.

\subsubsection{Candidate grouping}

Once single pulse candidates have been identified in the time series, some grouping should be performed to cluster candidates related to the same event. A bright pulse will be detected optimally in the DM trial and time bin most closely matching the true event, but also in other nearby DM trials and possibly in multiple matched filter trials. Grouping can be performed with a friends-of-friends algorithm that searches for clusters of points in a specified parameter space\footnote{\url{https://sourceforge.net/p/heimdall-astro/code/ci/master/tree/Pipeline/label\_candidate\_clusters.cu}}\footnote{See also \url{http://ascl.net/1807.014 for a machine learning-based approach}} \citep{2018MNRAS.480.3302P}. Alternatively, an acceptable proximity margin can be specified and two candidates within that margin are grouped together. 

\subsubsection{Post-processing RFI excision}

Additional RFI excision can be done using the list of candidates generated after grouping. This is particularly useful if multiple telescope beams have been recorded and searched separately. All previous steps are executed on individual beams of multi-beam receivers (in the case of a single dish, \S \ref{sec:singleDish}) or separate tied-array/compound beams (in the case of interferometers, \S \ref{sec:interferometer}). Candidates detected in many spatially separated beams can be rejected as RFI. In some cases, RFI can mimic the dispersive sweep of a genuine astrophysical source \citep[as in the case of the Perytons; ][]{2011ApJ...727...18B}. Multi-beam cross-checking can exclude candidates that might pass through a zero-DM RFI excision step.

The same grouping methods mentioned above can also be applied to candidates detected in multiple beams, and candidates with significant clustering in many telescope beams can be rejected as interference\footnote{\url{https://sourceforge.net/p/heimdall-astro/code/ci/master/tree/Applications/coincidencer.C}}\footnote{\url{https://github.com/ckarako/rrattrap}}\footnote{\url{https://github.com/danielemichilli/SpS}} \citep{2015ApJ...809...67K,2018MNRAS.480.3457M}.

\subsection{FRB search pipelines}\label{sec:pipelines}

The procedures outlined in \S \ref{sec:pipelineSteps} have been implemented in a number of search pipelines: i.e. software packages that read in telescope data and output a list of single-pulse candidates. Searches for FRBs at the Parkes telescope and with the UTMOST telescope in Australia have primarily been done with the \textsc{heimdall}\footnote{\url{https://sourceforge.net/projects/heimdall-astro/}} pipeline, which uses brute force dedispersion techniques on GPUs \citep{2016MNRAS.460L..30C,2017MNRAS.468.3746C}. FRB searches of survey data from Arecibo and Green Bank have been performed with the single-pulse search algorithms in \textsc{presto}\footnote{\url{https://www.cv.nrao.edu/~sransom/presto/}} \citep{2001PhDT.......123R}, which uses sub-band dedispersion techniques \citep{2014ApJ...790..101S}. FRBs detected with the ASKAP telescope have been found with the \textsc{fredda} pipeline using the FDMT algorithm \citep{2017ApJ...841L..12B}. Upcoming surveys at new telescopes are developing their own pipelines including the \textsc{amber} pipeline for the FRB search on the upgraded Westerbork Telescope\footnote{\url{https://github.com/AA-ALERT/AMBER}} \citep{2016A&C....14....1S,2018A&C....25..139M}, the \textsc{burst\_search} algorithm developed for archival GBT data\footnote{\url{https://github.com/kiyo-masui/burst\_search}}, and the \textsc{bonsai} algorithm for FRB searches with the CHIME telescope \citep{2018ApJ...863...48C}. 

The aforementioned pipelines have been developed independently by various groups. This independence is a strength, since no two pipelines should be subject to the exact same biases or errors. However, decisions at each step outlined above can affect the ultimate sensitivity of the pipeline. Additionally, each FRB search code has been developed and tuned to work on a specific survey configuration with data of a particular size or resolution. These differences can make each search code differently sensitive to FRBs, or less sensitive in certain areas of the parameter space \citep{2015MNRAS.447.2852K}. As yet, no standard metric has been developed to compare these codes and their effectiveness at finding FRBs. A `data challenge' with real and injected FRB signals would be ideally suited to this task.

\subsection{FRB searches with radio telescopes}

\subsubsection{Single-dish methods}\label{sec:singleDish}

\begin{figure}
\centering
\includegraphics[width=\linewidth]{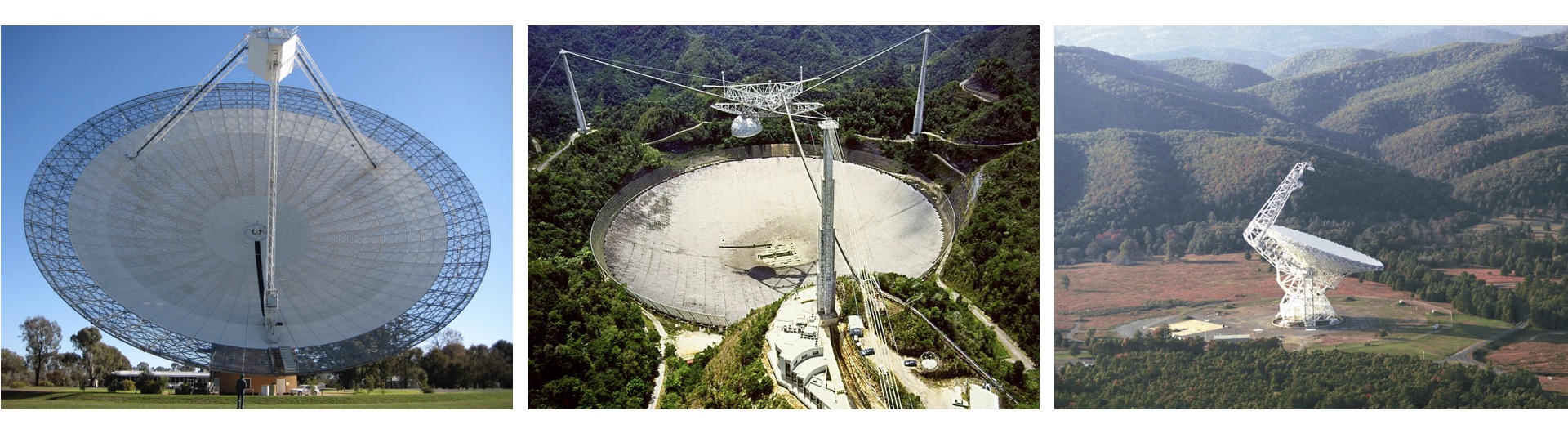}
\caption{Examples of single-dish radio telescopes used to search for FRBs (from left to right): the 64-m Parkes telescope in New South Wales, Australia, the 305-m Arecibo telescope in Puerto Rico, USA, and the 110-m Green Bank Telescope in West Virgina, USA.}\label{fig:singleDishes}
\end{figure}

Large single-dish telescopes that are searching for FRBs include Parkes (64~m), Lovell (76~m), Effelsberg (100~m), Arecibo (305~m), FAST (500~m), and GBT (110~m); see Fig.~\ref{fig:singleDishes}.  Roughly speaking, the limiting sensitivity of a radio dish is inversely proportional to its effective area. The diameter of the dish $D$ determines the size of the telescope half power beam width $\theta_\mathrm{HPBW} \simeq 1.22 \; \lambda/D$ where $\lambda$ is the wavelength of the observed light. To increase the field of view of single-dish telescopes, some are equipped with multi-beam receivers that sample a larger fraction of the telescope's focal plane. 

The primary advantages of single dishes in FRB searches come from their large collecting areas (and thus high sensitivity) and low signal processing complexity. Their large focus cabins also have space for several wide bandwidth, cooled receivers, which are useful for studying FRB emission and polarization. Their sensitivity also makes them ideal instruments to follow up known FRBs to search for repetition, particularly in cases where the original detection was made with a less sensitive instrument \citep{2018ApJ...861L...1C}.

The greatest disadvantage of current single dishes is their poor localization of an FRB discovery: the localization uncertainty is $\theta_\mathrm{HPBW}$ (often at least several arcminutes). Rejecting RFI in single-dish data can also be a challenge; however, this can be somewhat mitigated through multi-beam coincidence of candidates. 

Even as we move into an era of interferometric FRB searches (\S \ref{sec:interferometer}), single dishes still have an important role to play in the study of FRB emission and polarization. Single dishes offer the raw sensitivity and broad frequency coverage (using a suite of receivers) to study FRB emission. For example, breakthroughs in the study of the repeating FRB~121102 (\S \ref{sec:FRB121102}) have been made using receivers on single dishes at both higher and lower frequencies compared to the discovery observation. Future polarization studies of FRBs using sensitive single dishes are expected to provide further insights into the FRB emission mechanism (\S \ref{sec:emission}) and environment in their host galaxies. In the future, cooled phased array feeds (PAFs) installed on single dishes may result in better localization and increased survey speed \citep{2017PASA...34...26D}.

\subsubsection{Interferometric Methods}\label{sec:interferometer}

\begin{figure}
\centering
\includegraphics[width=\linewidth]{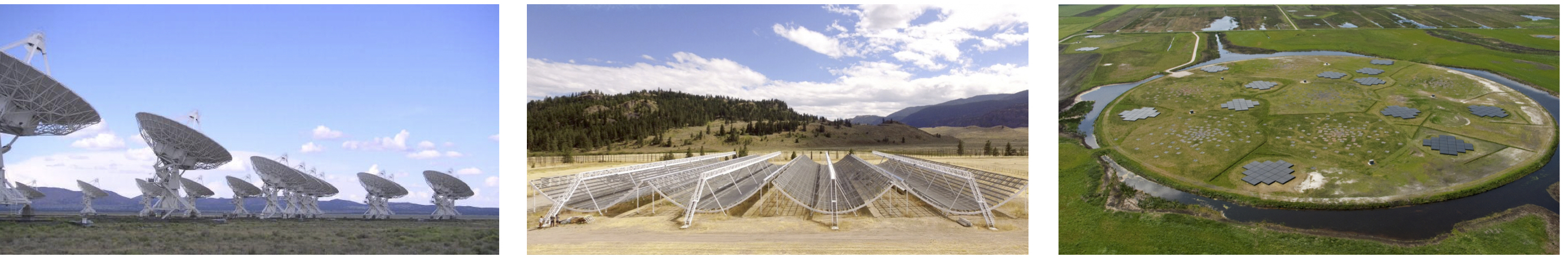}
\caption{Examples of radio interferometers used to search for FRBs (from left to right): the Jansky Very Large Array (JVLA) of 27 25-m dishes in New Mexico, USA, the Canadian Hydrogen Intensity Mapping Experiment (CHIME) with four cylindrical paraboloids each 100-m long and 20-m in diameter in British Columbia, Canada, and the core of the Low-Frequency Array (LOFAR) of dipoles in the Netherlands.}\label{fig:interferometers}
\end{figure}

Interferometric radio telescopes are composed of many antennas or dishes, whose signals are combined to achieve, roughly speaking, the resolution of a single large telescope with a diameter equivalent to the longest baseline. The field of view can be sampled more finely using many beams, each created by applying different weightings or delays between different elements of the array. Radio interferometers come in a wide variety of shapes and sizes (Fig.~\ref{fig:interferometers}). Some are made of smaller radio dishes such as the Jansky Very Large Array (JVLA, 27 25-m dishes), the Westerbork Synthesis Radio Telescope (WSRT, 14 25-m dishes), and the Australian Square Kilometre Array Pathfinder (ASKAP, 36 12-m dishes). 
Others consist of cylindrical parabaloids with many receivers sampling along the focal line, such as the Canadian Hydrogen Intensity Mapping Experiment (CHIME, 4 parallel 100-m long parabaloids) and the upgraded Molonglo Synthesis Telescope (UTMOST, 2 778-m long parabaloids). Others still are made from individual stationary dipole antennas such as the Low Frequency Array (LOFAR) and the Murchison Wide-field Array (MWA). 

FRB searches with interferometers can be done in a variety of ways \citep{2011PASA...28..299C}. Incoherent searches discard phase information and use a summation of the individual element intensities; these have the advantage of large fields of view (equal to the primary field of view of the elements), but sensitivity scales as $\sqrt{N}$ for N elements and localization precision is poor. Coherent searches create tied-array beams (TABs) by applying differential weights to different elements and summing the signals in phase; in this case sensitivity scales as $N$, thus providing both better sensitivity and better localization. However, beamforming with many elements can have high computational complexity requiring powerful backend hardware \citep{2017arXiv170906104M}. Image plane FRB searches look for short transients through difference imaging, which takes advantage of existing imaging hardware on many interferometers; however, short duration images may be low sensitivity or poor quality making the identification of genuine astrophysical transients difficult. Additionally, image plane FRB data may have lower time resolution and thus miss fine-scale temporal structure in the bursts. However, if the imaging time is short ($\sim$ms), it can still be possible to capture the basic information about the FRB such as DM and approximate pulse duration, as with the \textit{realfast} system \citep{2018ApJS..236....8L}.

General advantages to interferometric FRB searches are the flexible nature of interferometers in terms of pointing, localizing, and beam-forming, particularly if voltage data is recorded from each element upon detection of an FRB. The ability to track quickly, form sub-arrays, or do fly's eye surveys to increase field of view make interferometers quite dynamic facilities \citep{2018Natur.562..386S}. 

Interferometric FRB searches present substantial challenges. Combining data streams from many elements, coherently or incoherently, requires enormous computational power and large data rates. This becomes even more of a challenge when the goal is to search through incoming data for FRBs in real time. One dimensional arrays such as UTMOST and WSRT will also produce elongated beam shapes, making 2-D localization imprecise (though note that UTMOST is being upgraded to work as a 2-D array). Interferometers can also come with the penalty of reduced choice in observing band. Small dishes may lack the necessary space at the focus for multiple receivers at different frequencies, and dipole arrays may only be hardwired to operate at a specific set of frequencies. These may limit the information that can be gleaned from an individual FRB detection.

\section{Landmark FRB discoveries}\label{sec:indivFRBs}

In the following, we discuss some of the most influential FRB discoveries of the past 10 years. These include FRBs that extend the parameter space in one or more ways, as well as FRBs that have been the center of extended discussion in the literature.

\subsection{FRB~010724 $-$ The Lorimer Burst}\label{sec:LorimerBurst}

FRB~010724, also known as `the Lorimer burst', is considered to be the first FRB discovery \citep{2007Sci...318..777L}. It was discovered before the term `fast radio burst' was even coined \citep[the term was introduced by][]{2013Sci...341...53T}, and remains one of the brightest FRBs yet to be detected. The burst was initially reported as detected in three beams of the Parkes multi-beam receiver -- implying a location between the beams, which required an extremely high inferred peak flux density. The burst saturated the primary detection beam and was initially estimated to have a peak flux density of 30 Jy and a fluence of 200 Jy~ms \citep{2007Sci...318..777L}. Subsequent re-analysis of the data by \citet{2011ApJ...727...18B} detected the FRB signal weakly in a fourth beam of the receiver. Based on new beam pattern models of the Parkes multi-beam receiver, a revised analysis of FRB~010724 by \citet{2019MNRAS.482.1966R} localized FRB~010724 to a region of a few square arcminutes within the primary beam and the true fluence was estimated to be 800 $\pm$ 400 Jy~ms, further solidifying the Lorimer burst as one of the most luminous known FRBs.

While FRB~010724 remains an outlier in the Parkes FRB population, several FRBs in the ASKAP sample appear to have similar fluences \citep{2018Natur.562..386S}.  This is perhaps unsurprising given that the ASKAP surveys provided much larger sky coverage, but at lower sensitivity compared with Parkes.  Recent studies of the ensemble properties of FRBs have remarked that the Lorimer burst strongly affects the slope of the source counts and other statistics related to the brightness distribution of FRBs. \citet{2018MNRAS.474.1900M} have argued that FRB~010724 should be excluded from statistical analyses of the FRB population and that it is subject to discovery bias -- i.e. the idea that the first-discovered source in a new class may be easier to detect and have exceptional properties compared to the rest of the underlying population. 

\subsection{FRB 010621 $-$ The Keane Burst}\label{sec:KeaneBurst}

FRB~010621, also known as `the Keane burst' was the second candidate FRB reported in the literature.  Presented in \citet{2011MNRAS.415.3065K}, and further discussed in \citet{2012MNRAS.425L..71K}, the burst was discovered in a search of the Parkes Multibeam Pulsar Survey \citep[PMPS;][]{2001MNRAS.328...17M} for single pulses from RRATs and Lorimer-type bursts. The single bright pulse was reported with a DM of 745$\pm$10 cm$^{-3}$ pc along a sightline where the modeled DM contribution from the Galaxy is 523 cm$^{-3}$ pc according to the NE2001 model (although the line-of-sight DM$_\mathrm{MW}$ is only estimated to be 320 cm$^{-3}$ pc in the YMW16 model). The small fractional DM excess of the pulse relative to the expected DM of the Galaxy in that direction made it unclear whether the pulse was extragalactic in origin or from a Galactic source located along an overdense sightline through the Galactic plane. \citet{2014MNRAS.440..353B} studied the sightline of FRB~010621 in $H \alpha$ and $H \beta$ emission to determine a more precise electron density measurement and concluded with 90\% confidence that the burst was from a Galactic source along an overdense sightline. Unless repeating pulses, allowing a precise localization and a host galaxy association, are detected in the future, the true distance will remain uncertain.  FRB~010621 is thus considered a marginal case between the FRB and Galactic pulse source classes.

\subsection{FRB~140514}\label{sec:FRB140514}

FRB~140514, also known as `the Petroff burst', was discovered in a targeted search of the locations of previously detected FRBs, where the motivation was to search for repeating pulses from these sources \citep{2015MNRAS.447..246P}. It was found in the field of the previously reported bright FRB~110220 \citep{2013Sci...341...53T} in a receiver beam pointed 9$'$ away from the reported location of the previous FRB. Despite the similar sky location, the two FRBs were discovered with markedly different DMs: 944.38$\pm$0.05 cm$^{-3}$ pc for FRB~110220, and 562.7$\pm$0.6 cm$^{-3}$ pc for FRB~140514. \citet{2015MNRAS.447..246P} thus concluded that the bursts were not related and estimated a 32\% probability of finding two positionally similar but physically unrelated FRB in the survey as a whole. However, \citet{2015MNRAS.454.2183M}, using the argument that FRB~140514 occurred in the receiver beam pointed to the field of FRB~110220, concluded that the two bursts must be from the same source with 99\% confidence. Ultimately, the difference in statistical analyses between the two teams come from considering only a single beam of the Parkes multi-beam receiver or the entire receiver field of view \citep[see further discussion in Chapter 6, ][]{2016PhDT.......184P}. 

If FRB~110220 and FRB~140514 are indeed two bursts from the same source separated by 3 years, \citet{2017ApJ...841L..30P} argue that the source could be a neutron star embedded in a dense supernova remnant and the large change in DM could be explained by a shell of material expanding radially outward. In order to produce such a large fractional change they estimate that the supernova would have to have occurred within $\sim$10.2 years of FRB~110220. 

FRB~140514 was also the first discovery by a newly commissioned real-time search pipeline in operation at the Parkes telescope, which enabled the full polarimetric properties of the FRB to be recorded. The burst was found to be 20\% circularly polarized, with no detection of linear polarization. See \S \ref{sec:polarization} for a more detailed discussion.

\subsection{FRB~121102}\label{sec:FRB121102}

Discovered using the 305-m Arecibo telescope in Puerto Rico, \aofrb, also known as `the Spitler burst', was the first FRB to be detected with a telescope other than Parkes.  As such, it added credence to the astrophysical interpretation of the phenomenon in the early days of the field. \citet{2014ApJ...790..101S} discovered the burst in a single-pulse search of archival data from the PALFA Galactic plane survey \citep{2006ApJ...637..446C,2015ApJ...812...81L}.  It was the only burst seen in a 180-s observation, and no additional bursts were seen in a second survey scan coincidentally taken 2 days later.  \aofrb\ is in the Galactic anti-center at $l = -0.2^{\circ}$, $b = 175^{\circ}$.  The DM$ = 557$\,cm$^{-3}$\,pc is 300\% larger than that predicted by the NE2001 model \citep{2002astro.ph..7156C}, which suggested an extragalactic origin despite the low Galactic latitude of the source.  Curiously, the spectrum of the burst is inverted, following approximately $S_{\nu} \propto \nu^7$.  This led \citet{2014ApJ...790..101S} to hypothesize that the burst was detected in a side lobe of the ALFA 7-beam receiver.

At the time of discovery, it was unclear whether \aofrb\ was a genuine extragalactic burst, a RRAT with an anomalously high DM, or some type of pernicious RFI.  
While initial follow-up observations detected no additional bursts \citep{2014ApJ...790..101S}, a deeper campaign was planned to better establish whether \aofrb\ was truly a one-off event.  
Deep follow-up of the Lorimer and Keane bursts had made no additional detections and similar follow-up of other Parkes FRBs yielded no repeating pulses \citep{2015MNRAS.454..457P}. Thus it came as a surprise when Arecibo observations in May 2015 detected the first repeat bursts from \aofrb\ \citep{2016Natur.531..202S}.  These additional follow-up observations used the 7-beam Arecibo ALFA receiver to grid a large area around the original detection position.  Perhaps most surprising was how active \aofrb\ suddenly was.  Of the 10 new bursts detected by \citet{2016Natur.531..202S}, 6 were discovered within a 10-minute observation and some were substantially brighter compared with the first-detected burst.  The new detections showed that the original \aofrb\ burst had been detected in the sidelobe of one of the telescope beams; however, each new burst had a different spectrum, sometimes poorly modeled by a power-law and peaking within the observing band.  The strange spectrum was therefore something characteristic to the signal itself and not an instrumental artifact.

In terms of constraining theory, the detection of repetition provides a clear constraint: the FRB cannot come from a cataclysmic event and whatever is producing the bursts must be able to sustain this activity over a period of at least 7 years -- 2012 to present day.
The repeating pulses made it possible to study the source in greater detail and perform multi-wavelength measurements.  Most importantly, it became possible to precisely localize the source using a radio interferometer, without having to do this in real-time using the initial discovery burst. 

\citet{2016ApJ...833..177S} presented additional detections of \aofrb\ using Arecibo and the GBT.  They also performed a multi-wavelength study of the field around \aofrb\ and showed that it was unlikely that the source's high DM was produced by a Galactic H\textsc{ii} region along the line-of-sight.  

\begin{figure}
\subfloat[VLA localization of FRB 121102\label{fig:Chatterjee1}]{\includegraphics[width=0.6\textwidth]{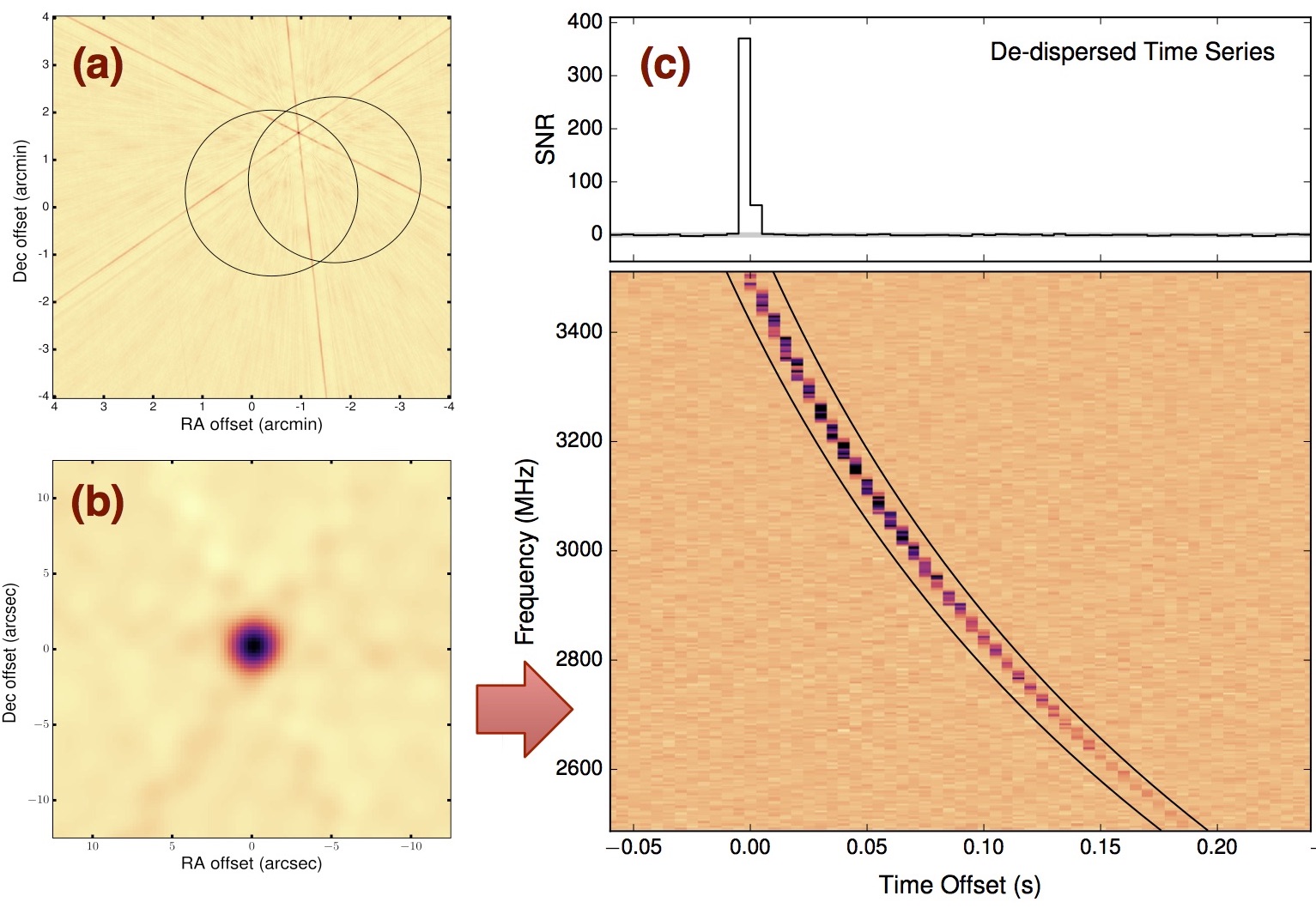}}
\hfill
\subfloat[Host galaxy identification for FRB 121102\label{fig:Chatterjee2}]{\includegraphics[width=0.4\textwidth]{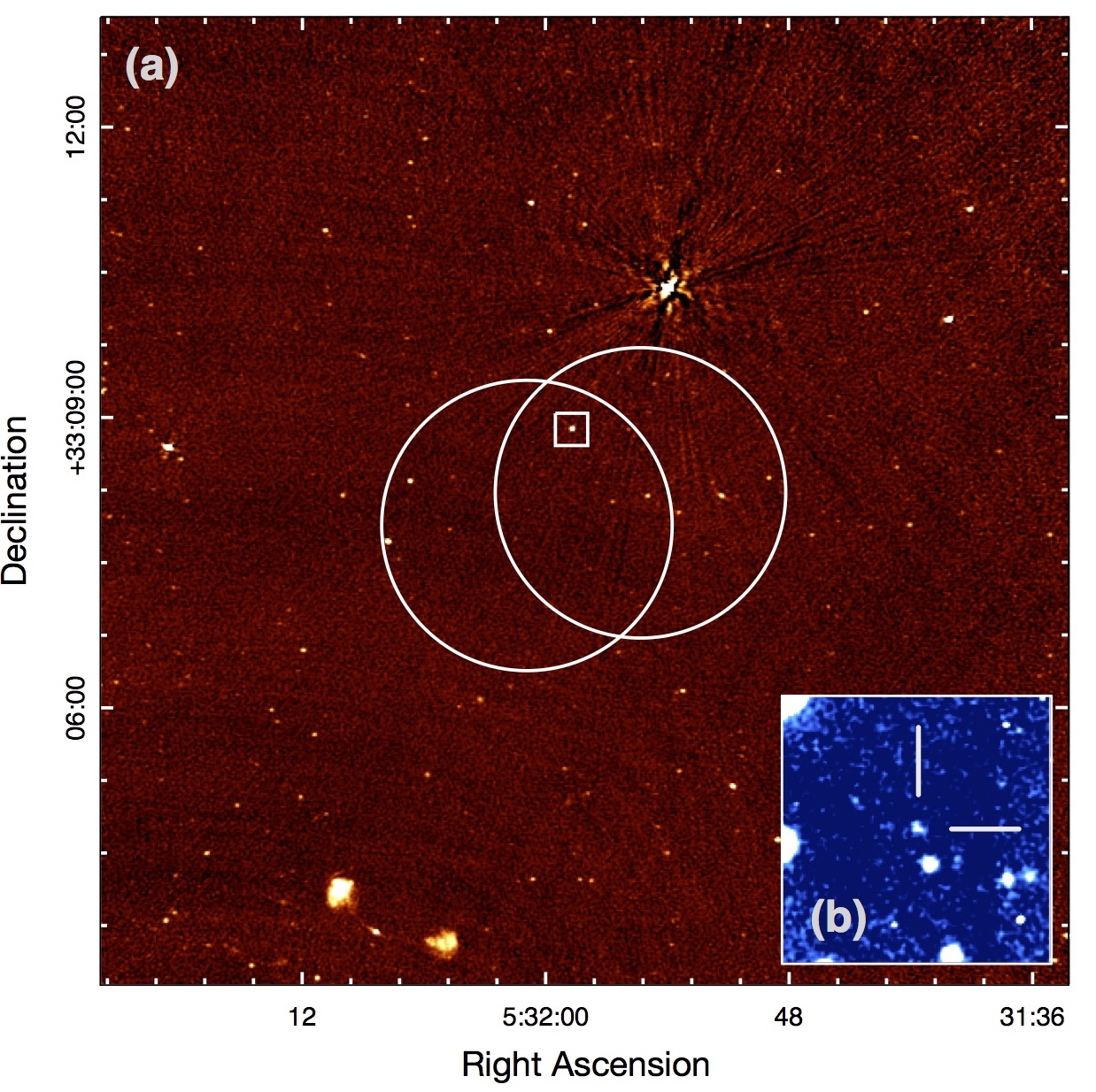}}
\caption{{\it Left}: Using the VLA, repeating bursts from FRB~121102 were localized to sub-arcsecond precision using interferometric techniques.  {\it Right:} The localization allowed for the identification of the host galaxy at radio and optical (inset) wavelengths. Figures 1 and 2 from \citet{2017Natur.541...58C}. \label{fig:Repeater}}
\end{figure}

At the same time, the VLA and European VLBI Network (EVN) were used to obtain a precision localization.  After tens of hours of observations with the VLA, 9 bursts were detected using high-time-resolution (5\,ms) visibility dumps \citep{2017Natur.541...58C}, which localized \aofrb\ to $\sim 100$\,mas precision (Fig.~\ref{fig:Repeater}, left). This allowed \citet{2017Natur.541...58C} to see that \aofrb\ is coincident with both persistent radio and optical sources (Fig.~\ref{fig:Repeater}, right).  Very-long-baseline radio interferometric observations using the EVN and Very Long Baseline Array (VLBA) showed that the radio source is compact on milli-arcsecond scales.  Archival optical images from the Keck telescope suggested that the optical source was slightly extended.

\citet{2017ApJ...834L...8M} managed to detect additional bursts using EVN data, providing another step in localization precision. \aofrb\ and the persistent source were found to be coincident to within $\sim 12$\,mas.  In parallel, \citet{2017ApJ...834L...7T} acquired Gemini North spectroscopic data that detected the optical source and measured its redshift: $z = 0.193$, which corresponds to a luminosity distance of $\sim 1$\,Gpc.  The extragalactic origin and host galaxy of \aofrb\ were thus conclusively established.

\aofrb's host galaxy turned out to be a low-metallicity, low-mass dwarf \citep{2017ApJ...834L...7T}.  Given that such galaxies are also known to be the common hosts of superluminous supernovae (SLSNe) and long gamma-ray bursts (LGRBs), this presented a tantalizing possible link between FRBs and these other types of extreme astrophysical transients \citep{2017ApJ...841...14M}.  Deeper observations of the host using the Hubble Space Telescope (HST) revealed that \aofrb\ is coincident with an intense star-forming region \citep{2017ApJ...843L...8B}. The EVN radio position is offset from the optical centroid of the star-forming region by 55\,mas, statistically significant, but within the half-light radius.

\begin{table}[h!]
\caption{Observed properties of FRB 121102 and their possible physical interpretations, from $^{1}$\citet{2016Natur.531..202S},  $^{2}$\citet{2018Natur.553..182M}, $^{3}$\citet{2018arXiv181110748H}, $^{4}$\citet{2018ApJ...863....2G}, $^{5}$\citet{2017ApJ...834L...7T}, and $^{6}$\citet{2017ApJ...843L...8B}.  \label{tab:FRB121102}}
\begin{tabular}{llll}
\hline
\hline
Description & Measurement & Interpretation\\
\hline
\hline
Bursts repeat$^{1}$ & $>10$ bursts detected & Non-cataclysmic origin \\
\hline
Bursts are polarized$^{2}$ & $\sim 100$\% linearly polarized & Related to\\ 
& $\sim 0$\% circularly polarized & emission mechanism\\
\hline
Bursts show complex & Sub-bursts drifting & Related to \\
time-frequency structure$^{3}$ & to lower frequencies & emission mechanism \\
                    & & or propagation effects \\
\hline
Large and variable & $\sim 147,000-100,000$\,rad\,m$^{-2}$ & Extreme and dynamic local \\ 
rotation measure$^{2,4}$ & within 7 months & magneto-ionic environment \\
\hline
Hosted in a low-metallicity & Host $M_* \sim 10^8$\,M$_\odot$ & Possible connection \\
 dwarf galaxy$^{5}$  & & with SLSNe \& LGRBs \\
\hline
Co-located with & SFR $\sim$ 0.23 M$_\odot$ yr$^{-1}$ & Possible late stellar \\
 star-forming region$^{6}$ & & evolution origin\\
\hline
\hline
\end{tabular}
\end{table}

Multi-wavelength observations also searched for prompt optical, X-ray and $\gamma$-ray flashes associated with the radio bursts.  No optical pulses were found in a campaign where the 2.4-m Thai National Telescope was shadowed by Effelsberg and 13 radio bursts were detected \citep{2017MNRAS.472.2800H}.  Similarly, despite the detections of multiple radio bursts, no prompt X-ray or $\gamma$-ray bursts were found in observations with simultaneous radio and \textit{Chandra}, \textit{XMM-Newton}, \textit{Swift}, and \textit{Fermi} coverage.  Nor is there any persistent X-ray/$\gamma$-ray emission detected \citep{2016ApJ...833..177S,2017ApJ...846...80S}.

In the absence of high-energy bursts, the radio bursts themselves become even more important for interpreting \aofrb.  The precision localization has allowed for observations at higher radio frequencies ($> 2$~GHz) and using higher time and frequency resolution. Given that the DM of the source is known, on-line coherent dedispersion can be used in order to avoid intra-channel dispersive smearing. The earliest high-frequency burst detections were made at 5~GHz with Effelsberg \citep{2018ApJ...863..150S} and at 3~GHz with the VLA \citep{2017ApJ...850...76L}. \citet{2018ApJ...863....2G} detected over a dozen bursts in only a 30-minute observing window using an ultra-wideband recording system from $4-8$~GHz.  Arecibo observations from $4-5$~GHz also detected over a dozen bursts, and the full Stokes recording mode allowed polarization to be detected for the first time.  The bursts were found to be $\sim 100$\% linearly polarized with a rotation measure of $1.46 \times 10^{5}$~rad~m$^{-2}$ that decreased to $1.33 \times 10^{5}$~rad~m$^{-2}$ within 7 months \citep[in the source frame; ][]{2018Natur.553..182M}.  This demonstrated that \aofrb\ is in an extreme and dynamic magneto-ionic environment.  It also distinguished the first repeater in a new way: no other FRB had been shown to have such a large RM.

Most recently, \citet{2018arXiv181110748H} used a sample of high-S/N, coherently dedispersed bursts to demonstrate complex time-frequency patterns in the signals from \aofrb.  This is discussed in more detail in \S\ref{sec:emission}, and it may represent a means to observationally separate repeating and non-repeating FRBs.  \citet{2019arXiv190302249G} studied a sample of low-S/N, low-energy ($10^{37-38}$~erg/s) \aofrb\ bursts and showed that their typically narrow-band spectra ($\sim 200$~MHz at 1400~MHz) are a significant impediment to detection when using standard search methods.  It is certain that current methods are sub-optimal and that bursts are being missed; one can speculate that this is true not only for \aofrb, but for FRBs in general.

Table~\ref{tab:FRB121102} summarizes the observational properties of FRB~121102 and its host galaxy.

\subsection{FRB~180814.J0422+73}\label{sec:R2}

In January, 2019 it was reported by the CHIME/FRB collaboration that a second repeating FRB was discovered in the pre-commissioning data from the telescope. This repeating burst source, FRB~180814.J0422+73 (also referred to colloquially as `R2', whereas FRB~121102 is `R1') was found at a very low dispersion measure DM = 189 cm$^{-3}$ pc.  Bursts were detected at 6 epochs between August and October 2018 \citet{2019Natur.566..235C}. The FRB source was found in a circumpolar region of the sky, meaning that it was visible to the CHIME telescope in both `upper' and `lower' transits. Using all detections, FRB~180814.J0422+73 was published with an estimated position: RA = 04:22:22, Dec = +73:40 with uncertainties of $\pm 4'$ in RA and $\pm 10'$ in Dec. 

\begin{figure}
    \centering
    \includegraphics[width=0.8\textwidth]{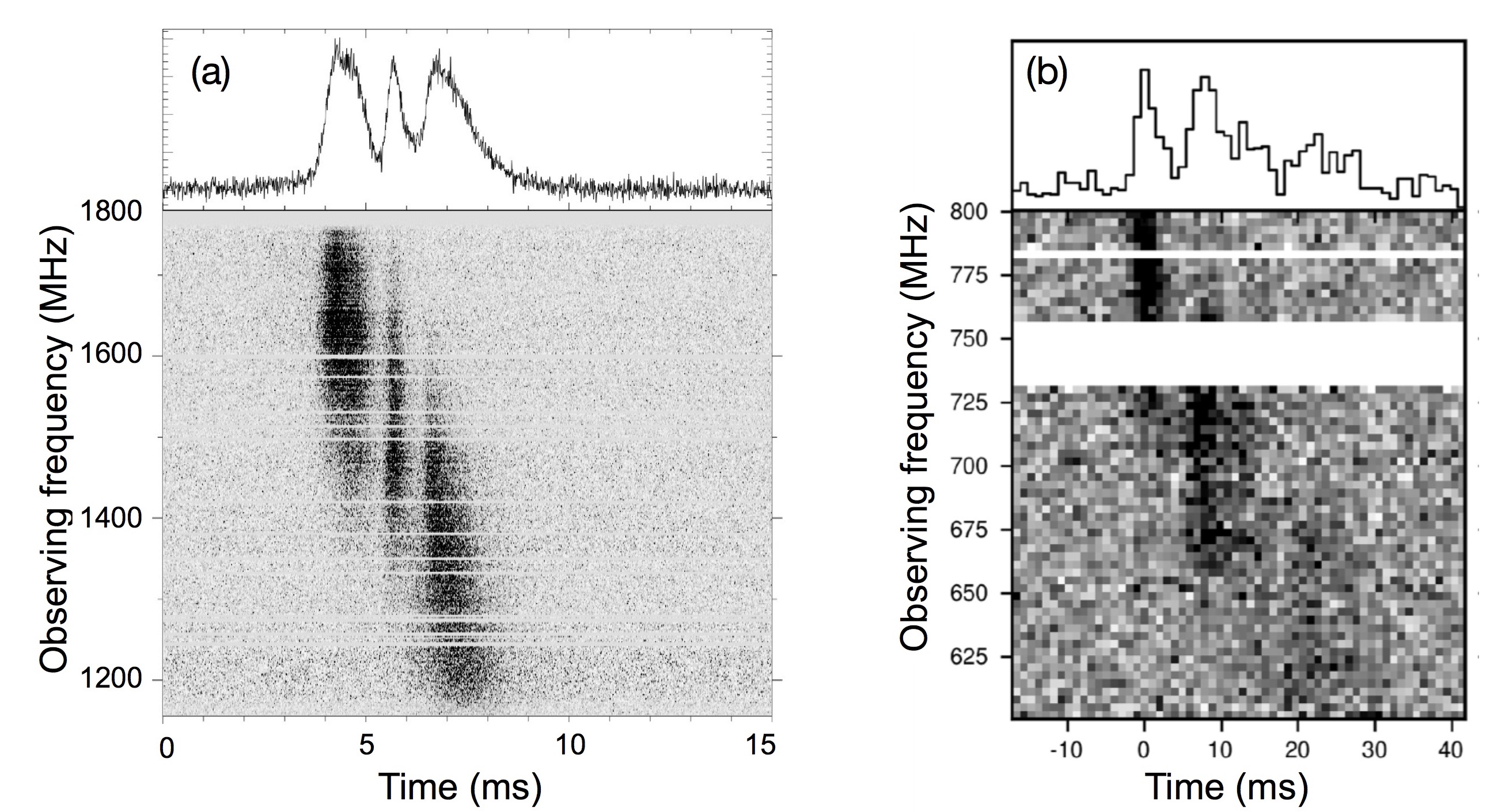}
    \caption{De-dispersed spectra of individual bursts from (a) the repeating FRB~121102 at 1.4~GHz using Arecibo, and (b) the repeating FRB~180814.J0422+73 discovered with CHIME at 700~MHz. Both repeating sources have some bursts that show distinct sub-burst structure with descending center frequencies over time. Horizontal bands in both spectra are due to narrow-band RFI excision in the data. FRB~121102 data from \citet{2018arXiv181110748H}.  FRB~180814.J0422+73 data from \citet{2019Natur.566..235C}.
    \label{fig:R1R2}}
\end{figure}

Interestingly, at least two bursts from R2 show complex time-frequency structure. These bursts show multiple sub-bursts, each with finite frequency bandwidth with earlier sub-bursts peaking in brightness at higher frequencies. The descending time-frequency structure within a total burst envelope is similar to structure seen in some pulses from FRB 121102 \citep{2018arXiv181110748H}. That this structure is seen in some pulses from both repeaters (Fig~\ref{fig:R1R2}) may indicate that the origin is intrinsic to the emission mechanism rather than an extrinsic propagation effect that requires a particular geometry, such as plasma lensing. 

Ultimately the full extent of similarities between the two repeaters is not yet known. Many properties of R2 remain un-probed as it has not yet been extensively studied. In the near future, with a more precise localization of R2 we may be able to make more comparisons between the two known repeating FRBs. The most important comparisons will be not only the polarimetric properties and rotation measures, but also whether FRB~180814.J0422+73 is associated with a persisent radio source, the properties of the host galaxy such as type, metallicity, star formation rate, and size, and the host redshift. 

\section{Population properties}\label{sec:popFRBs}

Here we describe the properties of FRBs as an ensemble.  Such considerations inform how we can optimize future FRB searches, whether there are observational sub-classes, and are a critical input for constraining theory.

\subsection{FRB polarization and rotation measures}\label{sec:polarization}

\begin{figure}
\includegraphics[width=\textwidth]{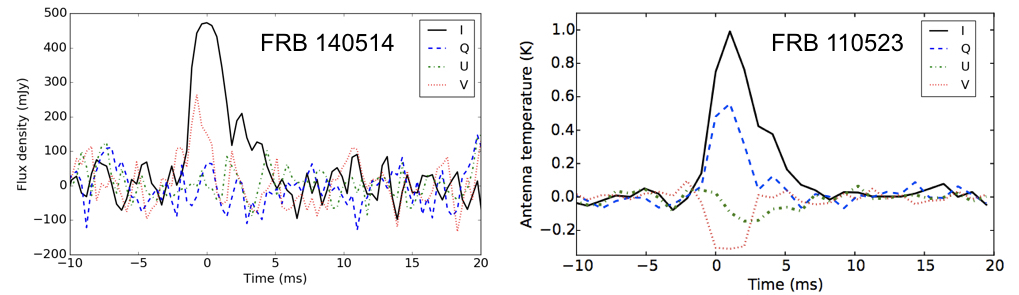}
\caption{Polarization profiles for FRB 140514 (left), the first FRB with measured circular polarization \citep{2015MNRAS.447..246P}, and FRB 110523 (right), the first FRB with measured linear polarization \citep{2015Natur.528..523M}. The Stokes parameters for total intensity I (solid), Q (dashed), U (dot-dashed), and V(dotted) are plotted for each burst. FRB~140514 profile from Fig.~1 of \citet{2015MNRAS.447..246P}; FRB~110523 profile from Fig.~3 of \citet{2015Natur.528..523M}.  \label{fig:polarization}}
\end{figure}

Currently only 9 of the more than 60 cataloged FRBs have polarimetric data available.  From this subset, we already see a heterogeneous picture emerging (Fig.~\ref{fig:polarization}): some FRBs appear to be completely unpolarized (e.g. FRB~150418), some show only circular polarization (e.g. FRB~140514), some show only linear polarization (e.g. FRBs~121102, 150215, 150817, 151230), and some show both (e.g. FRBs~110523, 160102).  A recent overview can be found in \citet[][see their Table~1 and references therein]{2018MNRAS.478.2046C}.  In one case, an FRB candidate (FRB~180301) has shown frequency-dependent polarization properties \citep{2019arXiv190107412P}, which may be indicative of a non-astrophysical progenitor if they cannot be explained through propagation effects \citep[e.g.][]{2019arXiv190201485G,2019MNRAS.tmpL..41V}. These varied polarization properties do not necessarily reflect different physical origins, however.  In analogy with pulsars, which show a wide variety of polarization fractions between sources, as well as individual pulses, a single type of emitting source could be responsible for the observed range of FRB polarization properties.  The heterogeneity in FRB polarization properties could thus arise from time-variable emission properties, different viewing geometries, or different local environments.

In the cases where linear polarization can be measured, the polarization angle as a function of time (across the burst duration) and frequency (across the observed bandwidth) can be measured (Eq.~\ref{eq:PA}). Though S/N is low in most cases, FRBs have thus far not shown large polarization angle swings.  Polarization swings are often, though not always, seen in radio pulsars, and are attributed to viewing different magnetic field lines from the neutron star polar cap as the radio beam sweeps past.  In radio pulsars, flat polarization swings are normally attributed to aligned rotators or large emission heights.

By measuring polarization angle as a function of frequency, Faraday rotation can be quantified (see \S \ref{sec:faraday}).  Here too, the known FRB population has presented a heterogeneous picture: while some FRBs  have rotation measures (RMs) $\sim$10 rad m$^{-2}$ that are consistent with that expected from the Galactic foreground (e.g. FRBs~150215, 150807), others have much higher RMs, which point to a dense and highly magnetised local environment. \citet{2015Natur.528..523M} presented the first detection of linear polarization from an FRB, and the derived RM = $- 186.1 \pm 1.4$\,rad\,m$^{-2}$ led them to conclude that the source is in a dense environment or surrounded by a nebula. Recently, FRB~160102 has also been found to have a relatively large RM ($-220 \pm 6.4$~rad~m$^{-2}$) \citep{2018MNRAS.478.2046C}.  Most strikingly, the repeating FRB~121102 was found to have an extremely high RM $\sim 10^5$\,rad\,m$^{-2}$ (see \S \ref{sec:FRB121102} and Fig.~\ref{fig:faraday} for more details). Such high RM values are difficult to detect given the limited frequency resolution in most FRB search experiments, and thus FRBs with apparently no linear polarization could potentially be high-RM sources de-polarized by intra-channel Faraday rotation smearing.  This could be the case for FRB~140514, which was the first FRB with detected polarization \citep[$\sim30$\% circular;][]{2015MNRAS.447..246P}.

Conversely, some FRBs show high linear polarization fraction, but low RM. \citet{2017MNRAS.469.4465P} showed that FRB~150215 (43$\pm$5\% linearly polarized) has an RM in the range $-9 < \mathrm{RM} < 12$\,rad\,m$^{-2}$ (95\% confidence level), i.e. consistent with zero and demonstrating a low Galactic foreground contribution.  Likewise, \citet{2016Sci...354.1249R} found RM $= 12.0 \pm 0.7$\,rad\,m$^{-2}$ for FRB~150807 (80$\pm$1\% linearly polarized), and used this to constrain the magnetic field of the cosmic web to $< 21$\,nG (parallel to the line-of-sight).  In both cases, the low RM points to negligible magnetization in the circum-burst plasma.

It is clear that measuring the RM provides an important way of characterizing FRB local environments, and may lead to clarity on whether there are multiple sub-classes of FRB. The increasing use of real-time triggering and full-polarization (or even voltage) data dumps should mean that a larger fraction of future FRB discoveries will have known polarimetric properties. Even the preservation of full-Stokes data for upcoming surveys with relatively narrow frequency channels may be sufficient to recover polarization profiles for many FRBs.

\subsection{Multi-wavelength follow-up of FRBs}\label{sec:multiwavelength}

Despite multi-wavelength searches, to date prompt FRB emission has only been convincingly detected at radio frequencies between 400\,MHz \citep{2019Natur.566..230C,2019Natur.566..235C} and 8\,GHz \citep[FRB 121102; ][]{2018ApJ...863....2G,2018Natur.553..182M}. 
Prompt emission outside of the radio band
has so far only been claimed in one source, FRB~131104, in a study by
\citet{2016ApJ...832L...1D} who searched archival {\it Swift} data around the times of several known FRB events. These authors claimed the detection of a gamma-ray transient associated with FRB~131104.  However, given the low significance of the X-ray signal ($3.2\sigma$), the association is arguably tenuous \citep[for a discussion, see][]{2017ApJ...837L..22S}. Further progress in this area can be made by dedicated experiments. One such study, currently in progress with a 20-m telescope at the Green Bank Observatory shadows the {\it Swift} daily source list for FRBs in the field of view (Gregg et al.~in preparation).

For longer-term emission akin to afterglows in GRBs, we note that since the FRB isotropic energy is about 10 orders of magnitude smaller than GRBs, the predicted FRB multi-wavelength afterglow is much fainter \citep[see, e.g.,][]{2014ApJ...792L..21Y}. In spite of these challenges, it is of great importance to continue to search for
longer-term emission. In one such study, \citet{2016Natur.530..453K} mounted an unprecedented multi-wavelength follow-up campaign triggered by FRB~150418.  This revealed a fading radio counterpart in the positional uncertainty region of the FRB.  Assuming an association with the FRB~150418, this led to the identification of a candidate host galaxy and its redshift.  However, this association has been disputed because of the non-negligible chance of a variable radio source in the field \citep{2015MNRAS.450.4221B}. \citet{2016ApJ...821L..22W} conducted additional radio follow-up and found that the candidate radio counterpart was continuing to vary and even re-brightened to the same levels as in the days following FRB~150418. They concluded that the source was a variable active galactic nucleus and could not be conclusively linked to the FRB source. \citet{2017ApJ...849..162E} and \citet{2018ApJ...860...73E} discuss the challenges of identifying FRB counterparts and show that, for
FRBs and hosts out to redshifts of $\sim 1$, positional determinations at the level of at least 20~arcseconds (and in some cases much better) are required in order to provide robust associations.

The repeating FRB~121102 has provided a great practical advantage for multi-wavelength follow-up (as described in detail in \S \ref{sec:FRB121102}). Other repeating FRB sources will be discovered in the future and followed-up in similar ways.  Importantly, the increasing use of real-time searches will also allow near-real-time triggering of multi-wavelength instruments to look for afterglows through machine-parsable automated mechanisms such as VOEvents \citep{2017arXiv171008155P}. Several experiments are also using multi-telescope shadowing, which could lead to the detection of multi-wavelength prompt emission -- e.g., the MeerLicht optical telescope shadowing radio searches with MeerKAT \citep{2016SPIE.9906E..64B}.

Recently, ever more detailed follow-up efforts have been undertaken after the discovery of new FRBs. \citet{2018MNRAS.475.1427B} undertook follow-up for FRBs~151230 and 160102 from X-ray to radio wavelengths including some of the first searches for associated neutrino emission with the ANTARES neutrino detector. Ultimately, without a precise localization of the sources from their radio bursts and the unknown multi-wavelength nature of FRB emission, it is difficult to pinpoint the location of an FRB from follow-up but these observations place limits that are useful for future targeted searches.

\subsection{Properties of the FRB population}

\begin{figure}
\subfloat[Width vs. DM\label{fig:DMWidth}]{\includegraphics[width=0.54\textwidth]{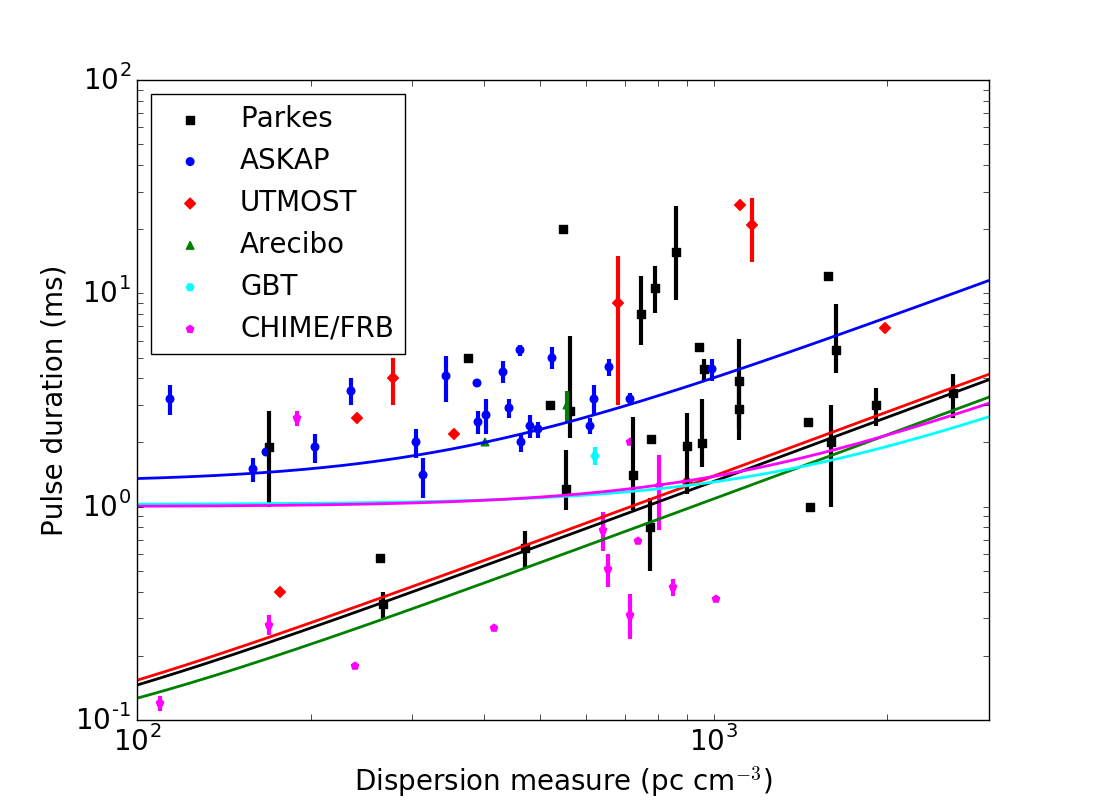}}
\hfill
\subfloat[Scattering vs. DM\label{fig:DMScattering}]{\includegraphics[width=0.48\textwidth]{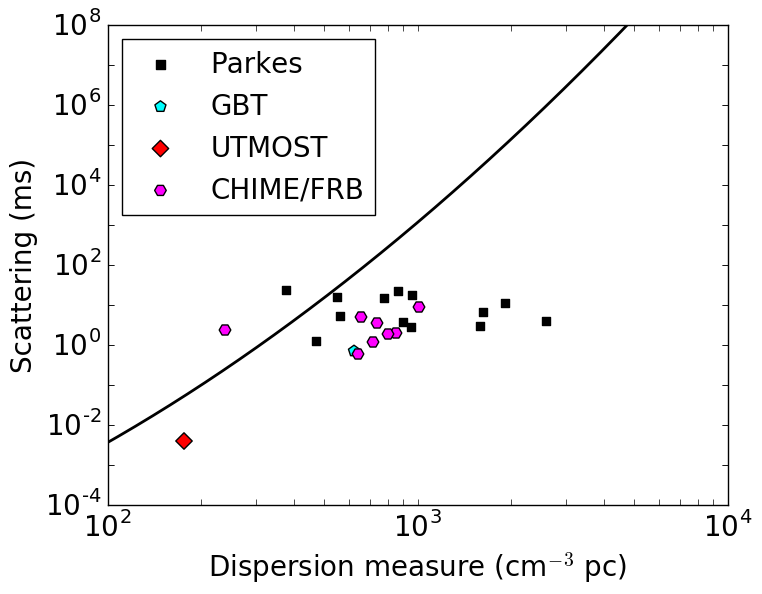}}
\vfill
\subfloat[DM excess histogram\label{fig:DMHist}]{\includegraphics[width=0.5\textwidth]{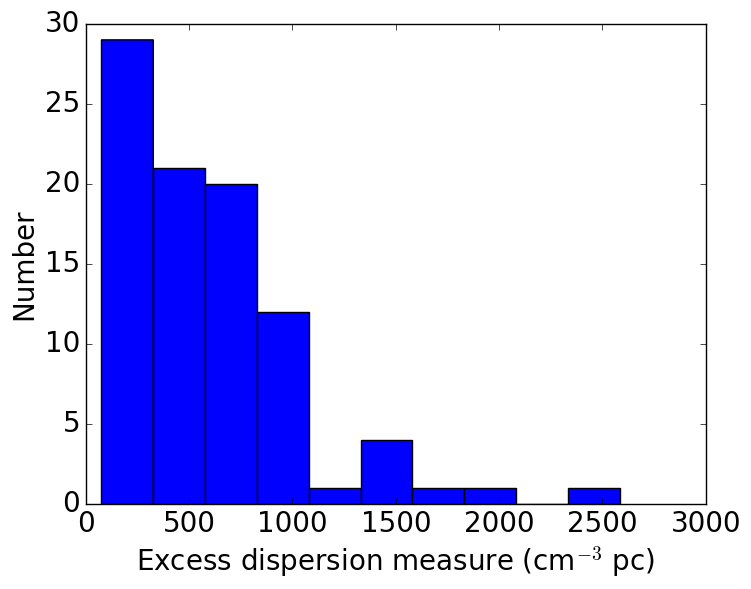}}
\hfill
\subfloat[Pulse duration histogram\label{fig:widthHist}]{\includegraphics[width=0.5\textwidth]{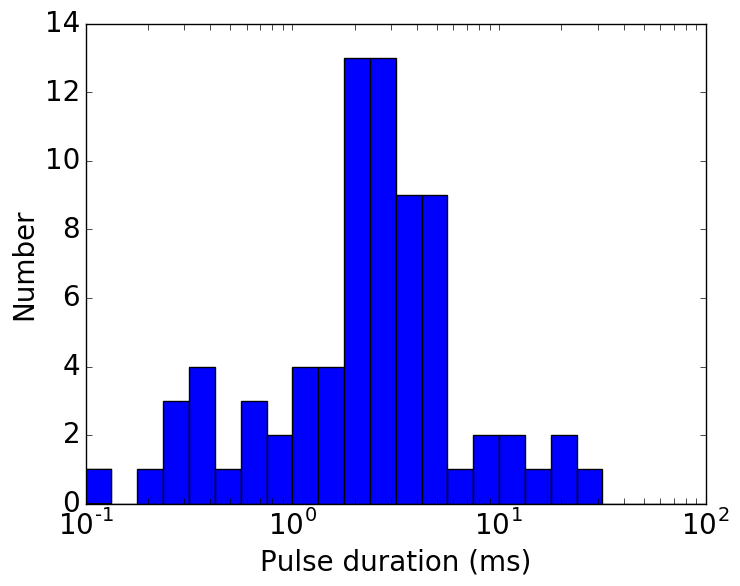}}
\caption{
The properties of the catalogued FRB population. (a) The pulse duration (width) versus DM. Solid lines represent temporal broadening from DM smearing in an individual frequency channel combined with the sampling time for different telescopes. In the case of FRBs from CHIME, plotted widths have been obtained through modeling and are not the observed FRB widths from the instrument. (b) Scattering timescale versus DM for all FRBs where scattering has been measured. The curve shows the DM-scattering relation for pulsars in the Galaxy derived by \citet{2004ApJ...605..759B}. FRBs are under-scattered relative to Galactic pulsars of similar DMs. 
(c) A histogram of the DM excess compared to the expected Galactic maximum along the line of sight. (d) A histogram of the pulse durations. For panels (a) and (b) colors correspond to the Parkes (black), ASKAP (blue), Arecibo (green), UTMOST (red), GBT (aqua) and CHIME (pink) telescopes.
\label{fig:Section6}}
\end{figure}

\begin{figure}
\centering
\includegraphics[width=\columnwidth, trim={0.5cm 0cm 0.5cm 0cm}]{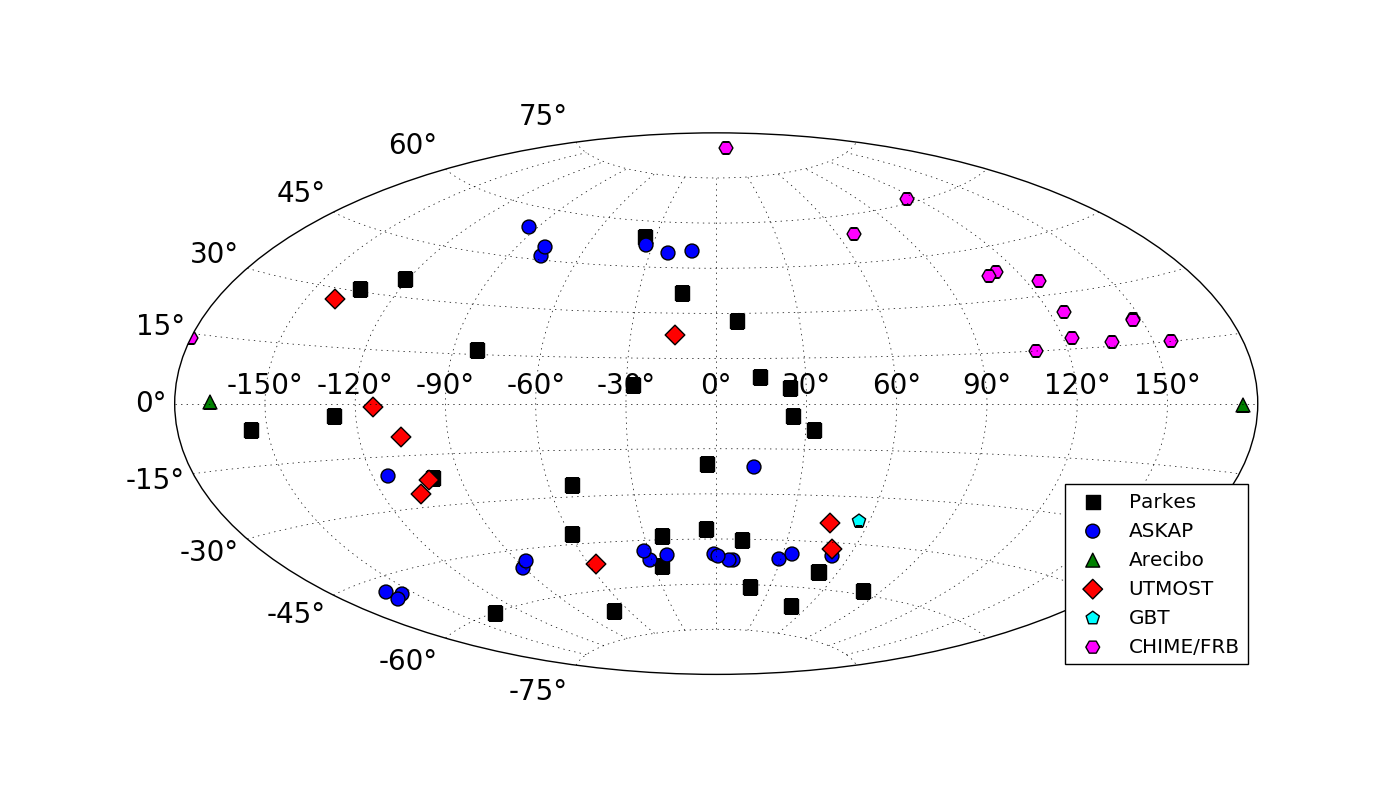}
\caption{An Aitoff projection map of the sky positions of all published FRBs as a function of Galactic longitude and latitude. As in Fig.~\ref{fig:Section6}, colors correspond to the Parkes (black), ASKAP (blue), Arecibo (green), UTMOST (red), GBT (aqua) and CHIME (pink) telescopes. \label{fig:glgbAitoff}}
\end{figure}

In \S \ref{sec:skyDist}, \S \ref{sec:DMdistribution}, and \S \ref{sec:widthDistribution} we consider the specific distributions of FRBs over the sky, in DM, and in pulse duration. First, however, we consider some of the two-dimensional distributions of the population as a function of various parameters. These are shown for some subsets of the known population in Fig.~\ref{fig:Section6}. 

In Fig.~\ref{fig:DMWidth} we show the pulse widths of FRBs versus their measured DM. Over-plotted are the curves per telescope showing the effects of instrumental smearing from Eq.~\ref{eq:smearing}, combined with survey sampling time as a function of DM. Some FRBs from each observing instrument closely follow this line, meaning that their intrinsic widths may in fact be much lower. In the range 500 cm$^{-3}$ pc $<$ DM $<$ 1500 cm$^{-3}$ pc pulse duration does seem to increase with DM, but this trend does not hold at the higher DMs where most FRBs are found with durations $<$ 10 ms. 

Fig.~\ref{fig:DMScattering} plots the scattering timescales, where measured for individual FRBs, versus their DMs. While currently only roughly 20 FRBs have published scattering timescales, the shape of this distribution may change as a larger population have measured scattering parameters. The existing data, however, do provide an intriguing picture of limits
on radio-wave scattering for FRBs. Most
notably, unlike the well-known correlation seen for Galactic pulsars \citep[see, e.g.,][]{2004ApJ...605..759B}, there does not appear to be a similar trend in the FRB distribution. As remarked by a number of authors \citep[see, e.g.,][]{2013MNRAS.436L...5L,2016arXiv160505890C}
for cases where most of the scattering
is produced at the source, a lever-arm
effect tends to minimize scatter broadening. The lack of any correlation with DM also
suggests that the IGM plays a very minor role in pulse broadening for FRBs
\citep{2016arXiv160505890C,2016ApJ...832..199X}.

Fig.~\ref{fig:DMHist} plots a histogram of FRB DMs in excess of the modeled Galactic contribution (see \S \ref{sec:DMdistribution}) and Fig.~\ref{fig:widthHist} plots a histogram of the FRB pulse durations (see \S \ref{sec:widthDistribution}). 

\subsection{The sky distribution}\label{sec:skyDist}

The sky distribution of all published FRBs is shown in Fig.~\ref{fig:glgbAitoff}. Early non-detections of FRBs at intermediate and low Galactic latitudes by the Parkes telescope led \citet{2014ApJ...789L..26P} to conclude that the FRB detection rate is greater at high Galactic latitudes. They found the HTRU results to be incompatible with an isotropic distribution at the 99\% confidence level based on 4 FRB detections at high Galactic latitudes and no detections at intermediate latitudes ($|b| < 15^{\circ}$) in a longer observing time. This was further supported by analysis from \citet{2014ApJ...792...19B}, upon the discovery of FRB~010125, which concluded that the high and low latitude FRB rates were strongly discrepant with 99.69\% confidence, although this confidence level may have been overstated even at the time \citep{2016MNRAS.458L..89C}. \citet{2015MNRAS.451.3278M} attributed the observed disparities found in these works to diffractive scintillation at higher Galactic latitudes, which boosts FRBs that might otherwise not be detected (see also \S\ref{sec:scint}).  The scintillation bandwidth is much wider along high latitude sight lines, and comparable to the observing bandwidth used by most surveys at Parkes. Conversely, in their study of the FRB rate, \citet{2016MNRAS.455.2207R} found no evidence to support a non-isotropic sky dependence of the distribution.

Recent studies have been somewhat more successful at higher Galactic latitudes, and some searches, such as the ASKAP Fly's Eye pilot study, have purposely concentrated their time on sky at high latitudes to maximize detections \citep{2017ApJ...841L..12B,2018Natur.562..386S}. As the population of FRBs grows, however, the statistical significance of the latitude-dependent detection rate has gotten much weaker and early indications of anisotropy may have been an artifact of small number statistics. Using 15 FRBs detected at Parkes in the HTRU and SUPERB surveys \citet{2018MNRAS.475.1427B} find no significant deviation of the sample from an isotropic distribution above the $2 \sigma$ level. 

As with many aspects of the FRB population, studies of the FRB sky distribution have been limited due to the small available FRB sample. With the new ultra-wide-field capabilities of CHIME as well as large-scale surveys from telescopes such as APERTIF, ASKAP, and UTMOST it may be possible to answer this question in the near future.

While the extragalactic nature of at least one FRB has been confirmed beyond doubt, a large and statistically isotropic population of FRBs would provide further weight behind the argument that FRBs are indeed extragalactic and possibly cosmological, similar to the early studies of GRBs \citep{1992Natur.355..143M,1993ApJ...413L.101K,2018NatAs...2..832K}. With a large enough population of FRBs it may also be possible to determine if there is any clustering on the sky associated with nearby galaxy clusters, if FRBs are extragalactic but non-cosmological.

\subsection{The DM distribution}\label{sec:DMdistribution}

A histogram of DM$_\mathrm{excess}$ for all FRBs is plotted in Fig.~\ref{fig:DMHist}. The true minimum and maximum values of dispersion measure possible for FRBs remain unknown; however, at the moment DM is one of the primary criteria that we use to distinguish an FRB from a Galactic pulse. Most searches for FRBs place a strict cut on DM. Real-time searches at the Parkes telescope only consider bright bursts with a DM value 1.5$\times$DM$_{\rm Galaxy}$ or greater \citep{2015MNRAS.447..246P} and deeper, offline searches may consider pulses with DMs $>0.9\times \mathrm{DM}_\mathrm{Galaxy}$. This requirement that the DM be larger than the expected contribution from the Milky Way makes it difficult to conclusively identify the minimum possible excess DM of an FRB. However, FRBs that occupy this border region between potentially galactic and extragalactic sources are beginning to be found \citep{2019MNRAS.tmp..718Q}. This dilemma will likely only be resolved once we have a more physical definition of an FRB that does not rely on DM. 

Thus far, the lowest DM measured for an FRB is 109.610$\pm$0.002 cm$^{-3}$ pc for FRB 180729.J1316+55 from the CHIME telescope \citep{2019Natur.566..230C}. In the context of the entire FRB population an FRB may be considered to have a low DM if $\mathrm{DM}_\mathrm{excess} \lesssim 350$ cm$^{-3}$ pc. There are now $>$15 FRBs in this category. Relative to the population discovered with each detection instrument, the low-DM FRBs tend to have higher peak flux densities and larger fluences than the overall sample, for example FRBs 110214 ($\mathrm{DM}_\mathrm{excess} = 130$ cm$^{-3}$ pc), 150807 ($\mathrm{DM}_\mathrm{excess} = 230$ cm$^{-3}$ pc), 180309 ($\mathrm{DM}_\mathrm{excess} = 218$ cm$^{-3}$ pc), and 010724 ($\mathrm{DM}_\mathrm{excess} = 330$ cm$^{-3}$ pc) are the four brightest FRBs detected at the Parkes telescope thus far, all with $S_\mathrm{peak} > 20$~Jy \citep{2018arXiv181010773P,2018ATel11385....1O,2016Sci...354.1249R,2007Sci...318..777L}. 

DM is often used as a rough proxy for distance (see \S \ref{sec:FRBproperties}) thus the maximum DM possible for an FRB is of great interest as it could tell us about the maximum possible redshift out to which we can see FRBs. High DM FRBs at $z > 3$ may even probe Helium reionization in the Universe \citep{2014ApJ...797...71Z,2018NatAs...2..836M}. The maximum DM pulse detectable by a telescope is dependent on several aspects of the observing configuration, including the time and frequency resolutions and the dedispersion algorithm used (see \S \ref{sec:pipelineSteps}). Thus far, the largest DM observed for an FRB is from FRB 160102 with DM $=2596.1 \pm 0.3$ cm$^{-3}$~pc, found using the Parkes telescope \citep{2018MNRAS.475.1427B}. If all the excess dispersion originates in the IGM, this FRB would be at a redshift $z = 2.10$, i.e. a comoving distance $D_L = 16$~Gpc. A larger sample will determine whether even higher-DM FRBs exist.

\subsection{The pulse width distribution}\label{sec:widthDistribution}

The observed FRB pulse width distribution is plotted in Fig.~\ref{fig:widthHist}. As with DM, the true minimum and maximum possible widths for FRBs are not yet known. However, the observed width distribution already spans several orders of magnitude. The known distribution peaks at a few milliseconds. The narrowest FRB single pulse yet measured is from FRB 121102 observed by \citet{2018Natur.553..182M} to have a width of $\lesssim 30$ $\upmu$s, although a sub-pulse of FRB 170827 revealed through voltage capture was measured to be 7.5 $\upmu$s in duration \citep{2018MNRAS.478.1209F}. The widest pulse reported in the literature is currently FRB 170922, which was detected with the UTMOST telescope at 835 MHz with $W = 26$ ms \citep{2017ATel10867....1F}. The width of an FRB can be heavily affected by scattering in the intervening medium, which broadens the pulse and reduces the peak flux density (see \S \ref{sec:scattering}). Thus, very wide, low peak flux density FRBs, even with equal fluence to short-duration easily detected FRBs, could exist but may be easily missed. 

Notably, FRBs are under-scattered compared with Galactic pulsars of comparable DM \citep[see Fig.~\ref{fig:DMScattering} and][]{2019MNRAS.482.1966R}.  This could be due to the significantly different relative distances between observer, scattering screen, and burst source.  In a simple one-screen toy model, scattering is maximized when the screen is half-way between source and observer.  In the case of FRBs, if the dominant scattering screen is in the host galaxy or Milky Way, then the temporal broadening of the signal will be comparatively modest.  Though FRBs may be less scattered compared to pulsars with similar DM, scattering may still be relevant for understanding the lack of FRBs detected at low frequencies \citep[e.g.][]{2015MNRAS.452.1254K}.

The minimum pulse width of an FRB is of interest as it probes the minimum physical scale on which these pulses can be generated. The $\lesssim 30$ $\upmu$s pulse from FRB~121102 already puts an upper limit on the emitting region for this burst at $\lesssim 10$km (in the absence of relativistic beaming effects). The maximum pulse width of an FRB would potentially tell us less about the emitting region and more about the propagation effects at play, as the widest pulse we detect is likely to be wide due to scatter broadening. Scattering has a larger effect at lower frequencies and FRBs found at lower frequencies ($< 600$ MHz) may be dominated by scattering effects. Recently reported FRBs between 400 and 800 MHz from CHIME show more scattering than might be explained by the normal ionized medium in a host galaxy, and \citet{2019Natur.566..230C} suggests that these bursts comes from special over-dense regions in their host galaxies, such as supernova remnants, star-forming regions, or galactic centers. However, other FRBs in the new CHIME sample exhibit very narrow pulse widths, such as a reported pulse duration of 0.08 ms for FRB180729.J0558+56. 

Finding the narrowest FRBs remains an instrumentation challenge, as narrow frequency channels (or coherent dedispersion) and fast time sampling are required to probe these regimes. Some FRBs detected at telescopes such as Parkes are unresolved in width due to insufficient frequency and time resolution and only upper limits can be placed on their intrinsic pulse duration \citep{2019MNRAS.482.1966R}. In the future, voltage capture systems on radio telescopes, either collected continuously as with Breakthrough Listen \citep{2018ApJ...863....2G} or triggered collection as with UTMOST \citep{2018MNRAS.478.1209F}, will help us probe this region of the FRB parameter space -- especially if we can observe at higher radio frequencies, where scattering is minimized.

\subsection{Repeating and non-repeating FRBs}

Clearly, an important diagnostic is whether an FRB has shown multiple bursts.  Conversely, it is less informative if an FRB has not yet been seen to repeat, because one can always argue that the burst rate is simply very low. 

In the current population, only two FRB sources have been seen to repeat (see \S \ref{sec:FRB121102} and~\S \ref{sec:R2}).  For these sources, only non-cataclysmic theories are viable, and it has been argued that perhaps all FRBs are capable of repeating. The locations of other FRBs have been re-observed to search for repeating pulses. Some FRBs have little to no follow-up published in the literature \citep[e.g. FRB 010125;][]{2014ApJ...792...19B} and others have been followed up for over 100 hours within $\pm$15 days of discovery \citep[e.g. FRB~180110 with $>$150 hours in the 30-day window around the FRB;][]{2018Natur.562..386S}. 
With only two repeaters in the FRB sample there are many outstanding questions about the potential for repetition from other FRBs. The repeat rate of FRB~121102 is highly non-Poissonian \citep{2018MNRAS.475.5109O} with epochs of high and low activity; FRB~180814.J0422+73 has not been studied sufficiently to constrain its repeat rate as a function of time. With the detection of more repeating FRB sources it may become clear that repeating FRBs come from an entirely different source class or progenitor channel compared to non-repeaters (see \S \ref{sec:subpopulations} and \S \ref{sec:progenitors}), but more data is needed and this issue may only be settled definitively with a very large sample of sources (hundreds to thousands).

\begin{figure}
\centering
\includegraphics[width=0.95\textwidth]{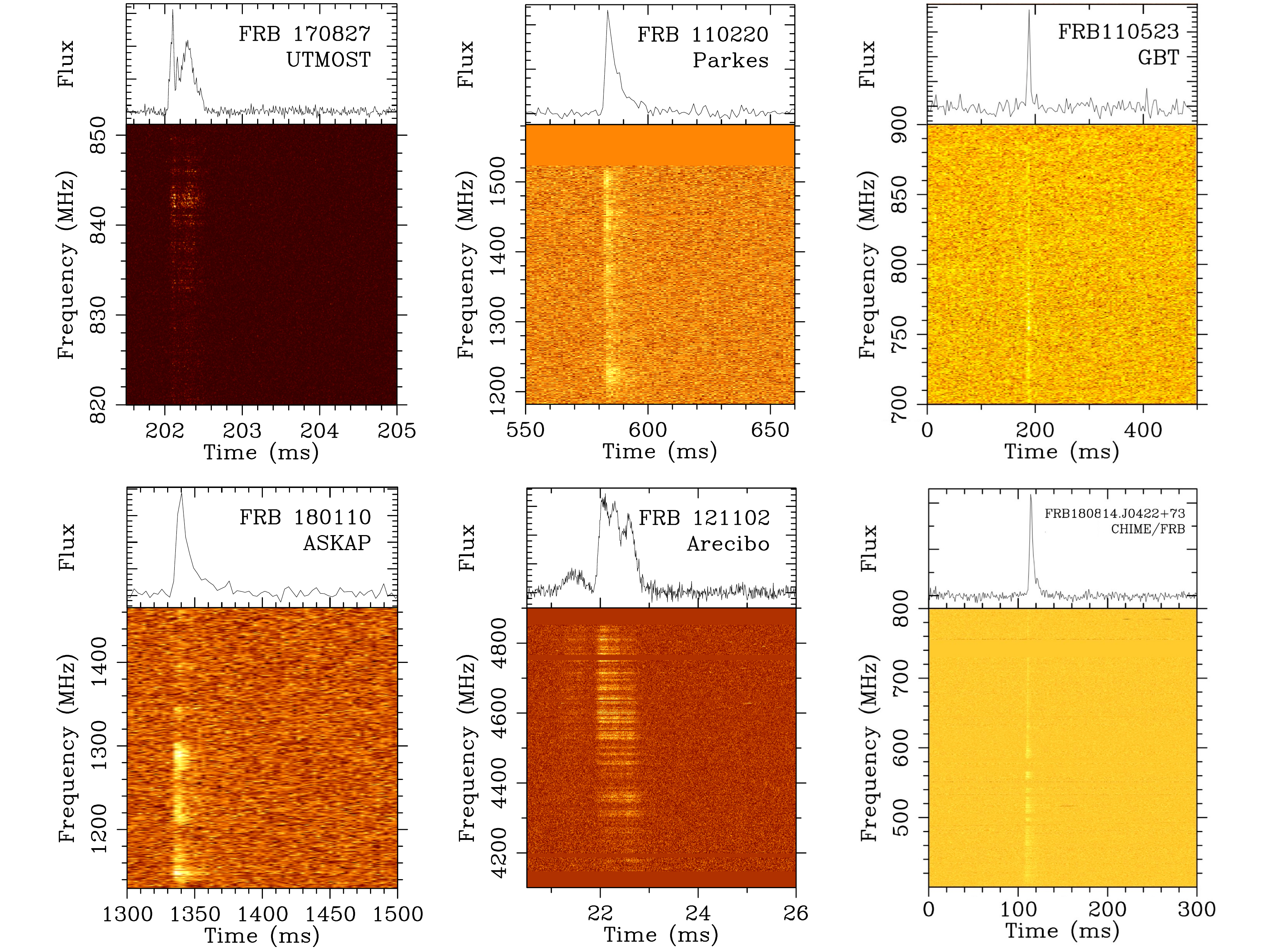}
\caption{De-dispersed pulse profiles and dynamic spectra of several FRBs. FRB~170827 (top, left) from \citet{2018MNRAS.478.1209F} detected with UTMOST at 835~MHz, FRB~110220 (top, middle) from \citet{2013Sci...341...53T} detected with Parkes at 1.4~GHz, FRB~110523 (top, right) from \citet{2015Natur.528..523M} detected with GBT at 800~MHz, FRB~180110 (bottom, left) from \citet{2018Natur.562..386S} detected with ASKAP at 1.3~GHz, a pulse from FRB~121102 (bottom, middle) from \citet{2018Natur.553..182M} detected with Arecibo at 4.5~GHz, and a pulse from FRB~180814.J0422+73 (bottom, right) from \citet{2019Natur.566..235C} detected with CHIME at 600~MHz. \label{fig:waterfall_profiles}}
\end{figure}

FRB~121102 is currently the only repeater that has been studied in great detail, but only a few of its properties are distinctive compared to the rest of the population: it has the highest observed rotation measure ($\sim 10^5$ rad~m$^{-2}$) of any FRB by several orders of magnitude, and it is capable of emitting bursts at a high rate (sometimes tens per hour), so it is clearly far more active compared to other sources. Some of the repeating pulses from both repeaters show complex frequency and time structure \citep{2018Natur.553..182M,2018arXiv181110748H,2019Natur.566..235C} but this may not be a distinctive trait to ``repeaters''. This structure may also be present in some one-off FRB detections with sufficient temporal and spectral resolution \citep[see Fig.~\ref{fig:waterfall_profiles};][]{2018MNRAS.478.1209F,2019MNRAS.482.1966R}. The pulses of FRB~121102 and FRB~180814.J0422+73 vary enormously in width (from $\sim$30 $\upmu$s to $\sim 10$~ms for FRB~121102 and $\sim$2~ms to $\sim$60~ms for FRB~180814.J0422+73) but in both cases the discovery pulse was not unusual in its duration \citep{2014ApJ...790..101S}.

However, for FRB~121102, the discovery peak flux density at discovery was much lower compared to previously discovered FRBs.  \citet{2018ApJ...854L..12P} argue that there is a growing body of evidence suggesting that FRB~121102 is fundamentally different compared with the other (so-far) non-repeating FRBs.  However, it may simply be an exceptionally active example, and not fundamentally different in physical origin.  

More observations of FRB~180814.J0422+73 -- including RM measurements, and identification of its host galaxy -- will elucidate further whether both repeaters have similar properties.  Additionally, even a few more repeating FRBs might help distinguish sources that are observed to repeat from those that remain one-off events. It is expected that ongoing CHIME observations, which sample the sky with daily cadence, will provide a much clearer picture of the population of repeating FRBs.

\subsection{Sub-populations emerging?}\label{sec:subpopulations}

With two repeating FRBs now known, both showing similar spectro-temporal structure, it may soon be possible to identify sub-populations in the overall distribution of FRBs. However, as both \citet{2019MNRAS.482.1966R} and \citet{2018NatAs...2..839C} conclude, besides the uniqueness of repeating pulses from FRB~121102 (before the publication of FRB~180814.J0422+73), the current sample offers no clear dividing lines over any other observed parameters. While FRB~121102 has a larger RM compared to measured values from other FRBs, RM has not been measured for the entire sample, making it difficult to draw definitive conclusions. The population of FRBs found with ASKAP are brighter than those at Parkes (see Fig.~\ref{fig:EDM}) but this is due to the different detection thresholds of these instruments. The majority of FRBs have durations $< 5$~ms, with a tail in the distribution towards longer pulse durations. However, no clear trends (such as the presence of a distinct short- and long-duration population) have yet emerged. 

With a larger population of FRBs, multi-modality in some observed parameters may indicate sub-populations in the way that a bi-modal duration distribution of short and long gamma-ray bursts became apparent as the population grew \citep{1993ApJ...413L.101K}. Some parameters may be more promising than others for investigation along these lines. Pulse duration (analogous to GRBs) may reveal information about the progenitor or emission mechanism, and the RMs of future FRBs may provide information about their origins in a dense and turbulent or clean and sparse local environment. The relationship between parameters such as fluence and DM (see \S \ref{sec:IntrinsicPopDist}) may also provide valuable clues.

However, once FRBs are more routinely localized to host galaxies, the types of galaxies and the specific regions thereof in which they reside may provide some of the most important clues for identifying sub-populations. The repeating FRB~121102 resides in a low-metalicity dwarf galaxy, and searches for galaxies of similar type have been done for other FRBs \citep{2018ApJ...867L..10M}. If some FRBs are found to reside in larger galaxies, or at different radii from their host galaxy centers, such as for GRBs \citep{2018NatAs...2..832K}, this may provide a valuable tool for distinguishing between types of FRB sources.

\section{The intrinsic population distribution}\label{sec:IntrinsicPopDist}

From the observed properties of the FRBs detected at various telescopes around the world, the next crucial but challenging step is to infer from observations the intrinsic physical properties of the population. Given that little is currently known about the progenitors and origins of FRBs this type of study is in its early stages. Nevertheless, efforts have already been made to extrapolate from the population of discovered FRBs to their population more globally. Here we summarize some results from FRB population studies and draw some conclusions from the publicly available sample of FRBs. 

\subsection{The fluence--dispersion measure plane}

\begin{figure}
\centering
\includegraphics[width=0.7\columnwidth]{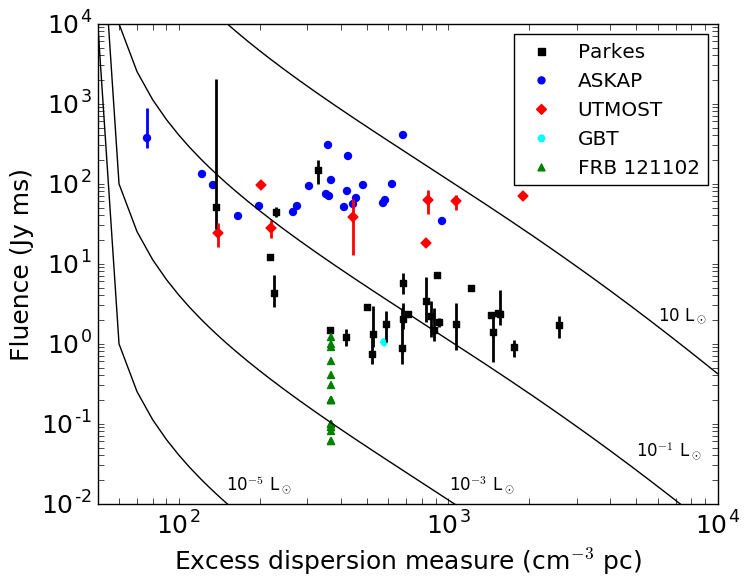}
\caption{Fluence--dispersion measure distribution for the currently observed sample overlaid with lines of isotropic equivalent luminosity values assuming a contribution of $\mathrm{DM}_\mathrm{Host} = 50$ cm$^{-3}$ pc. The recent detections from CHIME/FRB have been omitted (see text). 
\label{fig:EDM} }
\end{figure}

The current state of the FRB population as interpreted as a cosmological sample of sources is shown in Fig.~\ref{fig:EDM}. This sample includes the recent flurry of ASKAP discoveries and shows fluence
versus inferred extragalactic DM for the Parkes and ASKAP samples as well as the repeater FRB~121102. The CHIME/FRB detections from \citet{2019Natur.566..230C} have not been plotted due to the uncertain flux calibration of the instrument, as emphasised in the discovery publication. From this 
diagram, we see evidence for a change of fluence with DM, which is expected for a population of sources
at different distances. We note that the large scatter seen on this diagram is inconsistent with the idea of FRBs as standard candles. As can be seen from the overlaid curves, there is over an order of magnitude spread in the implied intrinsic luminosity. We also note that the distribution of pulse fluences for the repeater are dramatically different than the rest of the sample. Part of this 
difference could be due to a selection bias from the higher
sensitivity of FRB~121102 observations that has come from observations with Arecibo. As underscored in the previous section, further follow-up observations of all FRBs with as high sensitivity as possible are required.

Although there are clearly a lot of selection biases inherent in shaping this diagram, the process of FRB detection is reasonably well understood. As a result, it is possible to set up a Monte Carlo simulation of the FRB population that can mimic the properties shown in Fig.~\ref{fig:EDM} and allow us to infer the underlying, as opposed to the observed, distributions of population parameters. Population studies that attempt to account for these biases are now being used to help form a self-consistent picture on the FRB distribution and luminosity function. The Monte Carlo simulation process attempts to follow the process from emission of the signal to detection. A simulation typically proceeds by randomly drawing an FRB from intrinsic distributions of pulse widths and luminosities. Sources can be assigned distances based on an assumption about the underlying redshift distribution. Finally, assumptions about the electron content at the source, in the host galaxy, the IGM and the Milky Way can be made to infer the observed DM. With these ingredients, one can infer the observed pulse fluence and decide whether each model FRB is detectable or not. 

A pioneering study of this kind, based on a sample of only 9 FRBs from Parkes known at the time, is described in \citet{2016MNRAS.458..708C}. A key result from this study was that, although the source distributions could be reproduced by a cosmological FRB population, the sample was not large enough to discriminate between spatial distributions that resulted from uniform density with co-moving volume, or whether they follow the well-known peak in star formation that occurs around $z \sim 1$ \citep{2014ARA&A..52..415M}. Similar results were also found with a slightly larger sample by
\citet{2017PhDT........73R}. Both these studies estimate that sample sizes in the range 50--100 FRBs are needed to distinguish between different redshift distributions.

The overall form of the fluence--dispersion measure relationship in Fig.~\ref{fig:EDM} can be understood in terms of a range of luminosities and host DMs that produces FRBs that are detectable out to different DM
limits depending on telescope sensitivity. At the time of writing, the dominant contributions are ASKAP, which probes the bright low-DM part of the population, and Parkes, which is probing the fainter high-DM end. A recent population study by \citet{2018arXiv181100195L} suggests that this population will be extended to higher DMs in the future with the addition of sensitive FRB surveys with larger instruments. For example, FAST will more likely probe the FRB population with DMs above 2000~cm$^{-3}$~pc \citep{2018arXiv180805277Z}.

\subsection{The FRB luminosity function}

Although early analyses of the FRB population \citep[e.g.][]{2013MNRAS.436..371H,2013MNRAS.436L...5L} assumed,
in the absence of further constraints, that the population of FRBs is consistent with them being standard candles, as mentioned above and as seen in Fig.~\ref{fig:EDM}, a distribution of luminosities is required to model the emerging samples of FRBs.  This is perhaps unsurprising, given that FRBs may well have relatively narrow emission beams.  While the shape of the luminosity function is currently not well understood, more recent analyses \citep[see, e.g.,][]{2018MNRAS.481.2320L,2018ApJ...863..132F} seem to be favouring Schechter luminosity functions over power-law or normal distributions \citep{2016MNRAS.458..708C}. The Schechter function gives the number of FRBs per unit logarithmic luminosity interval
\begin{equation}
\phi(\log L) = \left(\frac{L}{L_*}\right)^{\beta+1} \exp \left(\frac{L}{L_*}\right),
\end{equation}
where the power-law index $\beta$ and cut-off luminosity $L_*$ are free parameters. This
empirical characterization is motivated by success in modeling extragalactic luminosities in
which very bright sources are rarer than expected from a straightforward power-law. Although, whether it will
serve as an accurate characterization of the FRB luminosity function remains to be seen.
Very recently, using a Bayesian-based Monte Carlo approach, \citet{2018MNRAS.481.2320L} prefer models with $-1.8 < \beta < -1.2$
and $L_* \sim 5 \times 10^{10}\,L_{\odot}$. Future progress in refining constraints on the form of this distribution, particularly
at the low-luminosity end, could be made by detections of FRBs in nearby galaxy clusters \citep{2018ApJ...863..132F}. At the high-luminosity end, constraints on the emission mechanism
may be possible from further studies of the fluence distribution \citep[for further discussion, see][]{2018MNRAS.tmpL.198L}.

\subsection{FRB rates and source counts}

The estimated rate of observable FRBs is typically given as an all-sky rate above some sensitivity limit rather than a volumetric or cosmological rate, since the redshift distribution of FRBs is unknown. Constraints on the all-sky rate
of FRBs, ${\cal R}$, have been carried out by a number of authors and are summarized in Table \ref{tab:FRBRates}.

All estimated rates are roughly consistent within the errors with $\gtrsim 10^{3}$ FRBs detectable over the whole sky every day above a fluence threshold of $\mathcal{F} \gtrsim 1$~Jy~ms. Under an assumption that these sources are distributed cosmologically out to a redshift $z \sim 1$ the implied volumetric rates of roughly $2\times10^{3}$ Gpc$^{-3}$ yr$^{-1}$ (of observable events) are two orders of magnitude lower than the estimated core-collapse supernova (CCSN) rate out to this redshift \citep{2004ApJ...613..189D}.

Rates of CCSN sub-classes vary considerably and the FRB rate may be consistent with Type Ib and Ic rates \citep{2012ApJ...757...70D} but is still one to two orders of magnitude larger than the estimated rate of super-luminous supernovae \citep{2017MNRAS.464.3568P}. While the all-sky GRB rate and the distribution of GRBs in redshift are highly uncertain, the observable FRB rate is still likely an order of magnitude larger than the total GRB rate in this redshift range, even when accounting for GRB events not beamed towards Earth \citep{2001ApJ...562L..55F}. The binary neutron star merger rate is also highly uncertain but $z = 0$ estimates from the detections of the LIGO Virgo Collaboration (LVC) give a rate of $R_{\rm BNS}~=~1540^{+3200}_{-1220}$~Gpc$^{-3}$~yr$^{-1}$ \citep{2017PhRvL.119p1101A}, broadly consistent with the merger rate as derived from Galactic BNS systems.
Extrapolating these rates to larger distances with no cosmological evolution gives an event rate within an order of magnitude of the estimated FRB event rate, although perhaps slightly lower. If there is significant evolution in the rate of BNS mergers over redshift the true rate of merger events may be much lower when integrating to high redshift.

The high all-sky rate of FRB events relative to many other types of observable transients already places some constraints on their progenitors. Even for a cosmological distribution of events, if FRBs are generated in one-off cataclysmic events their sources must be relatively common and abundant. 
This becomes even more important for progenitors only distributed in the nearby volume, such as young neutron stars in supernova remnants \citep{2016MNRAS.458L..19C}. However, if the high FRB rate is generated by a smaller population of repeating sources, the all-sky rate becomes slightly easier to account for and sources can be less common and far less numerous, but the engine responsible for repeating pulses must be relatively long-lived. Of the FRBs observed to-date, only two have been detected to repeat (see \S \ref{sec:FRB121102} and \S \ref{sec:R2}) and if all others repeat they are either infrequent, highly non-periodic, or may have very steep pulse-energy distributions.

\begin{table}
\caption{A summary of the various estimates for the all-sky FRB rate based on various surveys and analyses. Rates and ranges are quoted with confidence intervals (CI) above a fluence threshold (${\cal F}_{\rm lim}$) for observations at a given reference frequency.}\label{tab:FRBRates}
\begin{tabular}{cccccc}
\hline
\hline
Rate & Range & CI & ${\cal F}_{\rm lim}$ & Frequency & Reference \\
\multicolumn{2}{c}{(FRBs sky$^{-1}$ day$^{-1}$)} & (\%) & (Jy ms) & (MHz) & \\
\hline
$\sim225$ & ---  & --- & 6.7  & 1400 & \citep{2007Sci...318..777L} \\
10000 & 5000 -- 16000 & 68 & 3.0 & 1400 & \citep{2013Sci...341...53T} \\
4400 & 1300 -- 9600 & 99 & 4.4 & 1400 & \citep{2016MNRAS.455.2207R} \\
7000 & 4000 -- 12000 & 95 & 1.5 & 1400 & \citep{2016MNRAS.460L..30C} \\
3300 & 1100 -- 7000 & 99 & 3.8 & 1400 & \citep{2016MNRAS.460.3370C} \\
587  & 272 -- 924 & 95 & 6.0 & 1400 & \citep{2017AJ....154..117L}\\
1700 & 800 -- 3200 & 90 & 2.0 & 1400 & \citep{2018MNRAS.475.1427B} \\
37   & 29 -- 45 & 68 & 37 & 1400 & \citep{2018Natur.562..386S} \\
\hline
\end{tabular}
\end{table}

Determinations of the FRB rate from survey observations are very useful as they can be used to make predictions about other experiments without the need for a lot of assumptions about the spatial distribution of the population, or form of the luminosity function. The impact of the underlying population can be encapsulated within the cumulative distribution of event rate as a function of peak flux or, more generally, fluence. This dependence is usually modeled as a power law such that the rate above some fluence limit ${\cal F}_{\rm min}$ is given by 
${\cal R}(>{\cal F}_{\rm min}) \propto {\cal F}_{\rm min}^{\gamma}$, 
where the index $\gamma=-1.5$
for Euclidean geometry.
Since, for a survey with some instantaneous solid angle coverage $\Omega$ with given amount of observing time $T$ above some
${\cal F}_{\rm min}$, the number
of detectable FRBs $N(>{\cal F}_{\rm min})={\cal R}\Omega T \propto {\cal F}_{\rm min}^{\gamma}$ this same index is often used to describe the
source count distribution.

In event rate or source count
studies, there is currently a wide range of $\gamma$ values that have been claimed so far
beyond --1.5.
\citet{2018MNRAS.474.1900M}
estimate, based on a
recent maximum likelihood
analysis on the Parkes FRBs,
that $\gamma = -2.6_{-1.3}^{+0.7}$.
In contrast, based on essentially the same sample of
FRBs,
\citet{2017AJ....154..117L}
estimate $\gamma=-0.91 \pm 0.34$.
Very recently, a combined analysis of the source counts
for the ASKAP and Parkes samples
by \citet{2018arXiv181004357J}
has found evidence for a break
in the simple power-law dependence in which $\gamma = -1.1 \pm 0.2$ for the fainter and more distant Parkes
population and $\gamma = -2.2 \pm  0.5$ for the brighter and more nearby ASKAP population. If confirmed by future studies, this would signify a cosmologically evolving FRB progenitor population peaking in the redshift range 1--3. This issue is likely to be investigated by further, more detailed Monte Carlo simulations and a larger available sample of FRBs.

One issue that has not been discussed extensively in the literature so far is to what extent the above FRB rates need to be
scaled to account for beamed emission. This issue is well developed within
the field of radio pulsars \citep[see, e.g.,][]{1998MNRAS.298..625T} where it is
well known that this `beaming factor' is of order 10 for canonical pulsars, i.e.~we see only a tenth of the total population of active pulsars in the Galaxy. The rates shown in Table~\ref{tab:FRBRates} are, therefore, for potentially observable FRBs only. When computing volumetric rates, it is important to consider this correction. Given the uncertain nature of FRBs at the present time this is highly speculative, but we urge theorists to specify as far as possible the likely beaming corrections in emission models.

\subsection{Intrinsic pulse widths}

As noted originally by \citet{2013Sci...341...53T}, instrumental broadening of the pulses
in systems that employ incoherent dedispersion can often account for a substantial fraction
if not all of the observed pulse widths. In a recent study that carefully accounts for intrinsic 
and extrinsic contributions to FRB pulse profile morphology, \citet{2019MNRAS.482.1966R} demonstrates
that, after accounting for pulse broadening due to scattering, only five of the sample of seventeen
Parkes FRBs he analyzed have widths that exceed that predicted by dispersion broadening. Six FRBs
in this sample are temporally unresolved.
While a larger sample of FRBs in future will definitely help, it is true that current instrumentation
cannot resolve a significant number of currently detectable FRB pulses.

\subsection{Intrinsic spectra}

The difficulties in observing FRBs over large bandwidths have so far hampered attempts
to quantify their broadband spectra. As mentioned in \S\ref{sec:observed}, the simplest model
is to adopt a power-law dependence with flux density $S \propto \nu^{\alpha}$ for
some spectral index $\alpha$. Many statistical analyses either remain agnostic
about the spectrum and posit `flat spectra', i.e.~$\alpha=0$, or
assume (without strong justification) that FRBs have a spectral dependence similar to
that observed for pulsars, where $\alpha \sim -1.4$ \citep{2013MNRAS.431.1352B}. 

The most stringent constraints on $\alpha$ come from non-detections of FRBs at radio frequencies below 400~MHz. \citet{2017ApJ...844..140C} use the lack of FRB detections in the Green Bank
Northern Celestial Cap survey to limit $\alpha > -0.9$. Similarly, the lack of FRBs found with LOFAR limit $\alpha>+0.1$ \citep{2015MNRAS.452.1254K}. These constraint from lower frequencies are strongly at odds with a recent study of the ASKAP FRBs
\citep{2018arXiv181004353M} which finds
$\alpha=-1.6_{-0.2}^{+0.3}$, i.e.~similar
to the normal pulsar spectra.

These disparate results imply that the spectral behaviour of FRBs likely involves
a turnover at sub-GHz frequencies or may not follow a power-law at all but in fact be in emission envelopes \citep[e.g.,][]{2018arXiv181110748H,2019arXiv190302249G}. It is important to note that scattering may also play a significant role at low frequncies, artificially shallowing the measured spectral index of FRBs. \citet{2019Natur.566..230C} show that a large fraction of their sample at low frequencies exhibits significant scattering.

Spectral behaviour with a turnover might be observed in future for FRBs embedded in dense
ionized media. As pointed out by 
\citet{2014ApJ...797...70K} and
\citet{2017MNRAS.465.2286R}, and is known to be exhibited
in some radio pulsars, free-free absorption from a shell surrounding an FRB can
result in substantial modifications to the spectra, which might result in turnovers at decametric wavelengths. 
Very recently, in a comprehensive review of other propagation effects on FRB spectra,
\citet{2018arXiv181100109R} show that spectral turnovers might be ubiquitous, regardless of the emission mechanism.
Future surveys 
at these wavelengths might also observe
the effect of spectral turnovers occurring at higher frequencies that are being redshifted into the $<500$~MHz band \cite{2017MNRAS.465.2286R}.

\section{Emission mechanisms for FRBs}\label{sec:emission}

The high implied brightness temperatures of FRBs ($T_b > 10^{32}$\,K) and their short intrinsic durations (milliseconds or less) require a coherent emission process from a compact region.  The shortest-duration burst structures detected to date are $\sim 30$\,$\mu$s \citep{2018Natur.553..182M,2018MNRAS.478.1209F}, implying an emission site of $< 10$\,km (ignoring possible geometric and relativistic effects).  What creates this coherent emission, and what is the underlying energy source?
Does the same process that creates the radio burst also produce observable emission at other wavelengths?  Different radiation mechanisms will produce different observed properties, and the better we can characterize the radio bursts and multi-wavelength emission, the better the chance of identifying the underlying emission mechanism.  As described in \citet{2018arXiv181005836P}\footnote{www.frbtheorycat.org} \citep{2018arXiv181005836P}, one can consider the various radiation mechanisms relevant to astrophysics, as well as the necessary conditions for coherence, such as: bunched particles accelerating along electromagnetic field lines, simultaneous electron phase transitions (masers), and entangled particles collectively undergoing an atomic transition (Dicke's superradiance).

Here we briefly consider the nature of FRB emission before giving a more general survey of progenitor models in the following section.  The basic physical constraints on FRBs were investigated by \citet{2014ApJ...785L..26L}.  We also point the reader to \citet{2017RvMPP...1....5M}, who reviews the established coherent emission mechanisms in astrophysical plasmas in a general sense.  The FRB emission mechanism, specifically, has been addressed, e.g., by \citet{2014PhRvD..89j3009K}, \citet{2016PhRvD..93b3001R} and \citet{2018MNRAS.477.2470L}.  Given the possibility that there are multiple types of FRBs \citep[][]{2018NatAs...2..839C,2018ApJ...854L..12P}, we caution the reader that there could also be multiple types of emission mechanisms. Likewise, we caution the reader that separating intrinsic and extrinsic effects in the observed properties of FRBs adds significant uncertainty in investigating the emission mechanism, as we discuss below.

Pulsars (and magnetars) are well-established coherent radio emitters, and though the fundamental emission mechanism(s) are still a major open puzzle, they provide an important observational analogy.  In other words, it would already be helpful to know if FRB emission is from a similar physical mechanism as pulsars, even if that physical mechanism is still not well understood.  Firstly, it is important to note that radio pulsars show a wide range of observational properties that are similar to those seen in FRBs.  Like FRBs, pulsars have a variety of circular and linear polarization fractions, pulse widths, pulse structure, and spectra.  In canonical rotation-powered pulsars, emission is believed to originate a few tens to hundreds of kilometers above the neutron star polar caps \citep[e.g.][]{2012A&A...543A..66H}.  Some neutron stars, of which the Crab is the best-studied example, also show so-called `giant' pulses, which are brighter, shorter in duration, and may originate from a different region of the magnetosphere \citep{2016ApJ...833...47H}.  Magnetars also emit radio pulses \citep{2006Natur.442..892C}; their emission can be highly erratic, showing radio emission at a wide range of rotational phases, and with an average pulse profile that changes with time.  Hence, there are at least three types of radio emission seen from magnetised neutron stars.  Given the wide range of radio emission phenomena detected from Galactic neutron stars, it seems plausible that FRBs could be an even more extreme manifestation of one of these processes, or perhaps a fourth type of neutron star radio emission.  Unfortunately, however, the mechanisms responsible for creating neutron star pulsed radio emission are still not well understood.  Nonetheless, if we assume that FRBs originate in neutron star magnetospheres, or their near vicinity, the detection of a multi-wavelength counterpart (or lack thereof) could inform whether the bursts are rotationally or magnetically powered \citep{2016ApJ...824L..18L}.

FRBs and pulsar pulses have peak flux densities $\sim 1$\,Jy but the $\sim 10^6$ times greater distance of the FRB population implies a $\sim 10^{12}$ times greater luminosity (assuming the same degree of beaming), which corresponds to burst energies

\begin{equation}
\label{eq:burst_energy}
E_{\rm burst} = 4\pi D^2 (\delta\Omega/4\pi) \mathcal{F}_{\nu} \Delta\nu
\approx 10^{31}\, {\rm J}\,(\delta\Omega/4\pi) D_{\rm Gpc}^2  (\mathcal{F}_{\nu} / 0.1\ {\rm Jy\ ms}) \Delta\nu_{\rm GHz},
\end{equation}
where $\delta\Omega$ is the solid angle of the emission (steradians), $D_{\rm Gpc}$ the luminosity distance (Gpc), $\mathcal{F}_{\nu}$ the fluence (Jy~ms), and $\Delta\nu_{\rm GHz}$ the emission bandwidth (GHz).  All parameters are considered in the source frame.  The magnetospheres of canonical pulsars may have difficulty in providing this much energy \citep[e.g.][]{2016MNRAS.457..232C}.  FRBs might be powered instead by the strong $\sim 10^{14} - 10^{15}$\,G magnetic fields in magnetars \citep{2013arXiv1307.4924P,2017ApJ...843L..26B}.

A variety of works have considered whether FRBs could originate from rotationally powered super-giant pulses from rapidly spinning, highly magnetized young pulsars \citep[e.g.,][]{2016MNRAS.457..232C,2016MNRAS.462..941L}.  Because the available spin-down luminosity scales with the magnetic field strength $B$ and rotational period $P$ as $B^2/P^4$ \citep{2012hpa..book.....L}, it is conceivable that such a source could power giant-pulses that are orders of magnitude brighter than those seen from the Crab pulsar.  Importantly, we have not yet seen a cut-off in the brightness distribution of Crab giant pulses.  Beaming is also a critical consideration, and it is possible that the Crab giant pulses would appear substantially brighter if viewed from another angle.  In summary, the maximum possible luminosity of radio emission from a neutron star is not well established.

Furthermore, if a significant fraction of the observed DM can be contributed from a surrounding supernova remnant, then FRBs may be closer than we would otherwise infer \citep{2016MNRAS.458L..19C}, thereby reducing the energy requirement.  However, the precise localization of FRB~121102 led to a redshift measurement that places it firmly at $z = 0.193$ ($d_L \sim 1$\,Gpc) \citep{2017ApJ...834L...7T}.  \citet{2017ApJ...838L..13L} argue that this large distance rules out rotation-powered super-giant pulses like those from the Crab.  The Crab pulsar is a singular source in our Galaxy, however, and we do not know whether its giant pulses are at the limit of what a neutron star can produce.  Perhaps with more fortuitous beaming and a younger, more highly magnetized neutron star, the energy requirements imposed by FRB~121102 can be met with giant-pulse-like emission.

Magnetically powered bursts from neutron stars have also been considered in the literature.  Flares from magnetars were first proposed by \citet{2010vaoa.conf..129P} and \citet{2013arXiv1307.4924P}.  A flaring magnetar model for FRB~121102 was proposed by \citet{2017ApJ...843L..26B}, and was partially motivated in order to explain the source's compact, persistent radio counterpart \citep{2017Natur.541...58C,2017ApJ...834L...8M}.  In this model, the FRBs are from a giga-Hertz maser and originate in shocks far from the neutron star itself.

\citet{2018arXiv181110748H} show that FRB~121102 bursts have complex time-frequency structures.  This includes sub-bursts ($\sim 0.5-1$\,ms wide) displaying finite bandwidths of $100-400$\,MHz at 1.4\,GHz.  \citet{2018arXiv181110748H} also find that the sub-bursts have characteristic frequencies that typically drift lower at later times in the total burst envelope, by $\sim 200$\,MHz/ms in the $1.1-1.7$\,GHz band.  This differs from typical pulsars and radio-emitting magnetars, which have smooth, wide-band spectra \citep[even in their single pulses, e.g.,][]{2003A&A...407..655K,2018MNRAS.473.4436J}.  In pulsars, the only narrow-band modulation seen is from diffractive interstellar scintillation, which is augmented in some cases by constructive and destructive interference from multiple imaging due to interstellar refraction.

The spectral behaviour of FRB~121102 may be intrinsic to the emission process.  It could also be due to post-emission propagation processes, or some combination of intrinsic and extrinsic effects.  Dynamic spectral structures are seen in other astrophysical sources that emit short-timescale radio bursts: e.g., the Sun \citep[e.g.,][]{2015ApJ...808L..45K}, flare stars \citep[e.g.,][]{2006ApJ...637.1016O,2008ApJ...674.1078O}, and Solar System planets \citep[e.g.][]{1992AdSpR..12...99Z,2014A&A...568A..53R}.  Time-frequency drifts, qualitatively similar to those seen from FRB~121102 and the CHIME/FRB repeater FRB 180814.J0422+73, have been detected from such sources.  These drifts occur when the emission regions moves upwards to regions with lower plasma frequencies or cyclotron frequencies (these, in turn, are tied to the observed electromagnetic frequency).  Fine time-frequency structure in the radio emission is related to variations in the particle density \citep[e.g.,][]{2006A&ARv..13..229T}.  If we extrapolate similar processes to FRBs, it suggests that FRB~121102's (and FRB~180814.J0422+73's) emission could originate from cyclotron or synchrotron maser emission \citep{2014MNRAS.442L...9L,2017ApJ...843L..26B,2017ApJ...842...34W}, in which case relatively narrow-band emission in the GHz range could be expected. Antenna mechanisms involving curvature radiation from charge bunches have also been considered
\citep{2016MNRAS.457..232C,2018MNRAS.477.2470L}.  However, it is not clear if the energetics can be satisfied or how time-frequency structure is produced in this case. 

In the 100\,MHz to 100\,GHz radio frequency range, the Crab pulsar shows a remarkable range of emission features.  The Crab's rich and diverse phenomenology is thus potentially relevant to understanding FRB emission.
For example, as discussed in \citet{2018arXiv181110748H}, the giant pulse emission in the Crab pulsar's high-frequency interpulse \citep[HFIP;][]{2016ApJ...833...47H}, which is seen above $\sim 4$\,GHz radio frequencies, provides an interesting observational comparison to the burst features seen in FRB~121102.  Note that the polarimetric and time-frequency properties of the HFIPs are highly specific and differ significantly from those of the main giant pulses \citep[MP;][]{2010A&A...524A..60J,2016ApJ...833...47H}.

The Crab's HFIP spectra display periodic bands of increased brightness \citep{2007ApJ...670..693H} with separations $\Delta\nu$ that scale with frequency ($\Delta\nu/\nu = $ constant). In comparison, the drift rates in FRB~121102 potentially show a similar scaling \citep[see Figure 3 of][]{2018arXiv181110748H} but a larger sample is needed to be conclusive.  While the Crab HFIPs are microseconds in duration, the burst envelopes of FRB~121102 are typically milliseconds -- though with underlying $\sim 30$\,$\mu$s structure clearly visible in some cases \citep{2018Natur.553..182M}.  Searches for even finer-timescale structure in FRB~121102 should thus continue, using high observing frequencies to avoid smearing from scattering.
 
Lastly, the polarization angle of the $\sim 100$\% linearly polarized radiation from FRB~121102 at $4-8$\,GHz appears constant across individual bursts and is stable between bursts \citep{2018Natur.553..182M,2018ApJ...863....2G}. This phenomenology is also similar to that of the Crab HFIPs, which are $\sim 80-100$\% linearly polarized and have a constant polarization position angle across the duration of each pulse -- as well as between HFIPs that span $\sim 3$\% of the pulsar's rotational phase \citep[see Fig.~14 of][]{2016ApJ...833...47H}.  Lastly, the Crab HFIPs typically show no circular polarization, and thus far no circularly polarised emission has been detected from \aofrb. 

For now, the emission mechanism responsible for the coherent radio emission of FRBs remains a mystery.  As with pulsars, however, regardless of whether we eventually understand the detailed physical emission it should still be possible to identify the progenitors of FRBs and to use them as astrophysical probes.

\section{Progenitor models}\label{sec:progenitors}

At the time of writing there are at least 55 published progenitor theories for FRBs. Models for FRB progenitors can be grouped along several lines: repeating or non-repeating, long-lived or cataclysmic source, nearby or cosmological, rotationally or magnetically powered, etc. Many progenitor theories involve compact objects, the processes involved in their birth, or the medium surrounding them. Here we explore the models in more detail, grouped by the primary source involved, and in some cases splitting the category up further by looking at isolated or interacting/colliding mechanisms to generate the radio pulse. A tabular summary of existing FRB theories is maintained on the FRB Theory Catalogue.

\subsection{Neutron star progenitors}

The majority of current FRB progenitor theories involve neutron stars. Their large rotational energies and strong magnetic fields, as well as the often turbulent environments they occupy, make them plausible candidates for the progenitors of FRBs and some characteristics of FRB emission appear similar to radio pulsars (see also \S\ref{sec:emission}). Here we discuss the FRB progenitor theories that predict bright radio pulses from extragalactic neutron stars -- grouping by models that invoke isolated neutron stars (\S \ref{sec:isolatedNS}), neutron stars interacting with other bodies or their environment (\S \ref{sec:interactingNS}), and neutron stars colliding with other compact objects (\S \ref{sec:collidingNS}). 

\subsubsection{Isolated neutron star models}\label{sec:isolatedNS}

A number of theories argue that FRBs can be generated by isolated neutron stars, either via beamed radio emission from their magnetosphere, during the collapse of a supramassive neutron star due to its own gravity, or by relativistic shocks in the surrounding medium. 

Both \citet{2016MNRAS.457..232C} and \citet{2016MNRAS.458L..19C} theorize that some rotationally powered pulsars can produce FRBs as part of their normal emission process, from supergiant pulses from young neutron stars in the case of \citeauthor{2016MNRAS.458L..19C}, and from nano-shot giant pulses in the case of \citeauthor{2016MNRAS.457..232C}. \citet{2016MNRAS.462..941L} have proposed that young rotationally powered neutron stars with millisecond rotation periods could also produce FRBs from the open magnetic field lines at the poles that generate the normal radio emission. Additionally, \citet{2017MNRAS.467L..96K} has suggested that FRBs may originate from radio pulsars with unstable rotational axes that result in `wandering beams' on the sky. 
Other theories have argued that FRBs are generated from the magnetically powered neutron stars with ultra-strong magnetic fields known as magnetars. \citet{2010vaoa.conf..129P} proposed that an FRB might be generated during a magnetar hyperflare and \citet{2018ApJ...852..140W} theorized that FRBs are generated in starquakes on the surface of a magnetar. \citet{2017ApJ...834..199L} predicts a single bright radio pulse generated seconds after the birth of a magnetar with a millisecond rotation period, whereas \citet{2017ApJ...841...14M} predict repeating pulses from a stably emitting young millisecond magnetar in a dense supernova remnant. \citet{2019arXiv190201866M} theorize that FRBs are produced through maser emission in the ultra-relativsitic shocks through the ionized medium surrounding a young magnetar; this model also predicts a significant RM contribution from propagation through the highly magnetized outer layers of the mangetar wind nebula.

Cataclysmic models involving isolated neutron stars include the `blitzar' model, where an FRB is produced by a supramassive neutron star as it collapses to form a black hole decades or centuries after its creation in a supernova explosion \citep{2014A&A...562A.137F}. Similarly, \citet{2014ApJ...780L..21Z} proposed a comparable collapse mechanism, but happening in the seconds or minutes after the supramassive neutron star or magnetar is formed in a binary neutron star merger, coincident with a short GRB. \citet{2015MNRAS.450L..71F} have proposed that FRBs are generated by isolated neutron stars whose collapse is triggered by dark matter capture in the neutron star core. 

In almost all cases, the neutron star is not associated with any other observable stable body. In the case of a flare or collapse after birth in a supernova or binary neutron star merger, the FRB might be associated with multi-wavelength emission either in the form of an X-ray flare from a magnetar as is observed in our own Galaxy \citep{2017ARA&A..55..261K}, the multi-wavelength emission from a supernova such as an optical or radio afterglow \citep{2017ApJ...841...14M}, or the prompt emission from a binary neutron star merger such as a short GRB \citep{2014ApJ...780L..21Z}. For a young magnetar ejecta model, the supernova that created the magnetar may also produce an X-ray or $\gamma$-ray afterglow \citep{2019arXiv190201866M}.

\subsubsection{Interacting neutron star models}\label{sec:interactingNS}

Additionally, several models explaining FRBs invoke the interaction between a neutron star and its environment or a less massive orbiting body. In these cases, the FRB emission is generated in the neutron star magnetosphere or through a triggered reaction from the interaction of the two bodies. 

Similar to the theories involving isolated neutron stars, many such theories involve relatively normal rotationally powered neutron stars in other galaxies. \citet{2009AstL...35..241E} propose that FRBs are generated by magnetic reconnection of the neutron star after being struck by an energetic supernova shock and \citet{2017ApJ...836L..32Z} invoke a `cosmic comb' of fast-moving plasma hitting the magnetosphere of a neutron star, which triggers radio emission. Close approach between a neutron star and a supermassive black hole \citep{2017A&A...598A..88Z} or a pair of neutron stars in central stellar clusters of galactic nuclei \citep{2017arXiv170102492D} have also been proposed. 

In several models, FRBs are produced as the result of accretion onto a neutron star. \citet{2018arXiv180602352V} invoke magnetic reconnection after a magnetar accretes dark matter. \citet{2018MNRAS.478.4348I} proposed that FRBs are created as a neutron star accretes ionized plasma blown off of another body in a close approach, and \citet{2016ApJ...823L..28G} propose that FRBs are generated as a neutron star accretes material from a white dwarf companion that has overflowed its Roche lobe.
Neutron stars interacting with small bodies such as comets or asteroids are also a common theme: e.g., neutron stars traveling through asteroid belts \citep{2016ApJ...829...27D}, asteroids or comets impacting the surface of the neutron star \citep{2015ApJ...809...24G} or rocky bodies orbiting a neutron star within the magnetosphere \citep{2014A&A...569A..86M}. Finally, \citet{2014MNRAS.442L...9L} proposed that a magnetically powered hyperflare from a magnetar is released and then interacts with the surrounding medium to produce an FRB in the forward shock.

\subsubsection{Colliding neutron star models}\label{sec:collidingNS}

Lastly, a few neutron star theories predict that an FRB pulse is generated at the time of collision between a neutron star and another compact object. \citet{2013ApJ...768...63L} predicts an FRB from the precursor wind of a binary neutron star merger, \citet{2013PASJ...65L..12T} predicts an FRB from the magnetic braking associated with the same event, and \citet{2018PASJ...70...39Y} predicts FRBs from a neutron star produced in the merger. An FRB from coherent curvature radiation in a binary neutron star merger has also been predicted by \citet{2016ApJ...822L...7W}. \citet{2017arXiv170102492D} take this argument one step further and predict FRBs from binary neutron star collisions only in or near the center of densely packed stellar clusters in galactic nuclei, and \citet{2015PhRvD..91b3008I} theorizes that FRBs are generated in the collision between a neutron star and a dense axion star. Alternatively, \citet{2017arXiv171203509L} proposes that FRBs are produced in neutron star -- white dwarf collisions. 

\subsection{Black hole progenitors}

Although not as numerous as theories involving neutron stars, several theories have also been put forward proposing black holes as the engines of FRB production. Even before the identification of FRBs as a source class, \citet{1977Natur.266..333R} predicted observable millisecond-duration radio pulses from evaporating black holes both in the Galaxy and from other galaxies. 

Black holes interacting with their surrounding environment have also been proposed. \citet{2017A&A...602A..64V} predict FRBs from the interaction between the jet of an accreting active galactic nucleus and the surrounding turbulent medium. Similarly, \citet{2017arXiv170900185D} propose a model where a Kerr black hole produced from the collapse of a supramassive neutron star interacts with the surrounding environment to produce multiple repeating FRBs. Stellar mass black holes in binaries have been proposed to produce FRBs by \citet{2018arXiv181111146Y} through collisions of clumps in the jet produced during accretion.

Collisional progenitor theories involving black holes are limited since binary black hole mergers are thought to produce little or no emission in the electromagnetic spectrum. However, \citet{2016ApJ...827L..31Z} proposes that a binary black hole merger where one or both of the black holes carries charge could produce an FRB pulse at the time of coalescence. Additionally, \citet{2017arXiv170405931A} predict the production of an FRB through magnetic reconnection in the event of collisions between primordial black holes and neutron stars in galaxy dark matter halos, and \citet{2015ApJ...814L..20M} predict double-peaked FRBs as a precursor to some black hole--neutron star mergers. \citet{2018RAA....18...61L} predict FRBs from the accretion disk produced after a black hole -- white dwarf collision.

In the progenitor models above, which all invoke black hole engines for FRB emission, no additional observable emission is predicted either in the radio band or in other parts of the electromagnetic spectrum. Even black hole mergers are expected to be electromagnetically weak and in the progenitor theories included here the radio pulse is the only observable electromagnetic emission predicted from the interaction or merger. 

\subsection{White dwarf progenitors}

Only two models currently exist for the production of an FRB from one or more white dwarfs. The model of \citet{2016ApJ...823L..28G} mentioned in \S \ref{sec:interactingNS} predicts an FRB from the accretion of material from a Roche-lobe-filling white dwarf onto a neutron star.  White dwarfs alone have difficulty accounting for the energy budget required to generate a bright millisecond radio pulse visible at Mpc or Gpc distances. \citet{2016ApJ...830L..38M} has predicted an FRB from the accretion-induced collapse of a white dwarf where the burst is produced in the strong shock from the explosion ejecta colliding with the circum-stellar medium. \citet{2013ApJ...776L..39K} also predict that a single FRB could be produced at the polar cap of a massive white dwarf formed in a binary white dwarf merger. 

In the cases mentioned above, the FRB might also be associated with optical or radio synchrotron emission produced in the expanding ejecta from the stellar collapse or merger. However, these signatures may be too faint to detect in other galaxies as the energy budget for white dwarfs is much lower than that of typical neutron stars.

\subsection{Exotic progenitors}

There are a number of models for FRBs that do not neatly fall into the categories listed above. The only Galactic model currently proposed is that FRBs originate in activity from Galactic flare stars \citep{2014MNRAS.439L..46L} and that the excess DM from the FRB is accrued in the ionized stellar corona. All other theories propose an extragalactic origin and invoke rare or exotic phenomena to generate FRB pulses.

Some of these exotic models still feature dense compact objects and theorize that, for example, an FRB is generated when a primordial black hole explodes back out as a white hole \citep{2014PhRvD..90l7503B} or that the interaction between a strange star (a star made of strange quarks) and a turbulent wind might produce FRBs \citep{2018ApJ...858...88Z}. Others have proposed that the collapse of a strange star to form a black hole could generate an FRB similar to the model for a neutron star by \citet{2014A&A...562A.137F}, or that an isolated neutron star collapsing to form a quark star in a `quark nova' could produce a millisecond radio pulse \citep{2016RAA....16...80S}.

Still other models are arguably even more exotic, theorizing that FRBs come from superconducting cosmic strings \citep{2008PhRvL.101n1301V,2017EPJC...77..720Y,2018PhRvD..97b3022C}, the decay of cosmic string cusps \citep{2015AASP....5...43Z,2017arXiv170702397B}, superconducting dipoles either in isolation or orbiting around supermassive black holes \citep{2017ApJ...844...65T}, or the decay of axion miniclusters in the interstellar media of distant galaxies \citep{2015JETPL.101....1T}. Both \citet{2016PhRvD..93b3001R} and \citet{2018MNRAS.475..514H} theorize that clusters of molecules in other galaxies could produce FRBs: from cavitons in a turbulent plasma excited by a jet \citep{2016PhRvD..93b3001R} or through maser-like emission known as Dicke superradiance \citep{2018MNRAS.475..514H}. It has even been proposed that FRBs are the signatures of beamed emission powering light sails of distant spacecraft \citep{2017ApJ...837L..23L}.

\subsection{Differentiating between progenitor models}

A much larger sample of FRBs, with well characterized burst properties and robustly identified hosts, is needed to differentiate between the dozens of proposed progenitor theories described above.

CHIME and other wide-field FRB discovery machines will provide a large sample in the coming years, but it is also important to have detailed characterization of bursts -- e.g. full polarimetric information and time resolution that is not limited by instrumental smearing.  The shortest-possible timescale for FRB emission is currently poorly constrained.  It is also important to explore the detectability and properties of FRBs across the full possible range of radio frequencies and to continue to search for prompt multi-wavelength and multi-messenger counterparts.  Repeating FRBs provide a practical advantage for detailed characterisation via follow-up observations, but detailed characterization of the properties of apparently non-repeating FRBs is also required.  This means that real-time voltage buffers are highly valuable.

The statistics provided by a sample of hundreds to thousands of FRBs can better quantify how common repeaters are, and their range of activity level.  Through sheer statistics, it may be possible to convincingly show that there are distinct populations of repeaters and non-repeaters -- as opposed to a wide spectrum of activity levels from a population of FRBs that are all capable of repeating, in principle.  The distribution of dispersion measures will go some way towards quantifying the spatial distribution of FRBs, but this is still complicated by the unknown host contribution.

ASKAP and other precision-localisation machines will deliver a much larger sample of FRBs with unambiguous host galaxy associations.  The local environment and host galaxy type are powerful diagnostics, and precision localisations also enable deep searches for associated persistent emission from radio to high-energies.

As the distributions of FRB properties become better known, this will better inform observational strategies that optimize discovery rate, and it may even lead to the discovery of new FRB-like signals by exploring different areas of parameter space.

\section{Summary and conclusions}\label{sec:summary}

In this review we have aimed to capture the state of the FRB field as it stands at the beginning of 2019, with exciting prospects just around the corner. We have highlighted the major results from the past decade and summarized our current knowledge of FRBs and their properties. 
With a rapidly growing population of known sources, and more precision localizations on the near horizon, we expect to learn a lot more in the coming years.  Maximizing the information that can be gleaned from each FRB -- e.g. polarimetric properties, rotation measure, temporal structure -- will also continue to provide valuable clues.  Another critical piece of work in the coming years will be to fully understand our telescope and analysis systematics, in order to quantify incompleteness and biases in FRB searches. 

New FRB-finding machines are coming on-line with the first light of ASKAP, CHIME, APERTIF, and MeerKAT in 2018 and a enormous number of FRB discoveries expected in the coming years. These and other instruments already operating around the world -- such as UTMOST, Parkes, GBT, Arecibo, LOFAR, and the VLA -- are expected to find possibly hundreds of FRBs per year going forward. As the population of FRBs continues to grow we may expect to learn more about whether sub-populations of FRBs exist in different areas of the parameter space. Undoubtedly, as new interesting FRB observations are published, more theories about FRB progenitors and emission will emerge to explain what we see. New observations that may be particularly fruitful for theorists may be the discovery of several more repeaters in the next 100+ FRB discoveries, the detection of periodicity from any repeaters in the population, the presence of similar spectral structure in a large number of FRBs, and the discovery of FRB pulses at much higher or lower radio frequencies. 

\section{Predictions for 2024}\label{sec:predictions}

Looking back five years from the current state of the
field at the time of writing places us at the time of the announcement of four FRBs by \citet{2013Sci...341...53T}. At that time, predicting the current state of the field in 2019 would have been extremely challenging. One could argue, however, with the explosion of discoveries now taking place, that extrapolating five years into the future will be even more challenging. It is in this light, that we each advance our predictions for the field in the year 2024.

\subsection{EP}

It is hard to predict the FRB landscape in 2024 with any certainty. Since beginning this review only a year ago the field has already changed so much that multiple revisions were required. The only thing I can be absolutely certain about is that FRBs will continue to puzzle and delight us in new and exciting ways. I predict the population will be of order several thousands of sources dominated by the discoveries from wide-field interferometers, particularly from CHIME, but also from Apertif, ASKAP, the LWA, MeerKAT, and UTMOST. The community studying these many discoveries will also be much larger than it is now, and it is my hope that this review is useful for them. Single dishes with limited field of view and lower discovery power will still play a critical role in the field by helping us to understand the high and low radio frequency properties of FRBs. I anticipate that FRB emission will be discovered across several decades of radio frequency. By 2024, I predict that FAST will have detected an FRB at $z > 2$ and we will have found an FRB at $\sim$Mpc distances in a relatively local galaxy. Observationally, FRB polarization will be one of the most important properties we measure for a new source, and FRB rotation measures (and their changes over time for repeaters) will give us the greatest clues about the environments where FRBs reside. If FRBs are indeed produced by several source classes, I predict that RM will be one of the most important properties in distinguishing between FRB source types. The type of host galaxy for an FRB will also be an important indicator and by 2024 I expect that at least 50 FRBs will have identified host galaxies. The future is certainly bright, and there is no doubt that there will be plenty of surprises to keep both observers and theorists busy!

\subsection{JWTH}

I predict that observational efforts to detect FRBs and understand their origin(s) will continue to grow at a rapid pace, and will only be lightly constrained by the collective imagination of the community and its ability to acquire funding.  I see a strong role for both wide-field FRB-discovery machines, as well as high-sensitivity, high-resolution (spatial, time and frequency) follow-up initiatives.  New instruments, techniques and ever-expanding computational power will extend the search to new areas of parameter space, and will lead to surprises: e.g. (sub)-microsecond FRBs, FRBs at apparently enormous distance ($z > 3$), and FRBs only detectable at very high ($> 10$\,GHz) or low ($< 100$\,MHz) radio frequencies.  As we push into new parameter space it may become clear that there are many types of FRB sources, with fundamentally different origins (black hole vs. neutron star) and energy sources (magnetic, rotational or accretion).  We'll have to come up with new names that better link to an underlying physical process as opposed to an observed phenomenon; the community may even split into groups that specialize on specific source classes.  Low-latency follow-up of explosive transients like superluminous supernovae and long gamma-ray bursts at high radio frequencies ($> 10$\,GHz) will allow us to capture newly born repeating FRB sources.  At the same time, high-cadence monitoring of repeating FRBs will allow us to trace their evolution with time.  This includes the intrinsic source activity and energetics, as well as how evolving lensing effects, DM and RM probe the dynamic local environment.  I also think that very long baseline interferometry will continue to be an important tool not only for precision localization, but for constraining the size and evolution of FRB counterpart afterglows and/or nebulae.  Lastly, since it seems likely to me that we will have an observed population of $> 1000$ FRBs to work with by 2024, we may be able to start using FRBs to probe the intervening IGM, despite the challenges posed by the inaccuracies in modeling the Galactic foreground and local DM contributions.

Looking further down the road, I predict -- as with pulsars -- that the field will wax and wane, but that every time we think the field is exhausted, a stunning insight will be just around the corner.  See you at the `50 Years of FRBs' IAU Symposium.

\subsection{DRL}

I predict that the FRB sample will be dominated by CHIME discoveries and be at the level of 3000 high significance (S/N~$>10$) sources plus a much larger sample of weaker events. With the advent of sensitive searches in particular by FAST, the DM range of the sample will extend out to $10^4$~cm$^{-3}$~pc.  Repeating FRBs will make up only a small fraction  (1\%) of the sample but that localizations of these sources will have led to redshift determinations for a few dozen FRBs. Nevertheless, augmented by other observations, and detailed modeling, this small sample will have led to the development of an electron density map that is sufficient to be used to infer more meaningful distance constraints on the non-localized sources than is currently possible. Repeating FRBs will be linked to magnetars associated with central AGNs of their host galaxies, but far less will be known about the origins of non-repeating sources.

\begin{acknowledgements}
We thank Liam Connor, Griffin Foster, Evan Keane, Joeri van Leeuwen, Kenzie Nimmo, Vikram Ravi, Laura Spitler, Dan Stinebring, Samayra Straal, Dany Vohl, and Bing Zhang for their feedback on draft sections of this review. We thank Wael Farah for additional data from FRB 170827, and Hsiu-Hsien Lin, Kiyo Masui, and Cees Bassa for data reduction help for FRB 110523. EP acknowledges funding from an NWO Veni Fellowship and from the European Research Council under the European Union’s Seventh Framework Programme (FP/2007-2013)/ERC Grant  Agreement No. 617199. JWTH acknowledges funding from an NWO Vidi fellowship and from the European Research Council under the European Union's Seventh Framework Programme (FP/2007-2013) / ERC Starting Grant agreement nr.~337062 (``DRAGNET'').
DRL acknowledges support from the Research Corporation for Scientific Advancement and the National Science Foundation awards AAG-1616042, OIA-1458952 and PHY-1430284.

\newpage

\appendix
\section{Glossary}

\begin{table}[h!]
\centering
\begin{tabular}{ll}
Variable & Definition \\
\hline
$\alpha$ & Spectral index \\
$\beta$ & Luminosity power law index \\
$\gamma$ & Source counts index \\
$\lambda$ & Observing wavelength (m) \\
$\nu$ & Observing frequency (MHz) \\
$\Delta \nu$ & Observing bandwidth (MHz) \\
$\Delta \nu_\mathrm{Scint}$ & Scintillation bandwidth (MHz) \\
$\sigma_s$ & Root mean square fluctuations in the time series \\
$\tau_\mathrm{s}$ & Scattering timescale (s) \\
$\Theta$ & Position angle ($^\circ$) \\
$\Omega_m$ & Energy density of matter \\
$\Omega_\Lambda$ & Energy density of dark energy \\
$B(\ell)_{||}$ & Magnetic field parallel to the line of sight (G) \\
$D$ & Telescope diameter (m) \\
DM & Dispersion measure (cm$^{-3}$ pc) \\
DM$_\mathrm{E}$ & Dispersion measure excess (cm$^{-3}$ pc) \\
DM$_\mathrm{IGM}$ & Dispersion measure from the intergalactic medium (cm$^{-3}$ pc) \\
DM$_\mathrm{MW}$ & Dispersion measure from the Milky Way (cm$^{-3}$ pc) \\
$d_L$ & Luminosity distance (Gpc) \\
$\mathcal{F}$ & Fluence (Jy ms) \\
$G$ & Antenna gain (K Jy$^{-1}$) \\
$L$ & Luminosity (W) \\
$N_\mathrm{DM}$ & Number of DM trials \\
$N_t$ & Number of time trials \\
$N_\nu$ & Number of frequency channels \\
$n_e$ & Electron number density (cm$^{-3}$) \\
$\mathcal{R}$ & FRB event rate (FRBs sky$^{-1}$ day$^{-1}$) \\
RM & Rotation measure (rad m$^{-2}$) \\
SM & Scattering measure (kpc m$^{-20/3}$) \\
S/N & Signal-to-noise ratio \\
$S_\mathrm{peak}$ & Peak flux density (Jy) \\
$T_\mathrm{B}$ & Brightness temperature (K) \\
$T_\mathrm{sys}$ & System temperature (K) \\
$\Delta t$ & Dispersive delay (s) \\
$\Delta t_\mathrm{DM}$ & Dispersive delay across an individual frequency channel (s) \\
$\Delta t_\mathrm{DMerr}$ & Dispersive delay due to dedispersion at a slightly incorrect DM (s) \\
$t_\mathrm{samp}$ & Sampling time (s) \\
$x(z)$ & Ionization fraction as a function of redshift \\
$W$ & Pulse width (s) \\
$W_\mathrm{eq}$ & Equivalent width (s) \\
$W_\mathrm{int}$ & Intrinsic pulse width (s) \\
\end{tabular}
\end{table}

\end{acknowledgements}

\bibliographystyle{natbib}
\bibliography{AARV_FRB}   

\end{document}